\documentclass[12pt,a4paper]{article}  
 \usepackage[skins,theorems]{tcolorbox}

\newtcolorbox{whitebox}{colback=white,colframe=black,boxrule=0.5mm,arc=4mm,auto outer arc}

\tcbset{highlight math style={enhanced,
  colframe=red,colback=white,arc=0pt,boxrule=1pt}}
  \usepackage[bookmarksopen, bookmarksnumbered, bookmarksopenlevel=2]{hyperref}
  \usepackage{tikz}
  \usepackage{tikz-3dplot}
  \usepackage{multirow}
 \usetikzlibrary{calc}
 \usetikzlibrary{decorations} %
 \usepackage[UKenglish]{babel}
 \usepackage[toc,page]{appendix}
 \usepackage{amsmath}
 \usepackage{amssymb}
 \usepackage{graphicx}
 \usepackage{hhline}
 \usepackage[bf]{caption}
\usepackage{cite}
\usepackage[vcentermath]{youngtab}
\usepackage{geometry}
\usepackage{slashed}
\usepackage{color}
\usepackage{stackrel}
\usepackage{tikz-cd} 
\usepackage{tkz-euclide}
\usepackage{cancel} 
\usepackage{subcaption}
\usepackage[normalem]{ulem}
\usepackage{mdframed}
\usepackage{empheq}
\usepackage{arydshln}
\newenvironment{eqn*}{\begin{equation*}\begin{aligned}}{\end{aligned}\end{equation*}\noindent}
\newtheorem{conjecture}{Conjecture}
\newtheorem{fact}{Fact}

\newcommand{\bqa}{\begin{eqnarray}}
\newcommand{\eqa}{\end{eqnarray}}

\def\del{\partial}

\hypersetup{
    pdftitle={},
    pdfauthor={},
    pdfsubject={}
}
\numberwithin{equation}{section}
\numberwithin{table}{section}

\makeatletter

\definecolor{BF}{HTML}{f903d7}

\newcommand{\be}{\begin{equation}}
\newcommand{\ee}{\end{equation}}
\newcommand{\beq}{\begin{equation}}
\newcommand{\eeq}{\end{equation}}
\newcommand{\ba}{\begin{aligned}}
\newcommand{\ea}{\end{aligned}}

\newcommand{\bea}{\begin{eqnarray}}
\newcommand{\eea}{\end{eqnarray}}

\newcommand{\cE}{\mathcal{E}}

\newcommand{\cN}{\mathcal{N}}

\newcommand{\cB}{\mathcal{B}}

\newcommand{\cS}{\mathcal{S}}

\newcommand{\cV}{\mathcal{V}}

\newcommand{\cM}{\mathcal M}
\newcommand{\cQ}{\mathcal Q}

\newcommand\bi{\begin{itemize}}
\newcommand\ei{\end{itemize}}

\renewcommand{\a}{{\alpha}}

\newcommand{\g}{{\gamma}}

\def\Im{\mathop{\mathrm{Im}}\nolimits}

\def\unit{{1\kern-.65ex {\rm l}}}
\def\1{{1\kern-.65ex {\rm l}}}

\def\ii{{\rm i}}

\newcount\hour \newcount\minute
\hour=\time \divide \hour by 60
\minute=\time
\def\now{%
\ifnum \hour<13
  \ifnum \hour=0 \advance \hour by 12 \number\hour:\else \number\hour:\fi%
     \ifnum \minute<10 0\fi%
     \number\minute%
\ A.M.%
\else \advance \hour by -12 \number\hour:%
  \ifnum \minute<10 0\fi%
  \number\minute%
  \ P.M.%
\fi%
}

\makeatother

\begin{document}

\begin{flushright}
{\tt\normalsize ZMP-HH-25/5}\\
\end{flushright}

\vskip 40 pt
\begin{center}
{\large \bf
  Emergent Strings in Type IIB Calabi--Yau Compactifications
} 

\vskip 11 mm

Bj\"orn Hassfeld,${}^{1}$ Jeroen Monnee,${}^{2}$ Timo Weigand,${}^{2,3}$ and Max Wiesner${}^{2}$

\vskip 11 mm
\small ${}^{1}$\textit{Institute for Theoretical Physics, Heidelberg University,
Philosophenweg 19, \\ 69120 Heidelberg, Germany} \\[3 mm]
\small ${}^{2}$\textit{II. Institut f\"ur Theoretische Physik, Universit\"at Hamburg, Notkestrasse 9,\\ 22607 Hamburg, Germany} \\[3 mm]
\small ${}^{3}$\textit{Zentrum f\"ur Mathematische Physik, Universit\"at Hamburg, Bundesstrasse 55, \\ 20146 Hamburg, Germany  }   \\[3 mm]

\end{center}

\vskip 7mm

\begin{abstract}
We study infinite distance limits in the complex structure moduli space of Type IIB compactifications on Calabi--Yau threefolds, in light of the Emergent String Conjecture. We focus on the so-called type II limits, which, based on the asymptotic behaviour of the physical couplings in the low-energy effective theory, are candidates for emergent string limits.  However, due to the absence of
 Type IIB branes of suitable dimensionality,
 the emergence of a unique critical string accompanied by a tower of Kaluza--Klein states has so far remained elusive. 
 By considering a broad class of type II$_b$ limits, corresponding to so-called Tyurin degenerations, and studying the asymptotic behaviour of four-dimensional EFT strings in this geometry, we argue that the worldsheet theory of the latter describes a unique critical heterotic string on $T^2\times\mathrm{K3}$ with a gauge bundle whose rank depends on $b$. In addition, we establish the presence of an infinite tower of BPS particles arising from wrapped D3-branes by identifying a suitable set of special Lagrangian 3-cycles in the geometry.
The associated BPS invariants are conjectured to be counted by generalisations of modular forms.
 As a consistency check, we further show that in special cases mirror symmetry identifies the EFT strings with the well-understood emergent string limits in the K\"ahler moduli space of Type IIA compactifications on K3-fibred Calabi--Yau threefolds. 
Finally, we discuss the implications of the Emergent String Conjecture for type II limits which do not correspond to Tyurin degenerations, and  
predict new constraints on the possible geometries of type II degenerations which resemble those arising in the Kulikov classification of degenerations of K3 surfaces.  
\end{abstract}

\vfill

\thispagestyle{empty}
\setcounter{page}{0}
 \newpage
\tableofcontents
\vspace{25pt} 
\setcounter{page}{1}
\section{Introduction}
The exploration of moduli spaces in string theory has long been a cornerstone in understanding the interplay between low-energy physics, geometry, and quantum gravity. These spaces parametrise families of low-energy effective theories (EFTs), and their infinite-distance limits often reveal profound insights into the breakdown of the effective description and the emergence of new fundamental degrees of freedom. 
 Asymptotic regions in the moduli space furthermore allow for enhanced perturbative control because they typically correspond to weak coupling regimes. This explains why a classification of the possible infinite distance limits and their associated perturbative effective theories is of relevance also
 for applications of string theory to cosmology or particle physics, and a natural starting point for understanding the space of quantum gravity more generally.

Within this context, the Swampland program has unveiled universal principles that distinguish those low-energy effective gravitational theories that have a UV completion from those that do not \cite{Vafa:2005ui}, see \cite{Brennan:2017rbf,Palti:2019pca,vanBeest:2021lhn,Grana:2021zvf,Agmon:2022thq} for reviews. One of the pillars of this program is the Distance Conjecture \cite{Ooguri:2006in}, which asserts that every infinite-distance limit in the moduli space of a consistent low-energy effective theory coupled to gravity is necessarily accompanied by an exponentially light tower of states. This universal behaviour signals the breakdown of the original EFT, and suggests the emergence of a dual effective description. A central challenge, then, lies in characterising such limits both from a mathematical perspective --- in terms of the geometric degeneration of the underlying compactification manifold ---  as well as a physical perspective --- in terms of the new duality frame and the corresponding fundamental degrees of freedom. 

A proposal which precisely addresses this question is the Emergent String Conjecture \cite{Lee:2019oct}. This conjecture posits that every infinite-distance limit in the moduli space of a consistent low-energy effective theory coupled to gravity corresponds to either a decompactification limit, in which the leading tower of light states is a Kaluza--Klein tower (or multiple such towers), or an emergent string limit, in which the leading tower of light states corresponds to the excitations of a unique, critical, asymptotically weakly coupled string. The Emergent String Conjecture has been substantiated and tested in various corners of the string/M-theory landscape, namely in the K\"ahler moduli space of F/M/IIA-theory in 4d/5d/4d \cite{Lee:2018urn,Lee:2019oct,Lee:2019apr,Rudelius:2023odg}, in the complex structure moduli space of F-theory in 8d \cite{Lee:2021qkx,Lee:2021usk,Chen:2024cvc} and 6d \cite{Alvarez-Garcia:2023gdd,Alvarez-Garcia:2023qqj} and of M-theory in 5d \cite{Alvarez-Garcia:2021pxo}, in the 4d ${\mathcal{N}=2}$ hypermultiplet moduli space of Type II theories \cite{Marchesano:2019ifh,Baume:2019sry,Blumenhagen:2023yws}, in M-theory on $G_2$ manifolds \cite{Xu:2020nlh}, in 4d $\mathcal{N}=1$ F-theory \cite{Lee:2019jan,Klaewer:2020lfg}, and in non-supersymmetric \cite{Basile:2022zee} and non-geometric setups \cite{Aoufia:2024awo}. 
Recent works have additionally provided bottom-up arguments for the Emergent String Conjecture, relying on the properties of black hole thermodynamics \cite{Basile:2023blg,Basile:2024dqq,Herraez:2024kux}, the species entropy \cite{Cribiori:2023ffn}, gravitational scattering amplitudes \cite{Bedroya:2024ubj}, or by employing the consistency of supergravity strings in 5d $\mathcal{N}=1$ theories \cite{Kaufmann:2024gqo}.\footnote{See also \cite{Calderon-Infante:2024oed} for recent progress on the Emergent String Conjecture in AdS spacetime by studying the holographic dual.} The encouraging evidence emerging from these several corners of the landscape, together with independent bottom-up arguments, reinforce the potential universality of the Emergent String Conjecture as a guiding principle in the study of quantum gravity. 

Despite these advances, an important puzzle remains: From the top-down perspective, the Emergent String Conjecture has mainly been studied in setups where known string dualities can be leveraged to argue for the emergence of tensionless, critical strings in certain infinite distance limits in the moduli space. In these cases, the emergent string  can typically be identified with a higher-dimensional brane that wraps a suitable cycle in a compact manifold, thereby leading to a BPS string in the lower-dimensional theory. This is strikingly different for limits in the vector multiplet moduli space of Type IIB compactifications on Calabi--Yau threefolds. In the low-energy effective 4d $\mathcal{N}=2$ supergravity theory of these models, the vector multiplet moduli are encoded in the complex structure deformations of the Calabi-Yau. In this context there are no branes which, once wrapped on suitable cycles on the Calabi--Yau threefold, give rise to BPS strings with tension controlled by the moduli in the \textit{vector} multiplets. Identifying emergent, critical strings in asymptotic regions of the vector multiplet moduli space of these theories therefore requires a different approach than the ones underlying previous tests of the Emergent String Conjecture. The purpose of this work is exactly to provide such an alternative approach. As we will see, it will rely crucially on the \emph{geometric} properties of the underlying Calabi--Yau threefold in infinite distance limits of the complex structure. 

A potential solution to the above conundrum would be that there simply are no emergent string limits in the vector multiplet moduli space of Type IIB string theory, and instead all limits are decompactification limits. However, already from the perspective of the low-energy effective action there are clear signals that such emergent string limits do in fact exist. First, there are the mirror duals of the emergent string limits of Type IIA Calabi--Yau threefold compactifications identified in~\cite{Lee:2019oct}. These latter limits can be realised in the large volume regime of the mirror threefold which, via mirror symmetry, map to the large complex structure regime on the original threefold. Since the low-energy effective theory is invariant under mirror symmetry, there are hence limits in the large complex structure regime in which the relevant couplings behave as expected for emergent string limits. Moreover, there are also limits in the complex structure moduli space that do not intersect the large complex structure point, and which are also expected to feature emergent strings. A classic example are K-points in the one-parameter complex structure moduli spaces \cite{Joshi:2019nzi}. As shown in~\cite{vandeHeisteeg:2022btw,vandeHeisteeg:2023ubh}, in these limits the species scale given by the genus-one free energy behaves as expected for an emergent string limit, although the nature of the emergent string is unclear from a purely low-energy effective action point of view. 

A natural set of candidate objects in the 4d $\mathcal{N}=2$ supergravity theory that may give rise to emergent strings are the $\frac12$-BPS strings corresponding to the EFT strings of \cite{Lanza:2020qmt,Lanza:2021udy}. Indeed, it has been established that these cosmic-string-like solutions of the 4d $\mathcal{N}=2$ supergravity can become tensionless in infinite-distance limits in the complex structure moduli space. 
More generally, the structure of the low-energy effective theory is well-understood using techniques from asymptotic Hodge theory \cite{Grimm:2018ohb,Grimm:2018cpv,Grimm:2019wtx,Gendler:2020dfp,Bastian:2020egp,Calderon-Infante:2020dhm,Grimm:2021ikg,Palti:2021ubp,Grimm:2021vpn,Grimm:2022xmj,Bastian:2023shf}, building on the seminal works of \cite{schmid,CKS}, see also \cite{vandeHeisteeg:2022gsp,Monnee:2024gsq} for reviews and further references. In particular, this ``algebraic'' machinery provides detailed information about the physical couplings such as the moduli space metric and gauge-kinetic functions, together with the intricate network of finite- and infinite-distance limits possible in the complex structure moduli space. 
 Notably, a candidate for emergent string limits are the limits of type II in the nomenclature of \cite{Grimm:2018ohb}. In these limits, there can be asymptotically weakly coupled $\frac12$-BPS particles arising from D3-branes wrapped on special Lagrangian 3-cycles in the Calabi--Yau threefold whose mass-squared tends to zero at the same rate as the tension of the EFT string. If the existence of these particles can be established, they are natural candidates for the tower of Kaluza--Klein states that should accompany an emergent string limit, as required by the Emergent String Conjecture. Thus, already from the ``algebraic'' characterization of the couplings in the low-energy effective theory alone, we find strong hints that type II limits in the complex structure moduli space of Type IIB should indeed correspond to emergent string limits. 

However, establishing the validity of the Emergent String Conjecture in this setting requires information beyond the one accessible from a purely algebraic perspective. First, it is crucial to establish the \textit{criticality} of the emergent string, as well as its uniqueness. Furthermore, demonstrating the existence of a \textit{tower} of BPS particles requires a careful analysis of the BPS indices associated with special Lagrangian 3-cycles, which is a notoriously complicated problem \cite{Banerjee:2022oed}. Finally, to properly unveil the right duality frame that emerges in the infinite-distance limit, it is paramount to understand the worldsheet theory of the emergent string. To tackle these challenges, it is clear that a proper \textit{geometric} characterization of the underlying Calabi--Yau manifold in type II limits is of utmost importance. This is similar in spirit to the geometric analysis of complex structure limits for Weierstrass models \cite{Lee:2021qkx,Lee:2021usk,Alvarez-Garcia:2023gdd,Alvarez-Garcia:2023qqj} in the context of F-theory compactifications. To this end, we will focus our attention on a particular class of type II limits which are realised by a so-called Tyurin degeneration \cite{Tyurin:2003}, of which the aforementioned K-points \cite{Joshi:2019nzi} are a special case.\footnote{In Section \ref{sec:beyondTyurin} we address type II limits which do not correspond to Tyurin degenerations.} For an illustration, see Figure \ref{fig:Tyurin}. These are limits in the complex structure moduli space in which the Calabi--Yau threefold splits into two components that intersect over a K3 surface, such that complete control over the underlying geometry is ensured. In the context of mirror symmetry, these types of degenerations have been studied in the physics literature in  \cite{doran2016mirrorsymmetrytyurindegenerations,Doran:2024kcb}, but we stress that they can also occur away from the large complex structure point, as exemplified by the K-point.

The additional data obtained from the concrete geometries realised at a type II degeneration is the main input that we use in this work to study emergent string limits in the vector multiplets  of Calabi--Yau threefold compactifications of Type IIB string theory. In the following, we provide an overview over the main results obtained in this paper.

\subsection*{Summary of Results}

\paragraph{Criticality and worldsheet theory of the emergent EFT string.}
Given the geometry obtained at a Tyurin degeneration, the main idea is that the intersection locus, $Z$, of the two components into which the degenerate Calabi--Yau threefolds splits contains localised degrees of freedom which constitute both the sought-after emergent string and also the required particle towers. In this picture, the string as such is a gravitational $\frac12$-BPS solution in the sense of the EFT strings of \cite{Lanza:2020qmt, Lanza:2021udy}. The zero modes on its worldsheet are obtained by reducing the Type IIB 2-form and 4-form fields along $Z$. The key observation is that $Z$ contains localised 2-forms into which the Type IIB fields can be expanded. Since the intersection locus forms a K3 surface, these localised 2-forms can be decomposed into elements of the transcendental lattice (which in so-called type ${\rm II}_b$ limits is of rank $(2+b)$) and of the polarisation lattice of the K3 surface. The resulting string zero modes can be identified with the spectrum on a \emph{critical} heterotic string. Finally, the kinetic terms of the string localised zero modes allow us to distinguish between free and interacting modes. From this one deduces that the target space of the asymptotic string $\sigma$-model is 
    \begin{equation}
        \mathcal{M}=T^2\times \left(T^2\to\mathbb{C}\right)\times\mathbb{C}^*\,,
    \end{equation}
    where the middle factor can be understood as a local K3. In addition, the gauge group of the heterotic string, at the generic point on the Coulomb branch, is broken to a rank $b+2$ Cartan subgroup.\footnote{For a Tyurin degeneration the parameter $b$ is restricted by $0\leq b\leq 19$. Additionally, in the special case $b=19$ the worldsheet theory becomes left-right asymmetric and the interpretation of the theory is not as clear; possibly one should instead replace the $(T^2\to\mathbb{C})$ factor with a non-geometric CFT.   }

\paragraph{Tower of BPS states.}
Using the geometry of the Tyurin degeneration, we identify a set of special Lagrangian 3-cycles  $\Gamma_0$, each of which can be viewed as an $S^1$ fibration over a corresponding holomorphic curve $C_0\in H^2(Z)$, where $Z$ denotes again the K3 surface that arises in the Tyurin degeneration. Crucially, we argue that the curves $C_0$ have non-negative self-intersection on $Z$, such that their genera satisfy $g(C_0)\geq 1$. This ensures that the 3-cycles $\Gamma_0$ allow for multi-wrappings of D3-branes, and hence gives rise to  \textit{towers} of light BPS states. Furthermore, in view of the established heterotic dual interpretation, we propose that the BPS indices $\Omega_{\mathrm{BPS}}(n \Gamma_0)$ associated with $n\Gamma_0$ should correspond to the coefficients of a meromorphic Jacobi modular form. The precise statement is formulated in Conjecture \ref{conj-modular}. 

These two points indeed constitute our main findings:

\vspace{2mm}
\begin{mdframed}[backgroundcolor=white,shadow=true,shadowsize=4pt,shadowcolor=black,roundcorner=6pt]
In Type IIB string theory compactified on Calabi--Yau threefolds, a unique emergent, critical heterotic string with perturbative gauge group of rank $2+b$ becomes tensionless at \emph{any} Tyurin degeneration realising a type II$_b$ limit in the complex structure moduli space of the threefold. The string is always accompanied by a tower of BPS states corresponding to winding and momentum modes in accordance with the Emergent String Conjecture.
\end{mdframed}

\paragraph{Mirror symmetry.}
We also compare our findings with previous results concerning emergent strings arising in the K\"ahler moduli space of Type IIA string theory on K3-fibred Calabi--Yau threefolds \cite{Lee:2019oct}. In this setting, the emergent heterotic string arises from NS5-branes wrapping the generic K3-fibre $\widehat{Z}$ of the threefold, and taking the large-volume limit of the base while keeping the volume of the generic fibre finite. By comparing the worldsheet theories at various levels of detail, we find that they have the same massless spectrum, and that the number of free fields is the same if the two underlying K3s $Z$ and $\widehat{Z}$ are mirror to each other. Additionally, we argue that in setups with a perturbative heterotic dual string, the interactions on the two worldsheet theories agree if the underlying Calabi--Yau threefolds are mirror to each other.
Most notably, in a particular setting also considered in \cite{Braun:2017ryx}, a special Lagrangian $T^3$-fibration suitable for applying the SYZ conjecture can be identified and we explicitly show that the geometries on the IIA and IIB side are mapped to each other under mirror symmetry.


\paragraph{Beyond Tyurin degenerations and quantum gravity predictions for geometry.}
Finally, we discuss the question of whether there are more general kinds of type II degenerations which do not correspond to Tyurin degenerations. A simple modification is to instead consider a limit in which the Calabi--Yau threefold splits into two components intersecting over an Abelian surface instead of a K3. In this case, we argue via a similar reasoning as for the Tyurin degeneration that there is again an emergent string, which now corresponds to a critical Type II string. However, we do not delve into the details of the precise worldsheet theory. 

Perhaps of greater interest is to ponder on the possibility of type II limits in which the threefold splits into more than two components. For these to be consistent with the Emergent String Conjecture, we find that the non-vanishing intersections of the components are either all Abelian surfaces with the same complex structure, or all K3 surfaces that are polarised by the same lattice. Intriguingly, this implies that the integer $b$ characterising the type II$_b$ limit is bounded as $0\leq b\leq 19$. The precise statement is formulated in Conjecture \ref{conj1}. 

\subsection*{Structure of the Paper}
The article is organised as follows. In Section \ref{sec:emergent_strings} we start with an algebraic description of type II limits in the complex structure moduli space of a Calabi--Yau threefold, using the methods of asymptotic Hodge theory. Subsequently, we focus on a particular class of such limits realised by a Tyurin degeneration, and discuss their geometric properties. 
By combining these two perspectives, we establish in Section~\ref{sec:BPSstates} the existence of a tower of asymptotically massless BPS states that is accompanied by an asymptotically tensionless EFT string at a scale which is indicative of an emergent string limit. 
In Section \ref{sec:Worldsheet} we perform a detailed analysis of the worldsheet theory of the EFT string becoming tensionless in the Tyurin degeneration limit. In particular, we show that it corresponds to the worldsheet theory of a critical heterotic string, and furthermore identify the target space geometry and rank of the surviving heterotic gauge group. 
In Section \ref{sec:mirror_symmetry} we compare our findings with known results about emergent strings in the K\"ahler moduli space of Type IIA compactifications on Calabi--Yau threefolds. We argue that the two kinds of emergent strings correspond to the same heterotic string if the underlying Calabi--Yau threefolds are mirror dual to each other. We furthermore demonstrate how the two emergent strings are exchanged under mirror symmetry in a particular setting where a special Lagrangian $T^3$-fibre is available and we can apply the SYZ conjecture. 
In Section \ref{sec:beyondTyurin} we discuss how our results might generalise to type II limits which do not correspond to Tyurin degenerations. We argue that if the validity of the Emergent String Conjecture is assumed, this implies new constraints on the kinds of type II limits that can be realised geometrically.

Finally, some background material and technical details are deferred to the appendices. Appendix \ref{app:hodgestr} contains a brief summary of limiting mixed Hodge structures, Appendix \ref{App-MV} gives a brief review of the Mayer--Vietoris long exact sequence, and Appendix \ref{app_MV+CS} describes the geometric interpretation of mixed Hodge structures associated to degenerate varieties using the Mayer--Vietoris and Clemens--Schmid long exact sequences. 

\section{Type II Degenerations of Calabi--Yau Threefolds}
\label{sec:emergent_strings}

We consider the low-energy description of Type IIB string theory compactified on a Calabi--Yau threefold $V$. The resulting theory is a four-dimensional $\mathcal{N}=2$ supergravity theory containing a single gravity multiplet, along with $h^{2,1}(V)$ vector multiplets and $h^{1,1}(V)+1$ hypermultiplets. The moduli space $\mathcal{M}$ parametrised by the vacuum expectation values of the massless scalar fields  factorises into a direct product
\begin{equation}
\label{eq:moduli_space_decomp}
    \mathcal{M} = \mathcal{M}_{\mathrm{V}}\times\mathcal{M}_{\mathrm{H}}\,,
\end{equation}
where $\mathcal{M}_{\mathrm{V}}$ and $\mathcal{M}_{\mathrm{H}}$ denote the vector multiplet moduli space and the hypermultiplet moduli space, respectively. Geometrically, the former parametrises the complex structure deformations of the Calabi--Yau threefold $V$, while the latter incorporates the (complexified) K\"ahler deformations as well as the axio-dilaton. Due to the factorization \eqref{eq:moduli_space_decomp} the vector multiplet moduli space of Type IIB on Calabi--Yau threefolds is classically exact, i.e.~it is protected from loop corrections in $\alpha'$ and the string-coupling $g_s$.  

In this work, we will focus our attention on the gravity and vector multiplet sector. We denote by $u^i$, for $i=1,\ldots, h^{2,1}(V)$, the complex structure moduli forming the scalar fields in the vector multiplets. Furthermore, we denote by $A^I$, for $I=0,\ldots, h^{2,1}(V)$, the U(1) gauge fields residing in the gravity multiplet and the vector multiplets, and set $F^I=\mathrm{d}A^I$. Then the relevant bosonic part of the four-dimensional action is given by
\begin{equation}
\label{eq:action_4d}
    S_4 = \int \frac{1}{2}M_{\mathrm{Pl}}^2R\star1 +G_{i\bar{\jmath}}\,\mathrm{d}u^i\wedge\star\,\mathrm{d}u^{\bar{\jmath}}+\frac{1}{4}\mathcal{I}_{IJ}F^I\wedge\star\,F^J+\frac{1}{4}\mathcal{R}_{IJ}F^I\wedge F^J\,.
\end{equation}
To write down the explicit expressions for the various couplings appearing in \eqref{eq:action_4d}, it is useful to introduce some notation. For any $v,w\in H^3(V)$, we will write
\begin{equation}
\label{eq:pairings}
    (v,w) = \int_V v\wedge w\,,\qquad \langle v,w\rangle = \int_V v\wedge\star\,\bar{w}\,,\qquad \|v\|^2=\langle v,v\rangle\,,
\end{equation}
which respectively denote the symplectic pairing, Hodge inner product, and Hodge norm on $H^3(V)$. Furthermore, we will denote by $\Omega$ the holomorphic (3,0)-form on $V$. In terms of these quantities, the metric $G_{i\bar{\jmath}}$ on $\mathcal{M}_{\mathrm{V}}$ and its corresponding K\"ahler potential $K$ are given by
\begin{equation}\label{eq:metric}
    G_{i\bar{\jmath}} = \partial_i \bar{\partial}_{\bar{\jmath}}K\,,\qquad K=-\log\|\Omega\|^2\,.
\end{equation}
Furthermore, if $\alpha_I,\beta^J\in H^3(V,\mathbb{Z})$, for $I,J=0,\ldots, h^{2,1}$, define an integral symplectic basis, then the gauge-kinetic functions $\mathcal{I}_{IJ}$ and $\mathcal{R}_{IJ}$ are determined by the relations
\begin{equation}
    \langle\alpha_I, \beta^J\rangle = -\mathcal{R}_{IK}\mathcal{I}^{KJ}\,,\qquad \langle \beta_I,\beta^J\rangle = -\mathcal{I}^{IJ}\,.
\end{equation}
In particular, let us emphasise that all relevant couplings in the low-energy effective action \eqref{eq:action_4d} are determined by $\Omega$ and the Hodge inner product defined in \eqref{eq:pairings}. \\

\begin{figure}[t]
\centering
\begin{tikzpicture}[scale=0.7]

\draw[very thick](2,5) to[out=20,in=160] (10,5);
\draw[very thick](4,3) to[out=80,in=230] (7,8);

\node at (5.15,5.75)[circle,fill,inner sep=1.5,blue]{};

\tkzLabelPoint[left](2,5){ $u^1=0$};
\tkzLabelPoint[below](4,3){ $u^2=0$};

\draw[thick,dashed] (5.4,5) ellipse[x radius = 7cm, y radius = 3.5cm, start angle = 30,
end angle = 150];

\fill[gray,opacity=0.1] (5.4,5) ellipse[x radius = 7cm, y radius = 3.5cm, start angle = 30,
end angle = 150];

\tkzLabelPoint[right](10,5){ $\Delta_1$};
\tkzLabelPoint[right](7,8){ $\Delta_2$};
\tkzLabelPoint[above left](5.15,5.75){ \textcolor{blue}{$\Delta_{12}$}};
;
\end{tikzpicture}
\caption{A local patch in a complex two-dimensional vector multiplet moduli space $\mathcal{M}_{\mathrm{V}}$ containing two divisors $\Delta_1$ and $\Delta_2$ which intersect at a point $\Delta_{12}$. }
\label{fig:divisors}
\end{figure}
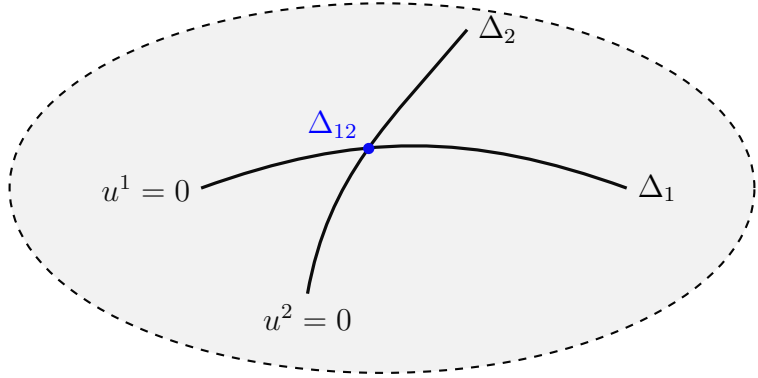

Throughout this work, we are concerned with studying infinite distance limits in the vector multiplet moduli space. Geometrically, such limits correspond to degenerations of the underlying Calabi--Yau threefold $V$. The set of points in $\mathcal{M}_{\mathrm{V}}$ at which $V$ becomes singular is referred to as the discriminant locus, and will be denoted by $\Delta$. It has been shown that one can always perform a resolution such that 
\begin{equation}
\label{eq:discriminant_locus}
    \Delta = \cup_k \Delta_k\,,
\end{equation}
where the $\Delta_k$ are divisors in $\mathcal{M}_{\mathrm{V}}$ that intersect normally \cite{Viehweg,Hironaka}. In the sequel, we will be interested in giving a local description near the intersection of $r$ divisors, which we denote by
\begin{equation} \label{eq:intDelta}
    \Delta_{k_1\cdots k_r} = \Delta_{k_1}\cap\cdots\cap\Delta_{k_r}\,.
\end{equation}
Correspondingly, we introduce local coordinates $u^{k_i}$ such that each divisor $\Delta_{k_i}$ is given by $u^{k_i}=0$ and the intersection \eqref{eq:intDelta} is described by 
\begin{equation}
\Delta_{k_1\cdots k_r}: \, u^{k_1}=\cdots=u^{k_r}=0 \,.
\end{equation}
The situation is depicted in Figure \ref{fig:divisors} for $r=2$.
Our main goal is to develop a physical and mathematical understanding of the low-energy effective description as one approaches $\Delta_{k_1\cdots k_r}$. Broadly speaking, there are two features of interest: 
\begin{enumerate}
    \item Since we are approaching a singularity in the moduli space, the various couplings appearing in the low-energy effective theory --- namely the metric $G_{i\bar{\jmath}}$ and the gauge-kinetic functions $\mathcal{I}_{IJ}$ and $\mathcal{R}_{IJ}$ --- will degenerate as well. The framework of asymptotic Hodge theory allows one to classify, in a precise sense, the kind of singular behaviour these couplings can undergo. In other words, the asymptotic Hodge structure determines the \emph{algebraic properties} of limits in the vector multiplet moduli space. 
    \item Complementary to the algebraic properties of the limit, the second feature of interest is the geometry of $V$ at the singular loci in the moduli space. In light of the Swampland Distance Conjecture and the Emergent String Conjecture, the \emph{geometric properties} of the limit are crucial in order to identify the relevant towers of states and the appropriate dual description. 
\end{enumerate}
In the remainder of this section, we first discuss in Section~\ref{subsec:Algprops} the algebraic properties of type II degenerations in the complex structure moduli space of Calabi--Yau threefolds based on the analysis of the asymptotic Hodge structure. In Section~\ref{subsec:Tyurin} we then review the geometric properties of a special class of type II limits given by Tyurin degenerations.\footnote{The geometric properties of more general type II limits will be the subject of Section~\ref{sec:beyondTyurin}.} 

\subsection{Algebraic Properties of Type II Degenerations}\label{subsec:Algprops} 
To characterise the behaviour of couplings in the low-energy effective theory near $\Delta_{k_1\cdots k_r}$, it is necessary to understand the asymptotic behaviour of the universal 3-form $\Omega$ and the Hodge inner product \eqref{eq:pairings}. Intuitively, as one approaches $\Delta_{k_1\cdots k_r}$, there are some elements $q\in H^3(V)$ whose Hodge norm $\|q\|$ grows, while for others $\|q\|$ shrinks or stays constant. The exact scaling behaviours can differ depending on the nature of the singularity.\footnote{Generically, the precise behaviour will also depend on the choice of path along which the singularity is approached.} The mathematical machinery that precisely classifies the different possibilities is the so-called limiting mixed Hodge structure associated with $\Delta_{k_1\cdots k_r}$. In the sequel, we will not review in detail how this object is constructed, but rather focus on its properties that are relevant for our purposes. We will first discuss the case of general degenerations and subsequently specify to type II degenerations, the main interest in this work. Some of the basic background material is collected in Appendix \ref{app:hodgestr}. For a more detailed account on various aspects of asymptotic Hodge theory from a physics perspective we refer the reader to \cite{Grimm:2018ohb,Grimm:2018cpv,Grimm:2019wtx,Gendler:2020dfp,Bastian:2020egp,Calderon-Infante:2020dhm,Grimm:2021ikg,Palti:2021ubp,Grimm:2021vpn,Grimm:2022xmj,Bastian:2023shf,vandeHeisteeg:2022gsp,Monnee:2024gsq} and references therein.

\subsubsection{Limiting Mixed Hodge Structures}
With each singularity, $\Delta_{k_1\cdots k_r}$, one can associate a (limiting) mixed Hodge structure $I^{p,q}(\Delta_{k_1\cdots k_r})$, which is simply a decomposition of the complexified middle cohomology $H^3(V,\mathbb{C})$ as
\begin{equation}
\label{eq:Ipq_splitting}
    H^3(V,\mathbb{C}) = \bigoplus_{0\leq p,q\leq 3} I^{p,q}(\Delta_{k_1\cdots k_r})\,,
\end{equation}
such that the various $(p,q)$-components satisfy a certain relation under complex conjugation.  

The dimensions 
\begin{equation}
i^{p,q}=\mathrm{dim}\,I^{p,q}\,
\end{equation}
are subject to various conditions that greatly restrict the number of possible limiting mixed Hodge structures.
First, only one of $i^{3,q}$ can be non-zero, leading to four different possibilities. These four possibilities are denoted by the Roman numerals I through IV, corresponding to $q=0,\ldots, 3$, respectively. One can roughly think of this as a characterisation of how severe the degeneration is, with type I being the mildest and type IV being the most severe degeneration. For example, it turns out that type I is always at finite distance in the metric $G_{i\bar\jmath}$ defined in \eqref{eq:metric}.\footnote{It is conjectured that, conversely, type I singularities are the \textit{only} ones at finite distance. This conjecture is proven for codimension one singularities, but is still an open question in general  \cite{wang1,wang2,lee2016hodgetheoreticcriterionfinite}.} Further symmetry relations reduce the number of independent $i^{p,q}$ down to only a single number, which we choose to be $i^{2,2}$. This additional piece of information is appended to the Roman numeral classification as a subscript. In other words, we write $\mathrm{II}_b$ for a type II singularity with $i^{2,2}=b$.

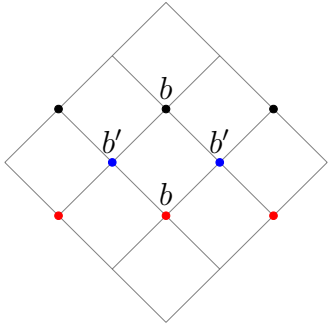
\begin{figure}[t]
    \centering
    \begin{tikzpicture}[scale=1,cm={cos(45),sin(45),-sin(45),cos(45),(15,0)}]
  \draw[step = 1, gray, ultra thin] (0, 0) grid (3, 3);

  \draw[fill,red] (0, 2) circle[radius=0.05];
  \draw[fill] (1, 3) circle[radius=0.05];
  \draw[fill] (2, 2) circle[radius=0.05];
  \node at (2.2,2.2) {$b$};
  \draw[fill,red] (1, 1) circle[radius=0.05];
  \node at (1.2,1.2) {$b$};
  \draw[fill,red] (2, 0) circle[radius=0.05];
  \draw[fill] (3, 1) circle[radius=0.05];
  \draw[fill,blue] (2,1) circle[radius=0.05];
  \node at (2.2,1.2) {$b'$};
  \draw[fill,blue] (1,2) circle[radius=0.05];
  \node at (1.2,2.2) {$b'$};
\end{tikzpicture}
    \caption{The Hodge--Deligne diamond corresponding to a type $\mathrm{II}_b$ singularity. The horizontal rows indicated in black, blue, and red correspond to the graded spaces $\mathrm{Gr}_4$, $\mathrm{Gr}_3$, and $\mathrm{Gr}_2$, respectively. Here we put $b'=h^{2,1}-b-1$.}
    \label{fig:II}
\end{figure}

It is convenient to package the information contained in the mixed Hodge structure \eqref{eq:Ipq_splitting} into a so-called Hodge--Deligne diamond. Similar to a Hodge diamond, this is a diagram in which the various $I^{p,q}$ spaces are depicted in a square as in Figure \ref{fig:II}, with a dot signifying that the corresponding space is non-trivial, and their labels indicating the dimension $i^{p,q}$. More information on the relations between the dimensions $i^{p,q}$ and the resulting Hodge--Deligne diamonds is reviewed in Appendix \ref{app:hodgestr}.

\subsubsection*{Growth theorem}

The reason why the asymptotic split \eqref{eq:Ipq_splitting} of the space of harmonic 3-forms is important for us is because it allows us to read off the asymptotic behaviour of the physical couplings appearing in the low-energy effective theory \eqref{eq:action_4d}. 
 This is because
the so-called \textit{growth theorem} of \cite{schmid,CKS} provides an asymptotic expression for the Hodge norm $\|\cdot\|$ defined in \eqref{eq:pairings}. To state the precise result, we introduce coordinates
\begin{equation}
\label{eq:def_t}
    t^j = \frac{\log u^j}{2\pi {\rm i}}=a^j+{\rm i} s^j\,,
\end{equation}
such that the limit $u^j\to 0$ corresponds to sending $t^j\to i\infty$. Here we have also introduced the axions $a^j$ and saxions $s^j$. Next, we introduce the \textit{growth sector}
\begin{equation}
\label{eq:growth_sector}
    R_{k_1\cdots k_n}=\left\{t^j: \frac{s^{k_1} }{s^{k_2}},\ldots, \frac{s^{k_{n-1}}}{s^{k_n}}, s^{k_n}>\gamma\right\}\,,
\end{equation}
for some constant $\gamma>1$, and where we set $n=h^{2,1}$. In words, the growth sector $R_{k_1\cdots k_n}$ parametrises the region in the (local patch of the) moduli space which is sequentially closest to $\Delta_{k_1}$, then $\Delta_{k_2}$, etc. Furthermore, the parameter $\gamma$ controls the precise hierarchy. Correspondingly, the growth theorem provides an asymptotic expression for the Hodge norm $\|\cdot\|$ with corrections being of order $\gamma^{-1}$.

We now fix a definite growth sector and corresponding enhancement chain of limiting mixed Hodge structures.  The latter is defined by a subsequent intersection of singular divisors --- following the hierarchy in the chosen growth sector --- such as to arrive at the singularity associated with 
$\Delta_{k_1\cdots k_n}$ and by following the changes in the spaces $I^{p,q}$. More information on the allowed enhancements is collected in Appendix \ref{app:hodgestr}. For a given $q\in H^3(V)$, the growth of the Hodge norm $\|q\|$ will depend on the spaces $I^{p,q}$ of which $q$ is an element. To quantify this, we introduce the so-called graded spaces\footnote{In principle, one can only make this identification after rotating to the $\mathbb{R}$-split limiting mixed Hodge structure, which can always be done.}
\begin{equation} \label{eq:Grl-1}
    \mathrm{Gr}_{\ell_r}(\Delta_{k_1\cdots k_r}) = \bigoplus_{p+q=\ell_r} I^{p,q}(\Delta_{k_1\cdots k_r})\,.
\end{equation}
In the Hodge--Deligne diamond, the space $\mathrm{Gr}_\ell$ corresponds to the $\ell$-th horizontal row, when counting from below. The central statement of the growth theorem is then the following. If
\begin{equation}
    q\in \mathrm{Gr}_{\ell_1}(\Delta_{k_1})\cap \cdots \cap\mathrm{Gr}_{\ell_n}(\Delta_{k_1\cdots k_n})\,,
\end{equation}
then 
\begin{equation}
\label{eq:growth_theorem}   
    \|q\|^2\sim \prod_{i=1}^n \left(\frac{s^{k_i}}{s^{k_{i+1}}}\right)^{\ell_i-3}\,.
\end{equation}
A similar result provides the growth of $\|\Omega\|$, and subsequently the K\"ahler potential and moduli space metric via the relation \eqref{eq:metric}. Indeed, define an integer $d_{i}$ such that $d_{i}=0,1,2,3$ if $I^{p,q}(\Delta_{k_1\cdots k_i})$ is of type I, II, III, IV, respectively. Then
\begin{equation}\label{Kgrowththeorem}
   e^{-K}= \|\Omega\|^2\sim \prod_{i=1}^n \left(\frac{s^{k_i}}{s^{k_{i+1}}}\right)^{d_{i}}\,.
\end{equation}
 
\subsubsection{Type II enhancement chains and the growth theorem}
For the purpose of this work, we will primarily be interested in type II degenerations. 
In these,
 \begin{equation} \label{eq:H3decomTypeII}
H^3(V,\mathbb{C}) = {\rm Gr}_2 \oplus {\rm Gr}_3 \oplus {\rm Gr}_4 \,, \qquad   {\rm Gr}_2 \cong  {\rm Gr}_4 \,,
 \end{equation}
where we drop the dependence on $\Delta_{k_1\cdots k_r}$ for notational simplicity, recall also Figure \ref{fig:II}. Of special importance will be the space 
 \begin{equation} \label{eq:Gr2-def}
    \text{Gr}_2 = I^{2,0} \oplus I^{1,1} \oplus I^{0,2} \,. 
\end{equation}
 The relations between the dimensions $i^{p,q}$ of the spaces $I^{p,q}$ reviewed in Appendix \ref{app:hodgestr} imply that $i^{2,2} = i^{1,1}$ so that
  \begin{equation}
 {\rm dim}\,I^{1,1} =b \quad \text{ in a II$_{b}$ degeneration,}
  \end{equation}
while $i^{2,0}=i^{0,2}=1$. Additionally, for a fixed $h^{2,1}$, the integer $b$ can only take finitely many values, namely $b=0,\ldots, h^{2,1}-1$. 

It will be relevant to know which singularities can enhance to a type II degeneration and, subsequently, to which singularities a type II degeneration can enhance further. 
The relevant rules for enhancements to type II singularities are given by \cite{Kerr2017,Grimm:2018cpv}
\begin{align}
    &\mathrm{I}_a\to\mathrm{II}_{b'}\,,\qquad \,\,\text{for $a\leq b',\,a<h^{2,1}$}\,,\\
    &\mathrm{II}_b\to\mathrm{II}_{b'}\,,\qquad \text{for $b\leq b'$}\,.
\end{align}
A particularly interesting case is the type II$_0$ degeneration, for which the only possible preceding degenerations are again a type II$_0$ degeneration, or a type I$_0$ degeneration. The latter is a very mild singularity where the monodromy is of finite order, so that it can be trivialised by sending $u^i\mapsto (u^i)^{k}$ for some integer $k$.

The relevant rules for enhancements starting from type II degenerations are given by
\begin{align}
    &\mathrm{II}_b\to\mathrm{II}_{b'}\,,\qquad \,\,\,\text{for $b\leq b'$}\\
    &\mathrm{II}_b\to\mathrm{III}_{c'}\,,\qquad \,\text{for $2\leq b\leq c'+2$}\,,\\
    &\mathrm{II}_b\to\mathrm{IV}_{d'}\,,\qquad \text{for $1\leq b\leq d'-1$}\,.
\end{align}
Again, the type II$_0$ degeneration is a particularly interesting case. This is because a type II$_0$ singularity can never enhance to a type III or a type IV singularity. This, in turn, implies that a type II$_0$ divisor can never intersect the large complex structure point, since the latter is always a type IV$_{h^{2,1}}$ singularity.\footnote{This is clear when the type II$_0$ singularity occurs in codimension one. One can always reduce to this case by the following argument. Suppose that the limit $u^1\to 0$ corresponds to a type II$_0$ singularity. Then one can simply take the diagonal limit where $u^2,\ldots, u^{h^{2,1}}$ all go to zero at the same rate. } As a result, these limits in the complex structure moduli space of Type IIB do not admit a mirror dual interpretation in the large-volume regime of Type IIA. Nevertheless, our analysis will apply equally well to these limits since we will not rely on mirror symmetry.

We now illustrate the use of the growth theorem for the case that will be of particular relevance in this work, namely when the enhancement chain only consists of type II degenerations:\footnote{One could also consider the most general case in which there are additional type I limits that precede the type II limits, for which it is intuitively clear that it will not alter our results since these limits are at finite distance. However, to keep the discussion simple we will not explicitly include this case.}
\begin{equation}
    \mathrm{I}_0\to \mathrm{II}_{b_1}\to\cdots\to\mathrm{II}_{b_n}\,.
\end{equation}
 Suppose we consider an element $q\in H^3(V)$ which lies in  $\mathrm{Gr}_2(\Delta_{k_1})$ defined in \eqref{eq:Gr2-def}.
 Then, because we are considering an enhancement chain in which the principal type does not increase, it turns out that actually $q\in \mathrm{Gr}_2(\Delta_{k_1\cdots k_r})$ for all $r=1,\ldots, n$. By employing the growth theorem \eqref{eq:growth_theorem} with $\ell_i=2$ for all $i=1,\ldots, n$, we thus obtain
\begin{equation} \label{eq:G2vanishing}
    \|q\|^2\sim (s^{k_1})^{-1}\,.
\end{equation}
In particular, the Hodge norm $\|q\|$ goes to zero, at a rate which is controlled by the saxion $s^{k_1}$ that is leading in the growth sector \eqref{eq:growth_sector}.

\subsection{Geometry of Tyurin Degenerations}\label{subsec:Tyurin}
So far, we have seen that the graded spaces $\mathrm{Gr}_\ell$ defined in \eqref{eq:Grl-1} are important in characterising the asymptotic growth of the Hodge norm. At the same time, these spaces are also closely related to the geometry of the degenerations of the underlying Calabi--Yau threefolds.  
In this paper, the main focus will be on type II degenerations which geometrically correspond to Tyurin degenerations \cite{Tyurin:2003}. The possible geometries arising in more general type II degenerations will be discussed in Section~\ref{sec:beyondTyurin}.
 
\begin{figure}[t]
    \centering
    \includegraphics[width=0.7\linewidth]{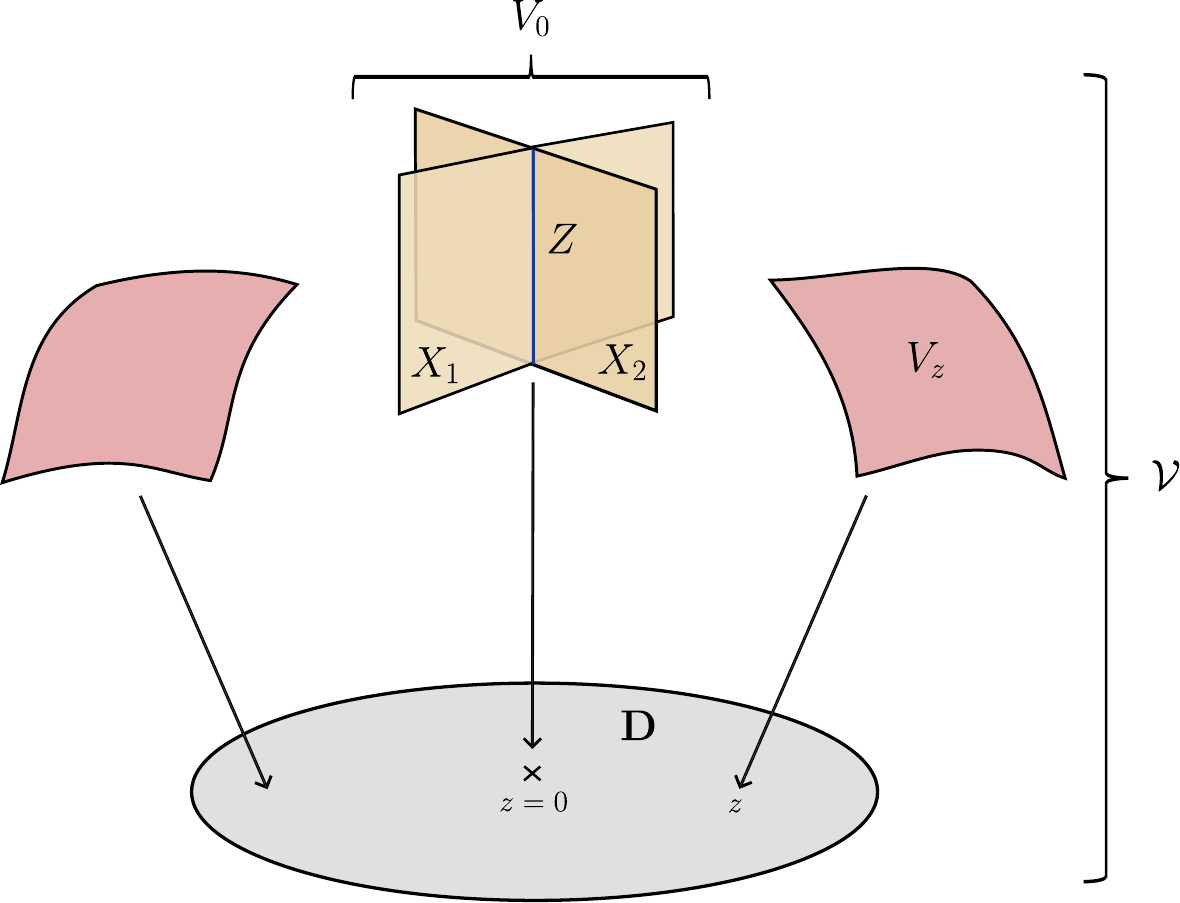}
    \caption{A schematic depiction of a Tyurin degeneration \eqref{eq:4foldV}. The fibre $V_z$ corresponding to the Calabi-Yau threefold over a generic point in the moduli space is depicted in red, while the central fibre $V_0$ is depicted in orange. For a Tyurin degeneration, the latter splits into two components $X_1$ and $X_2$ which intersect over a K3 surface $Z$, here indicated in blue.}
    \label{fig:Tyurin}
\end{figure}

Roughly speaking, a Tyurin degeneration corresponds to a limit in the complex structure moduli space in which the Calabi--Yau threefold $V$ splits into two components that intersect along a K3 surface. To be precise, we consider a family of Calabi--Yau threefolds varying over a disk $\mathbf{D}=\{z\in \mathbb{C}\big||z|<1\}$, with fibres $V_z$ for $z\in\mathbf{D}$:
\begin{equation}\begin{aligned} \label{eq:4foldV}
V_z \ \hookrightarrow & \  \ \mathcal{V} \cr 
&\ \ \downarrow\cr 
& \ \  \mathbf{D}\,.
\end{aligned}\end{equation}
Furthermore, we assume that the generic fibre $V_{z\neq 0}$ is smooth, whereas the central fibre $V_0$ corresponds to the union of two quasi-Fano threefolds 
\begin{equation} \label{eq:V0split}
    V_0= X_1 \cup_Z X_2 \,,
\end{equation} 
where we denote the intersection locus of $X_1$ and $X_2$ by $Z$. For a Tyurin degeneration the manifold $Z$ is a K3 surface. The situation is depicted schematically in Figure \ref{fig:Tyurin}. To compare the current situation with the setup described in Section  \ref{subsec:Algprops}, note that one can equivalently view the fibration \eqref{eq:4foldV} as an embedding of the disk $\mathbf{D}$ into the complex structure moduli space $\mathcal{M}_{\mathrm{V}}$ of the family $V_z$ of Calabi--Yau threefolds. In particular, one may identify the limit $z\to 0$ with some divisor $\{u^{k_1}=\dots = u^{k_r}=0\}$. To simplify the notation, we will denote the corresponding divisor by $\Delta$ without an additional subscript.\footnote{To be clear, this is not to be confused with the notation \eqref{eq:discriminant_locus} for the full discriminant locus.} Before we explain in general why Tyurin degenerations correspond to type II limits in the complex structure moduli space, we provide an illustrative example of a Tyurin degeneration. 

\subsubsection*{Example of a Tyurin Degeneration}
Consider the mirror of the Calabi--Yau threefold $\widehat{V}$ obtained as a degree-12 hypersurface in the weighted projected space $\mathbb{P}_{11226}$, denoted by $\mathbb{P}_{11226}[12]$. The Calabi--Yau threefold $\mathbb{P}_{11226}[12]$ with Hodge-numbers $h^{1,1}(\widehat{V})=2,h^{2,1}(\widehat V)=128$ admits a K3-fibration and is the main example in~\cite{Lee:2019oct} illustrating the Emergent String Conjecture in the vector multiplet moduli space of Type IIA/M-theory compactifications on Calabi--Yau threefolds. 
The mirror manifold $V$ to $\widehat{V}$ can be obtained by a Greene--Plesser \cite{Greene:1990ud} construction.
Consider the polynomial
\begin{equation}
    P = x_1^{12}+x_2^{12}+x_3^6+x_4^6 +x_5^2 - u_1^{-1} x_1x_2x_3x_4x_5-u_2^{-1} x_1^6x_2^6 \,.
\end{equation}
Then, the locus $\{P=0\}/G$ gives the mirror $V$ of $\mathbb{P}_{11226}[12]$ after resolving all orbifold singularities. Here $G=\mathbb{Z}_6^2\times \mathbb{Z}_2$ acts as a symmetry on $P$~\cite{Candelas:1993dm,Hosono:1993qy}. The complex structure moduli space of $V$ is spanned by $(u_1,u_2)$. Now, in the limit $u_2\to 0$, the defining equation of $V$ factorises as 
\begin{equation}
    P \stackrel{u_2\to 0}{\longrightarrow} u_2^{-1} x_1^6 x_2^6 \,,
\end{equation}
in two degree-6 hypersurfaces\footnote{Notice that since the first $\mathbb{Z}_6$-factor in $G$ acts as $x_1\to e^{-2\pi {\rm i}/6} x_1$, $x_2\to e^{2\pi {\rm i}/6}x_2$, the hypersurfaces $X_1$ and $X_2$ are irreducible in $\{P=0\}/G$ even though they are reducible in $\{P=0\}$.} 
\begin{equation}
    X_1= \{x_1^6=0\} \,,\qquad X_2=\{x_2^6=0\}\,. 
\end{equation}
The intersection of the two components is given by 
\begin{equation}
    Z=X_1 \cap X_2 = \{x_1^6=x_2^6=0\}\,. 
\end{equation}
The first Chern class of $Z$ can be computed using the adjunction formula
\begin{equation}
    c_1(Z)= 12 [H] - \left[\text{deg}(x_1^6) +\text{deg}(x_2^6)\right][H] =0 \,,
\end{equation}
where $H$ is the hyperplane class in $\mathbb{P}_{11226}$. The intersection $Z$ is therefore a K3 surface. Hence, already simple examples of Calabi--Yau threefolds obtained as hypersurfaces in weighted projective spaces can realise Tyurin degenerations. 
Further examples of Tyurin degenerations are given e.g.~in \cite{Doran:2024kcb,Berglund:2022dgb}.

\subsubsection*{Tyurin degenerations as type II limits} 
To understand why Tyurin degenerations correspond to type II limits in the language introduced in Section \ref{subsec:Algprops}, one proceeds as follows: First, one associates a mixed Hodge structure with $H^3(V_0)$, following the construction of Deligne \cite{Deligne:1971,Deligne:1974}. This is achieved with the help of the Mayer--Vietoris sequence, which relates the cohomology groups on $V_0$ to those on its components $X_1$, $X_2$ and their intersection $Z$. 
Subsequently, via the inclusion map 
\begin{equation}\label{def:iota}
    \iota: V_z \hookrightarrow \cV \,,
\end{equation}
one relates this mixed Hodge structure to the limiting mixed Hodge structure on $H^3(V_z)$ that was described in Section \ref{subsec:Algprops} in terms of the graded spaces \eqref{eq:Grl-1}. This is the content of the so-called Clemens--Schmid exact sequence. 
 We explain both of these constructions in Appendix \ref{app_MV+CS}. What is important for us is that the outlined procedure leads to a geometric characterisation of the 3-forms which, in a type II infinite distance limit associated with a Tyurin degeneration, take values in the space ${\rm Gr}_\ell(\Delta)$. 

\paragraph{Geometric interpretation of $\mathrm{Gr}_3$.}
First, consider the space $\mathrm{Gr}_3(\Delta)$, which contains the middle row of the Deligne diamond associated with a type II limiting mixed Hodge structure. For this space, the geometric construction reviewed in Appendix \ref{app_MV+CS} establishes the isomorphism
\begin{equation}
\begin{aligned}
    \mathrm{Gr}_3(\Delta)
    &\cong 
         H^3(X_1)\oplus H^3(X_2)\\
         & \cong  {\rm im}(k^\ast, l^\ast)   \,, 
    \label{eq:G2isos-2}
\end{aligned}
\end{equation}
where 
 \begin{equation}\label{def:kandl}
 k:X_1 \hookrightarrow  V_0 \,, \qquad  l:X_2 \hookrightarrow  V_0\,,
 \end{equation}
refer to the inclusions of the two components $X_i$ in $V_0$. In other words, the elements in $\mathrm{Gr}_3(\Delta)$ correspond to 3-forms on $V_0$ which restrict to 3-forms on $X_1$ or $X_2$.
\paragraph{Geometric interpretation of $\mathrm{Gr}_2$.}
Next, we consider the space $\mathrm{Gr}_2(\Delta)$, which contains the bottom row of the Deligne diamond associated with a type II limiting mixed Hodge structure. According to \eqref{eq:growth_theorem}, the 3-forms in $\mathrm{Gr}_2 (\Delta)$ are precisely those whose Hodge norm vanishes in the limit. For this space, one finds
\begin{equation}
\begin{aligned}
    \mathrm{Gr}_2(\Delta)
    & 
         \cong {\rm im}(d^\ast)  \\
         & \cong H^2(Z)/(\mathrm{im}(i^*)+\mathrm{im}(j^*))\,,
    \label{eq:G2isos-1}
\end{aligned}
\end{equation}
where 
\begin{equation} \label{ijemb}
i: Z \hookrightarrow X_1 \,, \qquad j: Z \hookrightarrow X_2\,,
\end{equation}
denote the embeddings of the K3 surface $Z$ into the two components $X_1$ and $X_2$ of the degenerate space $V_0$ and 
the map 
\begin{equation}\label{def:dstar}
d^\ast: H^2(Z) \to H^3(V_0)
\end{equation}
can be thought of as adding a leg normal to $Z$ in $V_0$ to a 2-form on $Z$.
 The first isomorphism states that the elements in $\mathrm{Gr}_2(\Delta)$ correspond to 3-forms on the degenerate space $V_0$ which can be built from 2-forms on $Z$
 by completing them in the direction normal to $Z$ via the map $d^\ast$.
 The second isomorphism identifies this space with the cohomology group of 2-forms on $Z$ which do not arise as the pullback of 2-forms from $X_1$ or $X_2$ to $Z$. 
The cohomology of such 2-forms on $Z$ gives rise to the transcendental lattice  of the K3 surface $Z$,\footnote{For a stand-alone K3 surface, the transcendental lattice is defined as the orthogonal complement of the Picard lattice. The geometry of the 2-cycles associated with the latter is governed by K\"ahler deformations, whereas the geometry of the 2-cycles associated with the transcendental lattice is governed by complex structure deformations. As the complex structure deformations of a CY threefold are associated with the geometry of 3-cycles, it must be such that the 2-cycles of $Z$ associated with the transcendental lattice must lift to 3-cycles of $V_0$.} 
\begin{equation} \label{eq:defLambdatrans}
{\rm Gr}_2(\Delta) \cong \Lambda_{\rm trans}  \,.
\end{equation}
By \eqref{eq:Gr2-def}, its signature is\footnote{Strictly speaking, at the level of 3-forms the space $\mathrm{Gr}_2$ does not immediately come equipped with a polarization, due to the fact that it is not primitive. The space $\mathrm{Gr}_4$, however, does carry a polarization which is induced from the polarization $(\cdot,\cdot)$ on $V$ together with the action of the log-monodromy operator, see for example \cite{Grimm:2018cpv} for further details. For our purposes, it is sufficient to use the fact that $\mathrm{Gr}_2\cong\mathrm{Gr}_4$ as vector spaces to speak of the polarization lattice on the K3 surface $Z$.}
\begin{equation} \label{eq_Transsign}
{\rm sgn}(\Lambda_{\rm trans}) = (2,b)   \quad \text{in a ${\rm II}_{b}$ Tyurin degeneration.}
\end{equation}
The remaining 2-forms on $Z$ which do not lie in $\mathrm{Gr}_2(\Delta)$ \textit{do} arise as the pullback of 2-forms from $X_1$ or $X_2$. These give rise to the polarization lattice $\Lambda_{\mathrm{pol}}$ of the K3 surface $Z$,
\begin{equation}
\label{eq:Lambda_pol}
    i^*(H^2(X_1))\oplus j^*(H^2(X_2)) = \Lambda_{\mathrm{pol}}\,.
\end{equation}
Its signature is
\begin{equation}
    \mathrm{sgn}(\Lambda_{\mathrm{pol}}) = (1,\rho)\,,\qquad \rho=19-b\,,
\end{equation}
such that the total lattice,
\begin{equation}
\label{eq:H2Z_decomp}
    H^2(Z) = \Lambda_{\mathrm{pol}}\oplus\Lambda_{\mathrm{trans}}\,,
\end{equation}
has signature $(3,19)$ as required for a K3 surface. As will be explained in Section~\ref{sec:Worldsheet}, the decomposition \eqref{eq:H2Z_decomp} has an important physical interpretation. Namely, it precisely distinguishes the interacting and free modes on the worldsheet theory of the emergent EFT string that becomes tensionless when approaching the Tyurin degeneration. For later reference, let us thus emphasise that the decomposition \eqref{eq:H2Z_decomp} means that a general 2-form $\omega_\alpha\in H^2(Z)$ may be decomposed as
\begin{equation}
\label{eq:decomp_omega}
    \omega_\alpha = (i^*-j^*)\bar{\omega}_i+\omega_a\,,
\end{equation}
where\footnote{Notice that due to the linearity of the maps $i^*$ and $j^*$ one finds $\text{im}(i^*-j^*)\cong\text{im}(i^*)+\text{im}(j^*)$.}
\begin{equation}
\label{eq:decomp_omega_2}
    \bar{\omega}_i\in H^2(X_1)\oplus H^2(X_2)\,,\qquad \omega_a\in H^2(Z)/(\mathrm{im}(i^*)+\mathrm{im}(j^*))\,.
\end{equation}
More generally, the geometric characterisation of the 3-forms in  $\mathrm{Gr}_2(\Delta)$ and $ \mathrm{Gr}_3(\Delta)$ summarised above is the basis for the physics interpretation of the type II limits given in this paper.  

\paragraph{Localised forms.}
Having understood the cohomology on the central fibre $V_0$, let us briefly discuss the interpretation from the perspective of the family $\mathcal{V}$. To this end, we note that there exists an isomorphism
\begin{equation} \label{HnV0}
 H^n({\cal V}) \cong  H^n({V_0})\,,
\end{equation}
between the cohomology groups on the fourfold ${\cal V}$ and its central fibre $V_0$ \cite{Morrison:1984}.  Importantly, the relation \eqref{HnV0} implies that cohomologically non-trivial forms which are ``localised'' to $z=0$ cannot give rise to cohomologically non-trivial forms on $V_0$.\footnote{Here by ``localised'' we mean that the form is not globally defined on the fourfold $\cV$ and only has support around $z=0$.}  This notion of localisation plays an important role in Section~\ref{sec:Worldsheet} in establishing the degrees of freedom which localise on the worldsheet of the emergent string. 

In particular, we will be interested in localised 2-forms $\omega_A$. These are either elements of $H^2(X_1)\oplus H^2(X_2)$ or $H^2(Z)$. 
 To be localised at $\{z=0\}$, $\omega_A$ cannot descend from a 2-form that is well-defined and cohomologically non-trivial as a 2-form on $V_0$, as just discussed. 
 In the notation of the inclusion maps $k,l$ defined in \eqref{def:kandl}, this means that $\omega_A$ cannot be written as $(k^* \bar \omega_A, l^* \bar{\omega}_A)\in H^2(X_1)\oplus H^2(X_2)$ for some $\bar \omega_A\in H^2(V_0)$.   
For 2-forms $\omega_A\notin \text{Im}(k^*,l^*)$,  the Mayer--Vietoris exact sequence reviewed in Appendix~\ref{App-MV}\footnote{Specifically, in the notation of \eqref{eq:MV-sequence1}, we know that ${\rm im}(\alpha_2) = {\rm ker}(\beta_2)$.} implies that also $\omega_A \notin \text{ker}(i^*-j^*)$. In other words, we conclude $(i^*-j^*)\omega_A \neq 0\in H^2(Z)$. The upshot of this discussion is the following
\begin{fact}[Localised 2-forms]\label{factcheck}
    All 2-forms $\omega_\alpha$ that localise to $\{z=0\}\subset \cV$ can be pulled back to 2-forms on $Z$ and admit a decomposition of the form \eqref{eq:decomp_omega_2}.
\end{fact}

\section{Type II Limits as Emergent String Limits}\label{sec:BPSstates}
Equipped with the algebraic and geometric characterisation of type II limits corresponding to Tyurin degenerations, we now discuss the physics associated with these limits. More precisely, we consider a Type IIB compactification on a Calabi--Yau threefold $V$ for which the asymptotic values of the scalar fields in the vector multiplet sector are tuned such as to realise a Tyurin degeneration. Concretely, we consider the 4d $\cN=2$ theories in asymptotically flat spacetime for which the scalars $t^k$ (recall the notation \eqref{eq:def_t}) in the vector multiplet sector are set to some constants at spatial infinity, $\bar{t}^k\equiv t^k(x^\mu \to \infty)$. The limit in moduli space $\cM_V$ then corresponds to tuning the values of $\bar{t}^k$ to some extreme value. In the sequel, we focus on \emph{one-parameter} limits in $\cM_V$, i.e.~on limits for which only the asymptotic value of a single scalar field is taken to infinity. Let us denote this scalar field by $t^{k_1}$. 

In these setups, we are interested in uncovering the physics of the effective theory of gravity and, in particular, in potential towers of particle-like states that become massless in the limit $\bar{t}^{k_1}\to i\infty$.
In our Type IIB setup, such particle-like states can have two origins. First of all, they can arise as BPS particles obtained from D3-branes wrapping special Lagrangian 3-cycles in $V$. In Section~\ref{subsubsec-BPStower}, we first use the algebraic properties of the limiting mixed Hodge structure close to Tyurin degenerations to identify those D3-brane charges that would lead to massless, weakly-coupled states if populated by BPS states. We then use the geometric features of Tyurin degenerations to argue for the existence of a tower of BPS states becoming light in these limits.
  
  Apart from BPS states arising from wrapped D3-branes, we also have to consider configurations of strings that asymptote to flat space. Based on the algebraic properties of the Hodge structure in type II limits, we argue in Section~\ref{sec:EFTstrings} for the existence of strings whose tension goes to zero at the same rate as the square of the mass of the lightest tower of BPS states identified in Section~\ref{subsubsec-BPStower}. The precise identification of the nature of this string is the subject of Section~\ref{sec:Worldsheet} which makes use of the geometric properties of Tyurin degenerations.

\subsection{BPS-Particle Tower from D3-Branes} \label{subsubsec-BPStower}

BPS particles in Type IIB compactifications on a Calabi--Yau threefold $V$ arise from D3-branes wrapping special Lagrangian 3-cycles. These BPS particles are charged under the ${\rm U}(1)$ gauge fields $A^I$ of the 4d $\cN=2$ supergravity theory obtained by expanding the RR-four form 
\begin{equation}
    C_4 = A^I \wedge \gamma_I \,,\qquad \gamma_I \in H^3(V,\mathbb{Z})\,. 
\end{equation}
 The BPS particles are hence characterised by a charge vector $q\in H^3(V,\mathbb{Z})$. Their BPS mass is given by the central charge as 
\begin{equation} \label{eq:mq-1}
    \frac{m_q}{M_{\mathrm{Pl}}} = |Z_q| = \frac{|\langle q,\Omega\rangle|}{\|\Omega\|}\,,
\end{equation}
where we recall the notation introduced in \eqref{eq:pairings}. The mass is measured in units of the four-dimensional Planck mass $M_{\mathrm{Pl}}$. Let us remark that the relation between $M_{\mathrm{Pl}}$ and the string scale $M_s$ depends on the string coupling as well as the volume of $Y_3$ and is controlled by scalars in the hypermultiplet sector of the moduli space.\footnote{More precisely, $M_{\mathrm{Pl}}$ is related to the string scale $M_s$ as $M_s \sim \frac{g_s}{\sqrt{\mathcal{V}_s}}M_{\mathrm{Pl}}$,
where $g_s$ denotes the string coupling and $\mathcal{V}_s$ denotes the volume of the Calabi--Yau threefold in units of the string length.} Since our analysis pertains to the vector multiplet sector, we may take $M_{\mathrm{Pl}}$ as constant and will henceforth set $M_{\mathrm{Pl}}=1$ without loss of generality. Hence, all the moduli-dependence of the mass will come from the periods of the holomorphic (3,0)-form.

Among the states obtained by wrapping D3-branes on special Lagrangian 3-cycles in $V$, we aim to identify an infinite tower of states that, as we approach a type II divisor $\Delta$ corresponding to a Tyurin degeneration, become light in Planck units. In addition, we require this tower of states to become \emph{weakly coupled} in the infinite distance limit. In this way, we exclude towers of states that are dyonic in nature: Even though such states can become light, we cannot write down a local action for them and they do not form the tower of states expected by the Distance Conjecture.\footnote{By a similar reasoning, the tower of light BPS states arising at SCFT boundaries in 5d $\cN=1$ theories of gravity (e.g.~obtained from M-theory compactified on  Calabi--Yau threefolds) do not correspond to the tower of states predicted in the context of the Distance Conjecture.} 
We hence require states with charge $q\in H^3(V,\mathbb{Z})$ for which the physical charge vanishes, i.e.
\begin{equation}
\cQ=\|q\| \to 0 \,,
\end{equation}
as we approach the type II limit. From our discussion of the growth theorem around~\eqref{eq:G2vanishing} a natural candidate state that can give a tower of weakly-coupled states in the asymptotic limit has charge $q\in \text{Gr}_2(\Delta)$. As we approach the type II divisor $\Delta$, an element $q\in H^3(V,\mathbb{Z})\cap \text{Gr}_2(\Delta)$ will generically have components in each of the subspaces of 
\begin{equation}
    \text{Gr}_2(\Delta) = I^{2,0}(\Delta) \oplus I^{1,1}(\Delta) \oplus I^{0,2}(\Delta) \,. 
\end{equation}
In this case, the mass of the state with charge $q$ scales like the Hodge norm 
\begin{equation}\label{massq}
    m_{q} \to \| q \| \sim \frac{1}{\sqrt{s^{k_1}}}\,, 
\end{equation}
where $s^{k_1}$ is the fastest-growing saxion in the chosen growth sector. If the type II divisor is approached along special paths in the moduli space, the intersection $H^3(V,\mathbb{Z})\cap \text{Gr}_2(\Delta)$ can contain elements $q_0$ that are entirely contained in $I^{1,1}(\Delta)$. In this case, the mass of $q_0$ is exponentially small in $s^{k_1}$.\footnote{The reason for this is that for such states the asymptotic coupling to the graviphoton $(q,\Omega_\infty)$ -- following the nomenclature of \cite{Bastian:2020egp} -- is zero. As a result, the scaling of the mass is determined by sub-leading corrections to the Hodge inner product, which for the type II singularity necessarily correspond to exponential corrections.} However, as we argue below, there can only be finitely many such states with exponentially small mass in $s^{k_1}$, and not an infinite tower of weakly coupled particle states with this property. Instead, to argue for the existence of a tower of weakly coupled states, we focus on generic charges $q\in H^3(V,\mathbb{Z})\cap \text{Gr}_2(\Delta)$. 

To identify such a tower of states we must find at least one charge 
\begin{equation} \label{eq:def-q0}
q_0 \in H^3(V,\mathbb{Z})\cap \text{Gr}_2(\Delta)
\end{equation}
such that there exists a bound state with charge $n q_0$ for $n\in \mathcal{J}_{q_0}$ and $\mathcal{J}_{q_0}$ an infinite order subset of $\mathbb{Z}$. Equivalently, if $\Gamma_{0}\in H_3(V)$ is the 3-cycle dual to $q_0\in H^3(V,\mathbb{Z})$, a sufficient condition for the existence of a tower of BPS particles corresponding to D3-branes wrapping $n\Gamma_0$ is the non-vanishing of the 4d BPS index $\Omega_{\rm BPS}(n \Gamma_0)$. For A-branes along special Lagrangians, an enumerative invariant that corresponds to such a BPS index has been defined in \cite{Banerjee:2022oed} as
\begin{equation}
    \Omega_{\rm BPS}(\Gamma) = (-1)^{{\rm dim}({\cal M}_\Gamma)} \chi({\cal M}_\Gamma) \,,
\end{equation}
where ${\cal M}_\Gamma$ is the moduli space of the A-brane on $\Gamma$ and $\chi({\cal M}_\Gamma)$ denotes its Euler characteristic.

In general, computing the BPS invariants for special Lagrangian 3-cycles in Calabi--Yau threefolds is notoriously difficult since the geometry of such cycles is in general unknown. However, much more can be said if we focus on the vicinity of the degeneration $\Delta$ and consider only the 3-cycles that are dual to elements $q\in H^3(V, \mathbb{Z})\cap \text{Gr}_2(\Delta)$. 

For a Tyurin degeneration,  the isomorphism 
\begin{equation}\label{eq:isom} 
\text{Gr}_2(\Delta) \cong H^2(Z)/(\text{Im}\,i^* +\Im j^*)\,,
\end{equation}
together with the relation \eqref{HnV0}, allows us to write a 3-form $q_0\in H^3(V,\mathbb{Z})\cap \text{Gr}_2(\Delta)$  as $q_0=\iota^* d^* \omega_0$, where $\omega_0\in H^2(Z)$ and the maps $\iota$ and $d^\ast$ are defined in \eqref{def:iota} and \eqref{def:dstar}, respectively. For the 3-cycle $\Gamma_0\in H_3(V)$ dual to $q_0$ this means that close to the degeneration locus, $\Gamma_0$ defines a curve class \
\begin{equation}  \label{eq:C0def}
C_0 \in H_2(Z)  
\end{equation}
with the following property:
 On $\Delta$ we can view $\iota_*\Gamma_0$ as an $S^1$ fibration over $C_0$ with the fibre corresponding to the circle that degenerates over $Z$. For the cycle $\Gamma_0\in H_3(V)$ to be special Lagrangian, $C_0$ must lie in the transcendental lattice of $Z$, i.e.~the restriction of the K\"ahler form of $Z$ must vanish, $J_Z|_{C_0} =0$, and $C_0$ is calibrated by the holomorphic 2-form $\Omega^{2,0}$ on $Z$.
 At least in the degenerate limit of vanishing $S^1$ fibre, which we interested in, this is sufficient to guarantee that $\Gamma_0$ is special Lagrangian.

To show the existence of a tower of states, we now consider multi-wrappings of D3-branes on $\Gamma_0$. By wrapping $n$ D3-branes on $\Gamma_0$ we obtain a $0+1$-dimensional super quantum mechanics (SQM) with eight supercharges and gauge group ${\rm U}(n)$. A bound state with charge $n q_0$ exists if it is possible to Higgs the ${\rm U}(n)$ gauge group to a single ${\rm U}(1)$. For this to be possible, the SQM obtained from the wrapped D3-branes must have sufficiently many scalar fields. For a special Lagrangian cycle $\Gamma_0$ the number of scalar deformation modes is counted by $b_1(\Gamma_0)$. The SQM has enough scalar modes to realise the Higgsing ${\rm U}(n)\to {\rm U}(1)$ if $b_1(\Gamma_0)\geq 3$.\footnote{To see this, we notice that the case $b_1(\Gamma_0)=3$ corresponds to $\Gamma_0$ being topologically a $T^3$ in which case all sixteen supercharges of the parent 4d $\cN=4$ ${\rm U}(n)$ SYM theory living on the worldvolume theory of the $n$ D3-branes are preserved. In this case the fields of the resulting SQM form complete multiplets of the super-algebra with 16 supercharges. As is well-known from the case of D0-branes, the scalar fields in the multiplets of $\cN=16$ are sufficient to completely break ${\rm U}(n)\to {\rm U}(1)$.} This requirement translates into a condition on $C_0$ since due to the fibration structure of $\Gamma_0$, $b_1(\Gamma_0)$ is bounded from below by the first Betti number of the base, $b_1(\Gamma_0)\geq b_1(C_0)$. For the curve $C_0$, the first Betti number is given by its genus $g(C_0)$, $b_1(C_0) = 2 g(C_0)$. Hence, for the special Lagrangian cycles $\Gamma_0$, bound states with charge $nq_0$ exist if the curve $C_0\in H_2(Z)$ has genus $g(C_0)\geq 1$.\footnote{Here, we include the case $g(C_0)=1$ since in this case the $S^1$-fibration over $C_0$ is trivial such that $b_1(\Gamma_0)=b_1(C_0)+1=3$.}

A tower of weakly coupled, light D3-brane states thus exists if $Z$ contains curves with genus $g(C_0)\geq 1$ that satisfy $J_Z|_{C_0} =0$. On $Z$ such curves are dual to 2-forms in the transcendental lattice $\Lambda_{\rm trans}\subset \Gamma_{3,19}$ introduced in \eqref{eq:defLambdatrans} of signature $\text{sgn}(\Lambda_{\rm trans}) = (2,b)$ for some $b\geq 0$. From the signature one infer that $\Lambda_{\rm trans}$ contains at least two integer 2-forms $\omega_a$, $a=1,2$, satisfying 
\begin{equation}
    \int_{Z} \omega_a \wedge \omega_a \geq 0\,.
\end{equation}
This implies that the 2-cycles $C_a$ dual to these 2-forms have non-negative self-intersection. By performing a hyper-K\"ahler rotation of the three complex structures on the K3-manifold $Z$, the curves $C_a$ can become holomorphic. Since for a holomorphic curve $C$ on a K3 surface, the adjunction formula relates the genus of a curve to its self-intersection on $Z$, $C \cdot_Z C =2g(C)-2$, the curves $C_a$ with non-negative self-intersection  have $g(C_a)\geq 1$. Thus on $Z$ there exist at least two curves $C_1$ and $C_2$ with genus $g\geq 1$. As before these give rise to special Lagrangian 3-cycles $\Gamma_{1,2}$ with $b_1(\Gamma_{1,2})\geq 3$ which therefore lead to an infinite tower of states upon multi-wrapping by D3-branes.

To summarise, our arguments suggest that for the 3-cycles $\Gamma_0$ dual to elements in ${\rm Gr}_2(\Delta)$ of a Tyurin degeneration,
\begin{equation}
    \Omega_{\rm BPS}(n \Gamma_0) \neq 0  \quad \forall n \in {\mathbb N} \qquad  \Longleftrightarrow \qquad C_0 \cdot_Z C_0 \geq 0 \,.
\end{equation}
In fact, our findings of Section~\ref{sec:Worldsheet} will make it natural to conjecture that for $C_0 \cdot_Z C_0 \geq 0$, $\Omega_{\rm BPS}(n \Gamma_0)$ are coefficients of a meromorphic (mock-)modular form. We will come back to this in Section \ref{sec_Hetdual}.

It remains to verify the mass scaling \eqref{massq} for the tower of states obtained in this way. To this end, we must argue that the charge $q_0$ does not exclusively lie in $I^{1,1}(\Delta)$ but has also components along $I^{2,0}(\Delta) \oplus I^{0,2}(\Delta)$. We show that this is indeed the case by contradiction. Therefore, assume that $q_0|_{I^{2,0}(\Delta)\oplus I^{0,2}(\Delta)}=0$. The isomorphism \eqref{eq:isom} identifies 
\begin{equation}
    I^{p,q}(\Delta) \cong H^{p,q}_{\rm prim.}(Z) \,,\quad \text{for}\quad p+q=2\,.  
\end{equation}
Hence, if $q_0|_{I^{2,0}(\Delta)\oplus I^{0,2}(\Delta)}=0$, we can write it as  $q_0 = \iota^* d^* \omega_0$ with $\omega_0\in H^{1,1}(Z)\cap H^2(Z,\mathbb{Z})$. This implies that 
\begin{equation}
    \int_Z \Omega^{2,0} \wedge \omega_0 = 0\,,
\end{equation}
and the 2-cycle $C_0$ dual to $\omega_0$ in $Z$ is shrinkable and hence has negative self-intersection and genus zero. We can then consider the special Lagrangian 3-cycle $\Gamma_0$ which in the limit reduces to an $S^1$ fibration over $C_0$. Since $C_0$ has genus zero, $\Gamma_0$ can at best have $b_1(\Gamma_0)=1$ in which case the fibration is trivial.\footnote{If the fibration is non-trivial, $\Gamma_0$ is topologically an $S^3$ which has $b_1(S^3)=0$.} Applying the same logic as above, we conclude that the SQM obtained by wrapping $n$ D3-branes on $\Gamma_0$ does not have enough scalar fields to break ${\rm U}(n)\to {\rm U}(1)$ such that no bound state with charge $nq_0$ exists. As a result, there is no tower of states with charge $n q_0$ in case $q_0|_{I^{2,0}(\Delta)\oplus I^{0,2}(\Delta)}=0$. Conversely,  any tower of states must have a non-trivial component along $I^{2,0}(\Delta)\oplus I^{0,2}(\Delta)$ such that the mass of the lightest tower of states in a type II limit defined by a growth sector with leading saxion $s^{k_1}$ scales as 
\begin{equation}\label{eq:mtower}
    \frac{m_{\rm tower}}{M_{\rm Pl}} \sim \frac{1}{\sqrt{s^{k_1}}}\,. 
\end{equation}

Before moving on, let us stress that the states furnishing the tower identified above are not mapped to each other upon successive application of the monodromy around the type II divisor. In particular, for a type II degeneration all states in Gr$_2$ are invariant under the monodromy. The tower of light BPS states that we identified is hence qualitatively different from the light towers of states identified in \cite{Grimm:2018ohb,Grimm:2018cpv} which are generated by successive application of the unipotent monodromy around a singular locus.\footnote{As pointed out in \cite{Grimm:2018ohb}, there is also the possibility that a tower of states could be generated by a monodromy around a different divisor. We do not exclude this possibility.} We further stress that in our above reasoning, the geometric understanding of the type II singularity is essential to establish the existence of the BPS towers. The precise scaling of the mass of these BPS states in turn follows by combining these geometric insights with the algebraic treatment of the singularity in terms of Hodge theory.

\subsection{Tensionless Strings}\label{sec:EFTstrings}
Apart from the BPS particle states discussed above, there can also be 4d BPS strings that can give rise to light states. To understand these, we consider BPS string solutions to the effective 4d $\cN=2$ supergravity action, whose bosonic part is given in \eqref{eq:action_4d}. Cosmic string solutions for this action have been discussed in detail in \cite{Lanza:2020qmt,Lanza:2021udy} for which the 4d metric takes the general form 
\begin{equation}
    {\rm d} s^2 = -{\rm d}t^2 + {\rm d}x^2 + e^{2D} {\rm d}z {\rm d}{\bar z}\,.
\end{equation}
Here $(t,x)$ are coordinates along the string whereas $z\in\mathbb{C}$ is the coordinate transverse to the string. The equations of motion for the complex structure fields read 
\begin{equation}
    G_{\bar{\imath}j} \del\bar{\del} t^j + G_{\bar{\imath} jk}\del  t^j \wedge \bar{\del} t^j =0\,,
\end{equation}
where $\del = (\del/\del z) {\rm d}z $. Simple BPS solutions to these equations of motions are obtained for holomorphic profiles ${\bar \del}t=0$. The Einstein equations then fix the warp factor to be 
\begin{equation} \label{warp-factor1}
    e^{2D} = |f(z)|^2 e^{-K}\,,
\end{equation}
for some non-vanishing, holomorphic function $f(z)$ which we can set to a constant. 

We are interested in string solutions that couple to the complex scalar fields $t^i$ spanning the complex structure moduli space. These strings are hence magnetically charged under the axions $a^i\equiv \text{Re}\,t^i$. Let $\mathtt{S}_{\mathbf{e}}$ be a string solution with magnetic charge vector $\mathbf{e}=(e^i)_{i\in \{1,\dots, h^{2,1}\}}\in \mathbb{Z}^{h^{2,1}}$. If we encircle the string in the $z$-plane, this shifts the complex scalar fields by
\begin{equation}
    t^i \to t^i + e^i\,.
\end{equation}
Such a shift is realised for logarithmic solutions \cite{Lanza:2021udy} of the form 
\begin{equation}
    t^i= \frac{e^i}{2\pi {\rm i}} \log \left(\frac{z}{z_0}\right)\,,
\end{equation}
for $z_0$ some constant that sets the overall scale of the solution. Infinitely extended string solutions of this kind change the asymptotics of spacetime, as they are codimension-two sources. In particular, due to the logarithm, $s^i\equiv \text{Im}\,t^i \to -\infty$ as $z\to \infty$. Therefore, infinitely extended string solutions cannot provide states that we should take into account when discussing the light spectrum in asymptotically flat Minkowski space with some fixed asymptotics. 

Instead, we can consider strings wrapping a loop of size $L$. Such configurations have also been discussed in \cite{Lanza:2021udy}, see also \cite{Marchesano:2022avb}. For such closed string configurations, the backreaction dies off at distances $r\gg L$ from the string core such that these strings can be viewed as excitations on top of a vacuum configuration characterised by some fixed asymptotic values for the saxions~$\bar{s}^i$. As discussed in~\cite{Lanza:2021udy,Marchesano:2022avb}, the tension of this string configuration is determined by the energy stored in the backreaction of the string. 
More precisely, the tension 
\begin{equation}
    T_{\mathbf{e}}(\bar{s}^i) = \mathcal{E}_{\rm back}(L, \bar{s}^i)
\end{equation}
is given by the linear energy density $\cE_{\rm back}$ of a string solution defined on the disk $\mathbf{D}(L)\subset \mathbb{C}$ with radius $L$, for which the value of the saxions at the boundary $\del \mathbf{D}(L)$ is 
\begin{equation}
    s^i\left.\right|_{\del \mathbf{D}(L)}=\bar{s}^i\,. 
\end{equation} 
The linear energy density of this configuration can be computed using a dual formulation, again following \cite{Lanza:2021udy}. Since we are interested in asymptotic limits of the moduli space in which the K\"ahler potential has a shift symmetry $t^i \to t^i+c^i$, the K\"ahler potential only depends on the saxionic components of $t^i$. In this case, we can define a dual saxion 
\begin{equation}
 \ell_i = -\frac{\del K}{\del s^i}\,.
\end{equation}
Without repeating the details of the computation, \cite{Lanza:2021udy} shows that in terms of the dual saxion $\ell_i$, the linear energy density for a string with charge $\mathbf{e}$ within a disk of radius $r$ is given by
\begin{equation}
    \mathcal{E}_{\rm back}(r) = M_{\rm pl}^2 \,e^i \ell_i(r)\,. 
\end{equation}
In the case of the loop-configuration of the string, the tension of the string is hence given by 
\begin{equation}\label{eq:tensionstring}
    T_{\mathbf{e}}(\bar{s}^i)= \left.-M_{\rm pl}^2\, e^i \left(\frac{\del K}{\del s^i} \right)\right|_{s^i = \bar{s}^i} \,. 
\end{equation}
With this preparation, we can now consider BPS-string configurations that become tensionless in a type II limit in the complex structure moduli space. To this end, consider a type II limit in the moduli space which is obtained by taking the $r$-parameter limit $s^{k_1},\ldots, s^{k_r}\to\infty$, for some indices $\{k_1,\ldots, k_r\}\subset\{1,\ldots, h^{2,1}\}$, and $1\leq r\leq h^{2,1}$. Furthermore, we denote by
\begin{equation}\label{def:I_II}
    \mathcal{I}_{\mathrm{II}}:=\{k\in\{1,\ldots, h^{2,1}\}: \text{$I^{p,q}(\Delta_{k})$ is of type II} \}
\end{equation}
those indices for which the one-parameter limit $s^{k}\to\infty$ corresponds to a type II limit. Then the K\"ahler potential becomes 
\begin{equation}\label{eq:kahlerpottypeII}
    e^{-K} \sim \sum_{k\in \mathcal{I}_{\rm II}} s^k \,. 
\end{equation}
In the above expression, we have suppressed exponentially small contributions coming from those saxions $s^l$ with $l\not\in\mathcal{I}_{\mathrm{II}}$. Any string whose tension is sensitive to this limit must be magnetically charged under an axion $a^k$ for $k\in \mathcal{I}_{\rm II}$. We thus have to consider strings with $e^k \neq0$ for at least one $k\in \mathcal{I}_{\rm II}$. Moreover, we are interested in identifying the \emph{lightest} string in the spectrum. Therefore, we focus on the strings with charge $e^{i}=0$ for $i\notin \mathcal{I}_{\rm II}$. The tension of these strings is given by 
\begin{equation}
    \frac{T_{\mathbf{e}}(\bar{s}^k)}{M_{\rm pl}^2} = \sum_{k \in \mathcal{I}_{\rm II}} \frac{e^k}{\bar s^k}\,. 
\end{equation}
We now consider a growth sector such that in the asymptotic limit
\begin{equation}\label{growthsector}
    \frac{\bar s^{k_1}}{\bar s^{k_a}} \to 0 \,,\qquad \forall k_a \in \mathcal{I}_{\rm II}\,,k_a\neq k_1\,. 
\end{equation}
In this case, the lightest string has charge $\mathbf{e}_{k_1} = (\delta^{ik_1})$ scaling as 
\begin{equation}\label{eq:stringtension2}
     \frac{T_{\mathbf{e}_{k_1}}(\bar{s}^k)}{M_{\rm pl}^2} = \frac{1}{\bar{s}^{k_1}}\,.
\end{equation}
Such strings whose charge vector corresponds to a generator of the cone of EFT strings were dubbed \emph{elementary} EFT strings in \cite{Lanza:2021udy}.
The scaling of the string tension can be compared to the scaling of the mass of the lightest tower of BPS states. From \eqref{eq:mtower} we then find 
\begin{equation}\label{eq:towervsstring}
    \frac{T_{\mathbf{e}_{k_1}}}{m_{\rm tower}^2}\to \text{const.} \,,
\end{equation}
which corresponds to the case $w=1$ in \cite{Lanza:2021udy}. This behaviour of the string tension in type II limits is as expected for an emergent string~\cite{Lee:2019oct}. 

Let us briefly summarise why the low-energy/algebraic treatment of type II singularities indicates that \emph{general} such limits correspond to emergent string limits. Consider again the limit $s^{k_1}\to \infty$ such that the mass of the towers of BPS states scales as in \eqref{eq:mtower}. The scaling of the tower mass can be compared with the scaling of the species scale, $\Lambda_s$ \cite{Dvali:2007hz}. In general, an estimate for the species scale is given by certain powers of the coefficients of the leading higher-derivative corrections to the Einstein--Hilbert action~\cite{vandeHeisteeg:2023ubh,vandeHeisteeg:2023dlw,Castellano:2023aum,Bedroya:2024ubj,Calderon-Infante:2025ldq}. In 4d $\cN=2$ theories the first non-vanishing correction corresponds to a specific $R^2$-term whose coefficient is given by the genus-one free energy, $F_1$. The inverse square root of $F_1$ then gives a good estimate for the vector multiplet moduli-dependence of the species scale in these theories~\cite{vandeHeisteeg:2022btw,Cribiori:2022nke}. It has further been argued in \cite{Martucci:2024trp}, that in EFT string limits in 4d supersymmetric theories of gravity, the leading  coefficient of the Gauss-Bonnet $R^2$-term is generically linear in the moduli that are sent to infinity. In the type II limit obtained as $s^{k_1}\to \infty$, the species scale thus scales as 
\begin{equation}
    \frac{\Lambda_s}{M_{\rm Pl}}\sim \frac{1}{\sqrt{s^{k_1}}}\,. 
\end{equation}
One thus concludes the species scale vanishes parametrically at the same rate as the mass of the lightest tower of BPS states above. 
 This is indeed a characteristic property of emergent string limits, where the accompanying particle tower is interpreted as a KK tower at the same scale as the string excitations, which defines the species scale \cite{Dvali:2009ks, Dvali:2010vm}. By contrast, in decompactification limits, the particle tower lies parametrically below the species scale, which now corresponds to the higher-dimensional Planck scale. 
 The above argument can straightforwardly be generalized to higher-codimension type II singularities for which $s^{k_r}\to \infty$ for $k_r\in \mathcal{I}_{\rm II}$ defined in \eqref{def:I_II}. Thus the IR perspective suggests that all type II limits are indeed emergent string limits.

However, at this stage we do not know much about the string itself since we only constructed it as a BPS solution of the 4d $\cN=2$ supergravity theory. In order to identify it as a fundamental string, we need to consider the worldsheet theory on this string, which is the subject of 
Section \ref{sec:Worldsheet}.

\subsection{Uniqueness of the Emergent String}

Before turning to the worldsheet theory on the lightest string, 
 we address an important consistency check of the emergent string limit. Namely, while so far we have assumed that in the limit only a single string becomes tensionless at the leading rate, one can also engineer situations in which two (or even more) such strings become light equally fast.
 This is realised if
 instead of a growth sector as in \eqref{growthsector}, leading to only one tensionless string, there is least one more $k_2 \in \mathcal{I}_{\rm II}$ such that in the asymptotic limit, $\bar{s}^{k_1}/\bar{s}^{k_2} = \text{const}$, whereas all other $\bar{s}^{k_i}$ still scale to infinity at a smaller rate. 
  In particular, such behaviour arises at the intersection of two type II divisors where the singularity type does not enhance beyond type II.
  In this regime the tension of the string with charge $\mathbf{e}_{k_2}=(\delta^{ik_2})$ is asymptotically of the same order as the tension of the string with charge $\mathbf{e}_{k_1}$; furthermore both tensions scale as the square of the BPS particle tower mass identified in Section \ref{subsubsec-BPStower}, and, importantly, this is the leading tower scale.\footnote{If the singularity type enhances to type III or type IV, there arise additional BPS towers whose mass lies below the mass scale of the two strings associated with each of the type II divisors. In this case, the coexistence of two different such strings at the same energy is not problematic because none of these strings defines the new perturbative duality frame.} One might hence conclude that there are two different emergent strings becoming weakly coupled at the same time. This would, however, be in conflict with the Emergent String Conjecture~\cite{Lee:2019oct}, according to which there can only be one emergent critical\footnote{A way out would be if the two strings of the same asymptotic mass scale are not critical; however, in Section~\ref{sec:Worldsheet} the criticality will be established for Tyurin degenerations irrespective of the details of the growth sector.} string at a given point in moduli space such that this string defines the perturbative duality frame.

To resolve this apparent contradiction, we now argue that the strings arising at the intersection of multiple type II divisors without enhancing beyond type II define the \emph{same} perturbative duality frame. In other words, the perturbative descriptions of the various strings becoming light at such intersections of multiple type II divisors are compatible with each other. To show this, we assume for now that the EFT string associated with a type II divisor   gives rise to a critical string with a well-defined supergravity low-energy approximation. For the special case of Tyurin degenerations, we argue that this is indeed the case in Section~\ref{sec:Worldsheet}.\\  

We first give a criterion to decide whether multiple perturbative, critical strings describe the same perturbative duality frame and then argue that this criterion is satisfied precisely if the singularity at the intersection of the type II divisors does not enhance beyond type II. 
For simplicity, we focus on the intersection of two $\text{codim}_{\mathbb{C}}=1$ divisors $\Delta_a$ and $\Delta_b$ associated with a type II$_a$ and a II$_b$ degeneration, respectively. The case of higher co-dimension intersections can then be studied by a straight-forward generalization of this basic building block. Let us denote the string associated with $\Delta_a$ ($\Delta_b$) by $\mathtt{S}_a$ ($\mathtt{S}_b$). By assumption, both $\mathtt{S}_a$ and $\mathtt{S}_b$ allow for a low-energy description in terms of a \emph{local} effective field theory corresponding to a perturbative gauge theory coupled to supergravity.

For $\mathtt{S}_a$ and $\mathtt{S}_b$ to be compatible, there must exist a local EFT that can simultaneously describe the gravity and gauge theory arising from both $\mathtt{S}_a$ and $\mathtt{S}_b$.
 This is the case if the states that are perturbative in either duality frame are mutually local. For the gauge sectors this means that there exists an electro-magnetic duality frame for which the states charged under the respective perturbative gauge group $G_a$ and $G_b$ of string $\mathtt{S}_a$ and $\mathtt{S}_b$ are purely electric. If the gauge sector contains a KK-U(1) and/or winding U(1) this also ensures that the graviton arising from the two strings can be identified.\footnote{In Section~\ref{sec:Worldsheet} we will see that the strings arising at Tyurin degenerations always contain KK- and/or winding U(1)s.} Let us denote by $q_a\in \Lambda_{Q,a}$ ($q_b\in \Lambda_{Q,b}$) a state in the charge lattice of the perturbative gauge group $G_a$ ($G_b$). The above compatibility requirement then amounts to the vanishing of the Dirac pairing 
\begin{equation}
    (q_a, q_b) = 0\,, \quad \forall q_a\in \Lambda_{Q,a},\, q_b\in\Lambda_{Q,b}\,. 
\end{equation}
For the string $\mathtt{S}_a$, the states that are electrically charged under $G_a$ lie in $\text{Gr}_2(\Delta_a)$. Since the log-monodromy operator $N_a$ associated with the monodromy around $\Delta_a$ induces an isomorphism (see Appendix \ref{app:hodgestr})
\begin{equation}
    N_a : \text{Gr}_4(\Delta_a) \to \text{Gr}_2(\Delta_a)\,,
\end{equation}
all elements $q_a\in \text{Gr}_2(\Delta_a)$ can be written as 
\begin{equation}
    q_a = N_a \bar q_{a}\,,
\end{equation}
for some $\bar q_{a}\in \text{Gr}_4(\Delta_a)$, and similarly for states $q_b$ electrically charged under the perturbative gauge group of string $\mathtt{S}_b$ associated with $\Delta_b$ and log-monodromy matrix $N_b$. The Dirac pairing for the states $q_a$ and $q_b$ then reads
\begin{equation} \label{eq:qaqboverlap}
    (q_a,q_b)=(N_a \bar q_{a},N_b \bar q_{b}) = -(\bar q_{a}, N_a N_b \bar q_{b})\,. 
\end{equation}
Since, by assumption, the intersection $\Delta_a \cap \Delta_b$ corresponds to a type II singularity, the log-monodromy matrix associated with the enhancement, $N_a + N_b$, must satisfy $(N_a + N_b)^2 = 2 N_a N_b = 0$, where we used that $N_a^2 = 0 = N_b^2$ for the individual type II log-monodromy matrices. By \eqref{eq:qaqboverlap}, this guarantees  $(q_a,q_b)=0$. Therefore, the two strings $\mathtt{S}_a$ and $\mathtt{S}_b$ are necessarily compatible with each other if the singularity at the intersection of their associated divisors in the moduli space does not enhance beyond type II. 

Conversely, if two divisors $\Delta_i$ and $\Delta_j$ give incompatible emergent strings, the singularity at their intersection is guaranteed to enhance to type III or IV because for incompatible strings $N_i N_j \neq 0$ and therefore $(N_i + N_j)^2 \neq 0$. In this way, the pattern of possible enhancements automatically avoids a conflict with the Emergent String Conjecture.\footnote{For degenerations in the K\"ahler sector, the analogous statement shown in \cite{Lee:2019oct} is that if a Calabi-Yau threefold $X_3$ admits two incompatible K3-fibrations, then there exists also a torus-fibration such that in the limit in which the base spaces of the K3-fibrations diverge at the same rate the limit reduces to a limit of Type $T^2$, which is a decompactification rather than an emergent string limit. See also \cite{Grimm:2019bey}.} 

\section{Type IIB/Heterotic Duality from the Worldsheet}
\label{sec:Worldsheet}
In the previous section, we have argued that for type II limits in the complex structure moduli space of Type IIB compactifications on Calabi--Yau threefolds, there exist asymptotically flat string configurations whose tension goes to zero as fast as the square of the mass of the lightest tower of BPS states. In order to establish that these strings correspond to critical strings, we now consider the worldsheet theory on the string.
To achieve this, the information about the degeneration provided by Hodge theory is not sufficient; rather our analysis makes full use of the detailed geometry of the degenerate Calabi--Yau threefold obtained in the type II limit. In the sequel, we therefore focus on the concrete case of Tyurin degenerations introduced in Section~\ref{subsec:Tyurin}.

To analyse the worldsheet theory on the asymptotically tensionless string, we assume a loop configuration of radius $L$ as discussed in Section~\ref{sec:EFTstrings}. Let us zoom in to the core of the string and restrict the string solution to a disk $\mathbf{D}$ of radius $r\ll L$. In this local treatment of the string solution, the loop configuration is well-approximated by a straight and infinitely extended string. Let us focus on the string with charge $\mathbf{e}_{k_1}$ defined below \eqref{growthsector}. Then, the string solution restricted to $\mathbf{D}$ is of the form 
\begin{equation}\begin{aligned}\label{eq:EFTstringsol}
     {\rm d} s^2 &= -{\rm d}t^2 + {\rm d}x^2 +e^{2D} {\rm d}z {\rm d}{\bar z}\,, \qquad \text{with} \quad e^{2D} = f_0 e^{-K} \,,\\
     t^{k_1}(z) &= {\rm i} \bar{s}^{k_1} + \frac{1}{2\pi {\rm i}} \log\left(\frac{z}{r}\right)\,,
\end{aligned}\end{equation}
where $z$ is the coordinate along $\mathbf{D}$. The above solution is a $\frac12$-BPS solution of the 4d $\cN=2$ supergravity theory such that the 2d worldsheet theory preserves four supercharges (half of the supercharges of the 4d $\cN=2$ supergravity theory). Equivalently, the string configuration can be viewed as a $\frac18$-BPS solution of the 10d Type IIB supergravity equations of motion. The superalgebra on the worldsheet of the string therefore descends from the chiral $\cN=(0,2)$ supersymmetry of Type IIB supergravity in 10d and is also chiral. This gives a string with $\cN =(0,4)$ worldsheet supersymmetry. 

From the perspective of the 10d Type IIB action, the spacetime can be viewed as a fibration of the Calabi--Yau threefold $V$ over the disk $\mathbf{D}$, which represents the two directions normal to the string in the four extended spacetime dimensions. Whereas the generic fibre $V_z$ of this fibration is smooth, the central fibre, $V_0$, realises the Tyurin degeneration described around \eqref{eq:V0split} and depicted in Figure \ref{fig:Tyurin}, i.e.~$V_0=X_1\cup_{Z} X_2$. The condition for the resulting configuration to preserve four supercharges in 2d is then equivalent to the total space $\cV$ of the fibration, as defined in \eqref{eq:4foldV}, to be a non-compact Calabi--Yau fourfold. 

\subsection{The Worldsheet Spectrum}\label{subsec:WSspectrum}
To infer the details of the worldsheet theory on the string, we begin with the spectrum of massless modes propagating along it. The first set of zero modes is associated with its position on the disk $\mathbf{D}$ and encodes the transverse fluctuations in the extended spacetime directions. We can view $\mathbf{D}$ as a subset of $\mathbb{C}$ and fix the string position to the centre of the disk. The universal zero modes of the string then correspond to the location $\mathbf{z}_0$ of the centre of $\mathbf{D}$ in $\mathbb{C}$. The scalar field $\mathbf{z}_0$ can be decomposed into two real scalars (left- and right-moving) of the 2d worldsheet theory accompanied by two right-moving fermions to form a 2d $\cN=(0,4)$ hypermultiplet. To keep the expressions simple, for the remainder of this section we set $\mathbf{z}_0=0$. \\

Additional bosonic zero modes arise from massless $p$-form fields of the 4d $\cN=2$ supergravity that localise on the string at $z=0$. The components of such $p$-form fields transverse to the string locus give rise to scalar fields on the worldsheet. To find these, we must hence  identify the $1$- and 2-form fields of the 4d $\cN=2$ theory that localise to $\{z=0\}\subset \mathcal{V}$.\footnote{Other $p$-form symmetries with $p>2$ would give rise to $(p-2)$-forms on the string worldsheet which are non-dynamical for $p>2$ in 2d.}

In the present Type IIB setup, the relevant $p$-form fields arise by expanding the higher-form fields of Type IIB supergravity in $n$-forms on the central fibre $V_0\subset \cV$. Let us start with the self-dual 4-form $C_4$ of Type IIB string theory. To obtain a 2-form in the 4d theory we can reduce $C_4$ along the 2-forms on the central fibre as 
\begin{equation}
    C_4 = B_2^A\wedge\omega_A \,,
\end{equation}
with $\omega_A$ either an element of $H^2(X_1)\oplus H^2(X_2)$ or $H^2(Z)$. In Section \ref{subsec:Tyurin} it was explained that if $\omega_A$ localises to $\{z=0\}$, then it can always be pulled back to $H^2(Z)$, cf.~Fact~\ref{factcheck}. As a consequence, $C_4$ gives rise to localised 2-forms when expanded in a basis $\omega_\a \in H^2(Z)$. Furthermore, as explained around equations \eqref{eq:decomp_omega}--\eqref{eq:decomp_omega_2}, we can distinguish two types of contributions 
\begin{equation}
\begin{aligned}
\label{eq:C4H2Z}
    C_4 = B_2^\a \wedge \omega_\a   
= B^i_2 \wedge (i^*-j^*) \bar\omega_i + 
    B_2^a\wedge \omega_a\,,
\end{aligned}
\end{equation}
where 
\begin{equation}
\bar{\omega}_i \in H^2(X_1)\oplus H^2(X_2) \,, \qquad \omega_a \in H^2(Z)/(\text{Im}(i^*)+\text{Im}(j^*)) \,,
\end{equation}
associated with the lattices $\Lambda_{\rm pol}$ and $\Lambda_{\rm trans}$ of $Z$, respectively. 
The scalar fields on the string worldsheet then arise from the modes of $B_2^\a$ transverse to $Z\times \{z=0\} \subset V_0\times \mathbf{D}$. 

As it turns out, for each element $\omega_\a\in H^2(Z)$ the associated $B_2^\a$ gives rise to only a single scalar mode on the string worldsheet. To see this, notice that the kinetic terms for the zero modes on the string are inherited from the kinetic terms for $C_4$ in the bulk Type IIB supergravity action 
\begin{equation}\label{eq:SC410}
    S_{\rm 10d} = \int {\rm d}C_4 \wedge \star_{10d} \,{\rm d} C_4\,.
\end{equation}

This action can be evaluated for the two types of 2-forms on $H^2(Z)$ that appear in the expansion \eqref{eq:C4H2Z} in turn.  Consider first the part of the expansion 
\begin{equation}
    C_4 \supset B^i_2 \wedge (i^*-j^*) \bar\omega_i\,, \qquad \bar{\omega}_i \in H^2(X_1)\oplus H^2(X_2)\,,
\end{equation}
associated with $\Lambda_{\rm pol}$.
When evaluating \eqref{eq:SC410} with this expansion,  the fibration structure of $\cV$ allows us to split the integration over $\cV$ into the integration over $V_z$ and $\mathbf{D}$. The only component of $B_2^i$ that contributes to the effective action will then be the ${\rm d}z\wedge {\rm d}{\bar z}$ component. Denoting this component by $b^i$, the effective action gives 
\begin{equation} \label{eq:action-bi}
    S_{\rm 10d} \to \int_\mathbf{D} {\rm d} z\, {\rm d}{\bar z} \int_{\mathbb{R}^{1,1}} \mathrm{d}t\, \mathrm{d}x (\partial b^i)(\partial b^j) e^{2D(z,\bar{z})} \int_{V_z} \bar{\omega}_i \wedge\star\,\bar{\omega}_j\,,
\end{equation}
where the factor of $e^{2D(z,\bar z)}$ arises from the metric on $\mathbf{D}$ given in~\eqref{eq:EFTstringsol}. The last integration yields an expression proportional to $\delta^{(2)}(z,\bar{z})$ because the 2-forms $\bar{\omega}_i$ only have support at $z=0$.

 It remains to analyse the modes from the expansion of $C_4$ along the 2-forms in $\Lambda_{\rm trans}$, 
\begin{align}
    C_4\supset B_2^a\wedge \omega_a\,,\qquad \omega_a \in H^2(Z)/(\text{Im}(i^*)+\text{Im}(j^*))\,.
\end{align}
The images of the forms $\omega_a$ under $d^*$ are non-trivial in $H^3(V_0)$ such that 
\begin{equation}
    0\neq d^* \omega_a = \gamma_a \in H^3(V_0)\,. 
\end{equation}
In fact, from the isomorphisms \eqref{eq:G2isos-1} we recall that the space of 3-forms writeable in this fashion is nothing but the space ${\rm Gr}_2(\Delta)$ given in \eqref{eq:Gr2-def}. 
In the vicinity of $Z$, the form $\gamma_a$ can locally be written as
\begin{equation}
    \gamma_a \to {\rm d}\phi \wedge \omega_a\,,
\end{equation}
with $\phi$ the coordinate transverse to $Z$ inside $V_0$. Accordingly, the only component of $B_2^a$ giving rise to a zero mode propagating along the string is the coefficient of ${\rm d}|z| \wedge {\rm d} \phi$. Denoting this coefficient by $b^a$, its kinetic term arising from the reduction of the kinetic term for $C_4$ in the 10d Type IIB action reads
\begin{equation} \label{eq:action-ba}
     S_{\rm 10d} \to \int_\mathbf{D} {\rm d} z\, {\rm d}{\bar z} \int_{\mathbb{R}^{1,1}} \mathrm{d}t \,\mathrm{d}x (\partial b^a)(\partial b^b) e^{2D(z,\bar{z})} \int_{V_z} \gamma_a \wedge\star\,\gamma_b\,.
\end{equation}
Since the 3-forms $\gamma_a$ exist on the generic fibre $V_z$ the last integral in the above expression does not yield a $\delta$-function. 

We thus conclude that each element of $H^2(Z)$ yields exactly one zero mode propagating along the effective string at $z=0$. In total this leaves us with $b_2(Z)=22$ scalar zero modes on the string, out of which $b_2^+(Z)=3$ are right-moving and $b_2^-(Z)=19$ are left-moving.\footnote{In addition, there could be 1-forms arising from the expansion of $C_4$ in 3-forms in $H^3(X_1)\oplus H^3(X_2)$ whose radial component can give additional zero modes on the string. However by \eqref{eq:MV-sequence1} (together with $H^3(Z)=0$), the map from $H^3(\cV) \cong H^3(V_0)$ to $H^3(X_1)\oplus H^3(X_2)$ is surjective. This implies that all 3-forms in the latter space are the pull-back of 3-forms on $H^3(\cV)$ and so do not localise on the central fibre. Thus all scalar zero modes on the string arising from localised modes of $C_4$ are counted by $H^2(Z)$.}\\

Another source of zero modes along the string is from the Type IIB supergravity fields $B_2$ and $C_2$, or, equivalently, from their magnetic dual 6-form fields ${B}_6$ and $C_6$. In fact, it is more straight-forward to work with these latter forms. According to our general discussion, if these forms give rise to localised massless $1$- or $2$-forms on the central fibre this in turn yields zero modes on the string. To obtain a 1-form from ${B}_6$ or $C_6$, we would need to expand these forms in harmonic 5-forms on either $X_1$, $X_2$ or $Z$. None of these spaces, however, supports any harmonic 5-forms.\footnote{For $Z$ this is obvious by dimensionality. On the other hand, since the $X_i$ are (quasi-)Fano, $H^5(X_i)=0$  by Hodge duality.} This leaves us with the possibility of obtaining 2-forms from reduction of the $6$-forms. To this end, we must consider harmonic 4-forms on $X_1$, $X_2$ and $Z$. Similar to the 2-forms, these are related via the Mayer--Vietoris sequence \eqref{eq:MV-sequence1}. By the same reasoning as in Section \ref{subsec:Tyurin}, all localised harmonic 4-forms arise from $H^4(Z)$ which, given that $Z$ is a K3 surface, is one-dimensional. Denoting the single 4-form on $Z$ by $\tilde\omega_{\rm K3}$ one obtains the massless, localised 2-forms from the expansion
\begin{equation}
\label{eq:B6_C6_reduction}
    B_6 = {\cal B}_{2}\wedge \tilde{\omega}_{\rm K3}\,,\qquad C_6 = {\cal C}_2\wedge \tilde\omega_{\rm K3}\,,
\end{equation}
for which the respective $(z,\bar{z})$ components yield two real scalar zero modes on the string. Since neither $C_6$ nor $B_6$ are self-dual the resulting scalar zero modes do not have definite chirality. This results in two real scalars on the worldsheet with left- and right-moving components.\footnote{Notice that reduction of $C_4$ along $\tilde \omega_{\rm K3}$ does not give rise to a real scalar mode localised along the string, nor does the reduction of $B_2$ and $C_2$ along the 2-forms localised on $Z$. The reason is that the scalar modes obtained in this way can also propagate transverse to the string along the disk $\mathbf{D}$ as is clear from the dimensional reduction of the 10d kinetic term for the Type IIB $p$-forms. }

\begin{table}[t]
    \centering
    \begin{tabular}{|c|c|c|c|}
    \hline &&&\\[-1em]
         \textbf{Field}& \textbf{Origin} & \textbf{free/interacting} & $(n_L,n_R)$\\&&&\\[-1em]\hline &&&\\[-1em]
         $|\mathbf{z}_0|$ & radial transverse coordinate $\in \mathbb{C}^*$ & interacting & $(1,1)$\\&&&\\[-1em]\hline &&&\\[-1em]
        $\arg(\mathbf{z}_0)$ & angular coordinate $\in \mathbb{C}^*$ & free & $(1,1)$\\&&&\\[-1em]\hline &&&\\[-1em]
         $\Phi$ & transverse coordinate $\in V_0$ & free& $(1,1)$ \\ &&&\\[-1em]\hline &&&\\[-1em]
         $\mathbf{b}^0$ & Reduction of $B_6$ over $H^4(Z)$ & interacting  & $(1,1)$ \\ &&&\\[-1em]\hline &&&\\[-1em]
         $\mathbf{c}^0$  & Reduction of $C_6$ over $H^4(Z)$ & interacting  &$(1,1)$\\ &&&\\[-1em]\hline &&&\\[-1em]
         $b^a$ & Reduction of $C_4$ over $\Lambda_{\rm trans}\subset H^2(Z)$ & free &$(19-\rho,2)$ \\&&&\\[-1em]\hline &&&\\[-1em]
          $b^i$ & Reduction of $C_4$ over $\Lambda_{\rm pol}\subset H^2(Z)$ & interacting & $(\rho,1)$ \\&&&\\[-1em]\hline\hline &&&\\[-1em]
         \multirow{2}{*}{\textbf{Total}} &&free & $(21-\rho,4)$ \\ 
         && interacting& $(\rho+3,4)$ \\&&&\\[-1em]\hline 
         
    \end{tabular}
    \caption{A summary of the scalar fields on the worldsheet of the EFT string realising a Tyurin degeneration in  compactifications of Type IIB string theory on Calabi--Yau threefolds including their 10d origin discussed in Section~\ref{subsec:WSspectrum}.  For degenerations of type $\mathrm{II}_{b}$,  $b= 19-\rho$. We also summarise whether or not there are worldsheet interactions for the scalar fields as analysed in Section~\ref{subsec:interactions}. In the last coloumn, $n_{L/R}$ denotes the number of left/right-moving scalar fields associated with the respective class of fields. }
    \label{tab:my_label}
\end{table}

To summarise, all zero modes on the string arising from massless higher-form fields of Type IIB supergravity are associated with harmonic forms on $Z$. In addition there are scalar modes arising from the deformation space of $Z$ inside $V_0$. If $V_0$ was smooth, there would be one complex deformation of the divisor $Z$ since $h^{2,0}(Z)=1$. However, $V_0$ is not smooth since it splits into two components $X_1$ and $X_2$ intersecting over $Z$. Geometrically, this degeneration corresponds to one of the directions normal to $Z$ shrinking to zero size. Hence, the deformation space of $Z$ has $\text{dim}_{\mathbb{R}}=1$, providing one real scalar on the string.\footnote{In the examples of Tyurin degenerations that will be discussed in Sect.~\ref{subsec:directomirror}, one can geometrically understand the deformation space of $Z$ as follows. For the degeneration depicted in Fig.~\ref{fig:Tyurin-degeneration_IIA-split}, it corresponds to the transverse direction of the cycles $\widehat{\gamma}_1,\widehat{\gamma}_2$ inside the base space. In the degenration depicted in Fig.~\ref{fig:Sen}, it corresponds to the transverse direction of $\gamma$ inside the fibre.} 
Together with the two modes transverse to the string in the extended 4d spacetime, in total we have eight right-moving scalars and 24 left-moving scalars. Since the worldsheet theory on the string preserves $\cN = (0,4)$ supersymmetry, the right-moving scalars have fermionic partners to form complete multiplets of the superalgebra on the worldsheet. The left-and right-moving central charges of the 2d theory are hence 
\begin{equation}
    c_L = 24 \,,\qquad c_R=12 \,,
\end{equation}
which are exactly the values expected for the critical heterotic string in light-cone gauge. 

\subsection{The Worldsheet Action}\label{subsec:interactions}

The next step in identifying the $\sigma$-model on the string is to analyse
 the action for the massless scalar fields.  In the sequel, we study the interactions between the zero modes on the string via the reduction of the 10d Type IIB action on $\cV$. If we collectively denote by $\phi^\mu$ the scalar fields on the string worldsheet, then the kinetic terms of the associated $\sigma$-model are of the form 
\begin{equation}
    S_{\rm WS, kin}=\frac12 \int {\rm d}^2\sigma \, g_{\mu \nu}(\phi) \, \partial_\sigma \phi^\mu \partial^\sigma \phi^\nu\,. 
\end{equation}
If the metric $g_{\mu\nu}$ does not depend on the $\phi^\rho$, all fields on the worldsheet are free; otherwise there are interactions between the corresponding scalar modes. We now extract the kinetic terms for the different modes on the string worldsheet, starting with the modes arising from $C_4$. Our general procedure is to insert the expansion for $C_4$ into \eqref{eq:SC410} to arrive at an expression of the form 
\begin{equation}\label{calSdef}
    S_{\rm 10d} \supset \int_\mathbf{D} {\rm d} z\, {\rm d}{\bar z} \,\mathcal{S}(z,\bar{z},b^A) \,, 
\end{equation}
from which we read off the action on the string worldsheet as 
\begin{equation}\label{eq:SWS}
    \delta^{(2)}(z-\mathbf{z}_0)\, S_{\rm WS, kin}(\mathbf{z}_0,\bar{\mathbf{z}}_0,b^A) = \delta^{(2)}(z-\mathbf{z}_0) \, \mathcal{S}(z-\mathbf{z}_0,\bar{z}-\bar{\mathbf{z}}_0,b^A)\,.
\end{equation}
Here we have re-introduced the variable $\mathbf{z}_0$ determining the location of the centre of the disk $\mathbf{D}$ in $\mathbb{C}$. Recall that, from the worldsheet perspective, $\mathbf{z}_0$ corresponds to a massless scalar field propagating along the worldsheet.

\paragraph{Kinetic Term for $b^i$.} We analyse the two classes of scalar fields arising from $C_4$ in turn. Starting with $b^i$ appearing in \eqref{eq:action-bi} we obtain 
\begin{equation}
    \cS(z-{\mathbf{z}}_0,\bar{z}-\bar{\mathbf{z}}_0, b^i) = e^{2D(z,\bar{z})} \delta^{(2)}(z-{\mathbf{z}}_0)\; \Omega_{ij}\;\int_{\mathbb{R}^{1,1}} {\rm d}^2 \sigma\; \partial_\sigma b^i \partial^\sigma b^j \,.
\end{equation}
Here we have introduced the notation 
\begin{equation}
    \int_{V_z} \bar{\omega}_i\wedge \star\, \bar{\omega}_j = \Omega_{ij}\; \delta^{(2)}(z-\mathbf{z}_0)\,,
\end{equation}
which reflects that the 2-forms $\bar{\omega}_i$ only have support on $V_0$. Using \eqref{eq:SWS} this gives 
\begin{equation} \label{eq:Sforbi}
    S_{\rm WS,kin}(\mathbf{z}_0,\bar{\mathbf{z}}_0,b^i)= \frac12 \int {\rm d}^2 \sigma \, g_{ij}(\mathbf{z}_0,\bar{\mathbf{z}}_0) \,\partial_\sigma b^i \partial^\sigma b^j\,,\quad \text{with} \quad g_{ij}(\mathbf{z}_0,\bar{\mathbf{z}}_0) = 2e^{2D(\mathbf{z}_0,\bar{\mathbf{z}}_0)} \Omega_{ij}\,.
\end{equation}
The metric for the fields $b^i$ thus depends non-trivially on the field $\mathbf{z}_0$ on the worldsheet which thereby induces non-trivial interactions between these fields. Note furthermore that the moduli space $g_{ij}(\mathbf{z}_0,\bar{\mathbf{z}}_0)$ diverges in the infinite distance limit:    
 To see this, recall that, for the BPS string solution \eqref{eq:EFTstringsol}, the warp factor is determined by the K\"ahler potential on the complex structure moduli space as 
\begin{equation}
    e^{2D} = f_0 e^{-K}\,,
\end{equation}
where we set the holomorphic function $f(z)$ to a constant. For a Tyurin degeneration, the relation \eqref{eq:kahlerpottypeII} then leads to a divergence for $z\to \mathbf{z}_0$ as
\begin{equation}
    e^{2D} \sim \log\left|\frac{z-\mathbf{z}_0}{r}\right|\,,\quad \text{for}
    \quad z\to \mathbf{z}_0\,,
\end{equation}
where we recall that $L$ is the radius of the disk as in \eqref{eq:EFTstringsol}. 
The divergence of the moduli space metric $g_{ij}(\mathbf{z}_0,\bar{\mathbf{z}}_0)$ in fact plays an important role for the physical interpretation of the Tyurin degeneration as an emergent string limit, as we will explain in Section \ref{sec:compvsnonc}.

\paragraph{Kinetic Term for $b^a$.} For the modes $b^a$ associated with the forms $\omega_a \in H^2(Z)/(\text{Im}(i^*)+\text{Im}(j^*))$, the kinetic term \eqref{eq:action-ba} takes the form
\begin{equation}
    \cS(z,\bar{z},b^a) = e^{2D(z,\bar{z})} \widetilde\Omega_{ab}(z,\bar{z})\;\int_{\mathbb{R}^{1,1}} {\rm d}^2 \sigma\; \partial_\sigma b^a \partial^\sigma b^b\,,
\end{equation}
where we have defined 
\begin{equation}
\widetilde\Omega_{ab}(z,\bar{z}) = \int_{V_z} \gamma_a \wedge \star\, \gamma_b\,,\quad \gamma_a = d^* \omega_a\,. 
\end{equation}
Notice that, unlike $\Omega_{ij}$, $\widetilde\Omega_{ab}$ depends on $z$ since the forms $\gamma_a$ exist on the generic fibre $V_z$. From \eqref{eq:SWS} we then find 
\begin{equation}\label{SWSba}
    S_{\rm WS}(\mathbf{z}_0,\bar{\mathbf{z}}_0, b^a) = \frac12 \int {\rm d}^2 \sigma \, g_{ab}(\mathbf{z}_0,\bar{\mathbf{z}}_0) \,\partial_\sigma b^a \partial^\sigma b^b\,, \quad g_{ab}(\mathbf{z}_0,\bar{\mathbf{z}}_0) = 2e^{2D(\mathbf{z}_0,\bar{\mathbf{z}}_0)} \widetilde\Omega_{ab}(\mathbf{z}_0,\bar{\mathbf{z}}_0)\,.
\end{equation}
The above expressions suggest that also the kinetic terms for $b^a$ depend on $\mathbf{z}_0$. However, as we now show, the $\mathbf{z}_0$-dependence of $\widetilde\Omega_{ab}(\mathbf{z}_0,\bar{\mathbf{z}}_0)$ precisely cancels the $\mathbf{z}_0$-dependence in $e^{2D}$ such that $g_{ab}$ is just a constant. 
 To determine the $z$-dependence of $\tilde\Omega_{ab}$ we must consider the growth of the Hodge norm $\|\gamma_a\|$ as $z\to \mathbf{z}_0$.
Given the isomorphism \eqref{eq:G2isos-1}, the forms $\gamma_a$ obtained from $\omega_a$ via the boundary map are mapped to elements of ${\rm Gr}_2$ upon pulling back with respect to the inclusion map $\iota: V_z \to \cV$. By the growth theorem \eqref{eq:growth_theorem} we  find
\begin{equation}
    \gamma_a\in\mathrm{Gr}_2:\quad \|\gamma_a\|^2\sim \frac{1}{\log|\frac{z-\mathbf{z}_0}{r}|}\sim e^{K}\,,\label{Hodge_norm_Gr2}
\end{equation}
where we used \eqref{eq:kahlerpottypeII} and the string solution \eqref{eq:EFTstringsol} subject to the shift $z\to z-\mathbf{z}_0$. 
As a consequence, the growth of the Hodge norm of a form $\gamma_a$ appearing in the definition of $\widetilde\Omega_{ab}$ exactly cancels the $\mathbf{z}_0$-dependence of the $e^{2D}$ factor appearing in $g_{ab}$ in \eqref{SWSba}.\footnote{To address the scaling of the Hodge inner product $\langle\gamma_a,\gamma_b\rangle$ for elements in $\mathrm{Gr}_2$ we may work in a basis in which the asymptotic Hodge star operator is diagonal.} Hence, the kinetic term of the modes $b^a$ along the string is independent of $\mathbf{z}_0$,
\begin{equation} \label{eq:Sforba}
 S_{\rm WS}(\mathbf{z}_0,\mathbf{\bar{z}}_0, b^a) = \frac12 \int {\rm d}^2 \sigma \, g_{ab} \,\partial_\sigma b^a \partial^\sigma b^b\,, 
 \end{equation}
and we can treat the modes $b^a$ as free modes on the string worldsheet.

\paragraph{Kinetic term for $\mathbf{b}^0$ and $\mathbf{c}^0$.} In addition to the zero modes arising from $C_4$, there are the modes along the string coming from $\cB_2$ and $\tilde{\cB}_2$, the 2-forms obtained by expanding $B_6$ and $C_6$ as described in equation \eqref{eq:B6_C6_reduction}. These 2-forms can be treated in analogy to the 2-forms $B^i_2$. Accordingly, the $z,\bar{z}$ components of $\cB_2$ and $\tilde{\cB}_2$ give real scalar modes along the string which we denote by $\mathbf{b}^0$ and $\mathbf{c}^0$, respectively. From the kinetic term in 10d, we can define the function $\cS$ in complete analogy to \eqref{calSdef} and find 
\begin{equation}
    \cS(z-\mathbf{z}_0,\bar{z}-\bar{\mathbf{z}}_0,\mathbf{b}^0,\mathbf{c}^0) = e^{2D(z,\bar z)} \delta^{(2)}(z-\mathbf{z}_0) \eta \int_{\mathbb{R}^{1,1}}\mathrm{d}^2\sigma \left[\partial_\sigma \mathbf{b}^0 \partial^\sigma \mathbf{b}^0+\partial_\sigma \mathbf{c}^0\partial^\sigma \mathbf{c}^0\right]\,,
\end{equation}
where we have introduced 
\begin{equation}
    \int_{V_z} \bar{\omega}_{\rm K3} \wedge\star\, \bar{\omega}_{\rm K3} = \eta \delta^{(2)}(z-\mathbf{z}_0) \,\qquad \tilde\omega_{\rm K3} = (i^*-j^*)\bar{\omega}_{\rm K3}\,. 
\end{equation}
The worldsheet kinetic term for the fields arising from $C_6$ and $\tilde{B}_6$ is then given by 
\begin{equation}
    S_{\rm WS, kin}(\mathbf{z}_0,\bar{\mathbf{z}}_0,\mathbf{b}^0,\mathbf{c}^0)=\frac12 \int {\rm d}^2 \sigma\, g_{00}(\mathbf{z}_0,\bar{\mathbf{z}}_0) \left[\partial_\sigma \mathbf{b}^0 \partial^\sigma \mathbf{b}^0+\partial_\sigma \mathbf{c}^0\partial^\sigma \mathbf{c}^0\right] \,,
\end{equation}
with 
\begin{equation}
    g_{00} = 2e^{2D(\mathbf{z}_0,\bar{\mathbf{z}}_0)} \eta\,,
\end{equation}
such that the kinetic term for the fields $\mathbf{b}^0$ and $\mathbf{c}^0$ depends on $\mathbf{z}_0$ and introduces interactions between these fields on the worldsheet. 

\paragraph{Kinetic term for transverse modes.}
This leaves us with the kinetic term for the transverse modes to the string. On the one hand, we have the single real field $\Phi$ describing the normal deformations of $Z$ in $V_0$. The modes of $\Phi$ completely decouple from the other modes on the string, since it is a modulus of the central fibre $V_0$ determining the intersection locus of the two (quasi-)Fano threefolds $X_1$ and $X_2$. Therefore, the kinetic term of $\Phi$ is simply 
\begin{equation}
    S_{\rm WS, kin}(\Phi) = \frac12  \int_{\mathbb{R}^{1,1}} \mathrm{d}^2\sigma \,\partial_\sigma \Phi \partial^\sigma \Phi\,,
\end{equation}
i.e.~$\Phi$ is a free field on the worldsheet.  Finally, we have the modes corresponding to the complex field $(\mathbf{z}_0,\bar{\mathbf{z}}_0)$. The worldsheet fields must arrange into full supermultiplets of the 2d $\cN=(0,4)$ supersymmetry, and in particular the number of free right-moving scalars must come in multiples of four.\footnote{By contrast, the purely right-moving supersymmetry algebra does not force the left-moving scalars within a hypermultiplet to be either all interacting or all free.} This implies that the modes $(\mathbf{z}_0,\bar{\mathbf{z}}_0)$ must give a free and an interacting real scalar field. Since the kinetic terms for the interacting fields only depend on $|\mathbf{z}_0|$ we conclude that $|\mathbf{z}_0|$ is an interacting scalar whereas $\text{arg}(\mathbf{z}_0)$ remains free, see (\ref{tab:my_label}) for a summary.
\subsection{Heterotic Dual} \label{sec_Hetdual}

To analyse the target space of the string $\sigma$-model we start from the grouping of the  string  zero modes  into free and interacting fields achieved in the previous section and summarised in Table \ref{tab:my_label}. 
  The number of free versus interacting fields depends on the embedding of the K3 surface $Z$ into the degenerating threefold $V_0$:
 There are the  free fields $b^a$, $\Phi$, and ${\rm arg}({\bf z}_0)$, 
   where the scalars $b^a$ are associated with the 2-forms in $H^2(Z)$ that form the transcendental sub-lattice $\Lambda_{\rm trans}\subset \Gamma_{3,19}$ with signature $(2,b)$, see \eqref{eq:defLambdatrans}. 
  All remaining modes in Table \ref{tab:my_label} are interacting. Of these, 
  the interacting modes $b^i$ are associated with the polarization lattice $\Lambda_{\rm pol}$, the orthogonal complement of $\Lambda_{\rm trans}$ in $\Gamma_{3,19}$  induced by $H^2(X_1)$ and $H^2(X_2)$ on $H^2(Z)$, see \eqref{eq:Lambda_pol}, which has signature $(1,\rho)$, with $\rho=19-b$.

As we will discuss now, the interpretation of the $\sigma$-model target space depends on the rank of the lattice $\Lambda_{\rm trans}$ (or of its complement $\Lambda_{\rm pol}$).

\subsubsection{$20\geq \text{rk}(\Lambda_{\rm trans})\ge 4$ ($1\le \rho\le 17$)} We begin with the simplest case, corresponding to $1\le \rho\le 17$ (i.e. $2 \leq b \leq 18$). In this situation $\Lambda_{\mathrm{trans}}$ contains two copies of the hyperbolic plane $U$, i.e.~$U\oplus U\subseteq\Lambda_{\mathrm{trans}}$, such that the two right-moving scalars arising from $C_4$ are always paired with two left-moving scalars to give two real, free scalars on the worldsheet. These scalars are compact. Let us denote the two right(left)-moving scalars by $(b^1_{+ (-)}, b^2_{+(-)})$ and the real scalars by 
\begin{equation}\label{eq:mathbfbdef}
    \mathbf{b}^{1,2} = b^{1,2}_{+} + b^{1,2}_{-}\,.
\end{equation}
Since the 2-forms associated with the $b^a$ can be extended to 3-forms on $V_z$, the compact scalars $\mathbf{b}^{1,2}$ give rise to gauge fields of a $U(1)^4$ gauge group in the bulk arising from the expansion 
\begin{equation}
    C_4 \supset A_\pm^{a} \wedge \gamma_a^\pm   \,,\qquad \gamma_a^\pm = \iota^* d^* \omega_a^\pm \,, \quad a=1,2\,. 
\end{equation}
The real scalars $\mathbf{b}^{1,2}$ can be viewed as coordinates on the target space of the string $\sigma$-model. Furthermore, since they give rise to four ${\rm U}(1)$ gauge fields in the bulk, the corresponding target space factor must be a copy of $T^2$ with two of the Abelian gauge factors arising as KK-${\rm U}(1)$'s and the remaining two ${\rm U}(1)$'s coming from the winding of the string. More precisely, the winding and KK-${\rm U}(1)$'s of the two $S^1$ factors in $T^2$ can be identified with the fields $A^a_\pm$ as 
\begin{equation}\label{eq:gaugefields}
    A^a_{\rm KK} = A^a_+ + A^a_- \,,\qquad A_{\rm wind.}^a = A^a_+ - A^a_-\,. 
\end{equation}

The remaining free scalars arising from $C_4$ are purely left-moving and realise a left-moving $U(1)^{17-\rho}$ algebra on the string worldsheet that gives rise to a $U(1)^{17-\rho}$ gauge group in the bulk.\footnote{Subleading corrections to the leading order behaviour of the Hodge norm \eqref{Hodge_norm_Gr2} would trigger (some of) the scalars $b^a$ to be interacting. On the heterotic side, this is associated with a perturbative breaking of the gauge group.} 
On the other hand, the free real scalars $\Phi$ and ${\rm arg}({\bf z}_0)$ on the worldsheet do not lead to gauge fields in the bulk. The target space for $\Phi$ is non-compact $\mathbb{R}$, and for ${\rm arg}({\bf z}_0)$ it is an angular coordinate that will be interpreted momentarily.

Let us now turn to the interacting fields on the string worldsheet and start again with the modes coming from $C_4$. The interacting modes are associated with 2-forms in the polarization lattice $\Lambda_{\rm pol}$ of the K3-surface $Z$. Since we assume $1\le \rho \le 17$, the single right-moving interacting scalar is always paired with a left-moving scalar to give a real scalar on the worldsheet. In analogy to \eqref{eq:mathbfbdef}, let us define 
\begin{equation}\label{def:b3}
    \mathbf{b}^3 = b^3_+ + b^3_-\,. 
\end{equation}
$\mathbf{b}^3$ is a compact real scalar on the worldsheet, whose target space is an $S^1$. However, since it does not give rise to a ${\rm U}(1)$ gauge field in the bulk, there is no one-cycle in the target space associated with $\mathbf{b}^3$. Instead, the $S^1$ is non-trivially fibred and degenerates at the location of the string. 

Similarly, the remaining left-moving scalar fields $b^{i\neq 3}$ do not give rise to massless gauge fields in the bulk. If we turned off the interactions for these fields, these would give rise to ${\rm U}(1)$ gauge fields in the bulk. Turning on interactions for the worldsheet fields corresponds to Higgsing this  Abelian gauge group in the bulk. The bulk hence contains a $U(1)^{\rho-1}$ gauge group that is completely Higgsed. 

What remains to be discussed are the fields $\mathbf{b}^0$ and $\mathbf{c}^0$. Being compact real scalars, they can be viewed as coordinates on the target space of the $\sigma$-model and give rise to $S^1$'s that are non-trivially fibred over the space transverse to the string due to the non-trivial interaction of these fields with $\mathbf{z}_0$. Thus, the fields behave like the field $\mathbf{b}^3$ and together form a $T^3$. From the target space perspective, two of these three scalar fields form a $T^2$ that is non-trivially fibred over the $\mathbf{z}_0$-plane whereas the remaining coordinate can be viewed as the angular coordinate in the space transverse to the string. In the following, we view $(\mathbf{b}^0, \mathbf{c}^0)$ as the coordinates on the torus $T^2$. The interaction between $|\mathbf{z}_0|$ and the fields $\mathbf{b}^0, \mathbf{c}^0$ and $\mathbf{b}^3$ is manifest as a non-trivial fibration of the torus parametrised by $(\mathbf{b}^0, \mathbf{c}^0)$ over the $|\mathbf{z}_0|e^{\ii \mathbf{b}^3}$-plane. On the other hand, the combination 
\begin{equation}
    u \equiv \Phi\, e^{\ii\arg(\mathbf{z}_0)}\,,
\end{equation}
of worldsheet fields parametrises the complex plane transverse to the string, i.e., $u\in \mathbb{C}^*$. 

To summarise, the target space manifold $\cM$ of the string $\sigma$-model has the form 
\begin{equation}\label{eq:targetspace}
    \cM = T^2_{(\mathbf{b}^1,\mathbf{b}^2)} \times \left(T^2_{(\mathbf{b}^0, \mathbf{c}^0)} \to \mathbb{C}_{|\mathbf{z}_0|e^{\ii {\mathbf{b}}^3}}\right) \times \mathbb{C}^*_{u=\Phi e^{\ii \arg(\mathbf{z}_0)}}\,,
\end{equation}
where the subscripts denote the worldsheet scalar fields that are the coordinates on the respective factors. To preserve $\cN= (0,4)$ supersymmetry on the worldsheet the second factor must in fact be a local K3 manifold. 
 We will come back, in Section \ref{sec:compvsnonc}, to the question why the base of its torus fibration is non-compact here.
 In addition to the geometry the left-moving sector realises an Abelian algebra on the worldsheet with rank $19-\rho$. 

In total, we can identify the $(c_L, c_R)=(24,12)$  $\sigma$-model with target space  $\cM$ as the theory of a heterotic $G= E_8\times E_8$ or $\mathrm{Spin}(32)/\mathbb Z_2$ string on $\cM$ with a gauge bundle on a local 
\begin{equation}
    K3_{\rm het} = \left(T^2_{(\mathbf{b}^0, \mathbf{c}^0)} \to \mathbb{C}_{|\mathbf{z}_0|e^{\ii {\mathbf{b}}^3}}\right)\,.
\end{equation}
 The gauge background breaks the original $G \times U(1)^4$ gauge group of the heterotic string to a rank $21-\rho$ group\footnote{Recall that in this section we assume $\rho\geq 1$ such that the rank of the surviving gauge group is at most 20 which can hence be a subgroup of $E_8\times E_8\times U(1)^4$ (or $Spin(32)/\mathbb Z_2 \times U(1)^4$).} at a generic point on the Coulomb branch where the surviving gauge group is given by its Cartan subgroup.\footnote{Since the $b^{i\neq 3}$ fields only interact with $|\mathbf{z}_0|$, the bundle breaking the heterotic gauge group has to be on the local K3.}

\subsubsection{$\text{rk}(\Lambda_{\rm trans}) = 3$ or $2$ ($\rho = 18$ or $19$)}\label{sec:rksmall}
The discussion so far was restricted to the case that $\text{rk}(\Lambda_{\rm trans})\geq 4$, where $\Lambda_{\rm trans}$ contains two copies of the hyperbolic plane $U\oplus U \subseteq \Lambda_{\rm trans}$. This ensures that the expansion of $C_4$ in harmonic forms along $Z$ gives at least two free left-moving zero modes along the string that pair up with the two right-moving modes to form two real, free scalars $(\mathbf{b}^1,\mathbf{b}^2)$ that we interpreted as the coordinates on a $T^2$ as in \eqref{eq:targetspace}. 

Now, if $\text{rk}(\Lambda_{\rm trans})=3$ (i.e. for $b=1$), the signature of $\Lambda_{\rm trans}$ is $(2,1)$. According to our previous discussion this means that one of the left-moving scalars $b^{1,2}_-$, which w.l.o.g. we take to be $b^2_-$, has a non-trivial interaction with $\mathbf{z}_0$. From the bulk perspective the ${\rm U}(1)$ gauge theory associated with the gauge field $A^2_-=A^2_{\rm KK}-A^2_{\rm wind}$ is Higgsed, while the combination $A^2_+ = A^2_{\rm KK}+A^2_{\rm wind}$ remains massless.

A natural way to achieve this in the target space is if the radius of one of the circles of the $T^2$ in \eqref{eq:targetspace} is fixed to its self-dual value of string scale size so that the ${\rm U}(1)$ gauge field with potential $A^2_-$ enhances to $\rm{SU}(2)$; the latter is subsequently broken by a gauge background, resulting in the observed rank reduction. 
 For $\text{rk}(\Lambda_{\rm trans})=3$, a candidate $\sigma$-model target space is thus
\begin{equation}
\cM = T^2_{U=T} \times \left(T^2_{(\mathbf{b}^0, \mathbf{c}^0)} \to \mathbb{C}_{|\mathbf{z}_0|e^{\ii {\mathbf{b}}^3}}\right) \times \mathbb{C}^*_{u=\Phi e^{\ii \arg(\mathbf{z}_0)}}\,,
\end{equation}
where the subscript in the first factor indicates that the complex structure, $U$, and the  K\"ahler parameter, $T$, of the torus agree if one of the torus radii is at the self-radius. 

If $\text{rk}(\Lambda_{\rm trans})=2$ (i.e.~$b=0$), the lattice $\Lambda_{\rm trans}$ has signature $(2,0)$. Now, also the zero mode $b^1_-$ on the worldsheet is interacting with $\mathbf{z}_0$. In the spacetime this means that $A^1_-$ is Higgsed as well. By the same logic as above, this suggests that both one-cycles of the torus factor in $\cM$ are now fixed to string size and a non-abelian gauge background breaks the associated gauge group factor. The candidate target space of the $\sigma$-model becomes 
\begin{equation}
    \cM = T^2_{T=U=\tau_0} \times \left(T^2_{(\mathbf{b}^0, \mathbf{c}^0)} \to \mathbb{C}_{|\mathbf{z}_0|e^{\ii{\mathbf{b}}^3}}\right) \times \mathbb{C}^*_{u=\Phi e^{\ii\arg(\mathbf{z}_0)}}\,,
\end{equation}
where the complex structure and K\"ahler parameter of the torus are fixed to either $\tau_0=i$ or $\tau_0=(-1)^{1/3}$. For $\text{rk}(\Lambda_{\rm trans})=2$ we hence obtain a $\sigma$-model corresponding to the heterotic string on a local K3 times a torus with K\"ahler and complex structure fixed to one of the orbifold points in the fundamental domain of $\mathrm{SL}(2,\mathbb{Z})$. The heterotic dual frame of models with $\rho=19$ including non-perturbative sectors arising from NS5-branes are investigated in more detail in~\cite{Monnee:2025msf}.

\subsubsection{$\text{rk}(\Lambda_{\rm trans}) =21$ ($\rho =0$)}\label{sec:rk21}
The final case to discuss is $\rho=0$ (i.e. $b=19$), corresponding to $\text{rk}(\Lambda_{\rm trans})=21$ and $\text{rk}(\Lambda_{\rm pol})=1$. The real scalar field $\mathbf{b}^3$ defined in \eqref{def:b3} now has a free left-moving and an interacting right-moving piece. This leads to a left-right asymmetric worldsheet theory for $\mathbf{b}^3$, which  cannot correspond to the coordinate on a geometric  K3 surface. Accordingly, in $\cM$ the geometric $\left(T^2_{(\mathbf{b}^0, \mathbf{c}^0)} \to \mathbb{C}_{|\mathbf{z}_0|e^{{\ii\mathbf{b}}^3}}\right)$-factor must be replaced by a non-geometric CFT. Since $b_-^3$ is now a free field, this CFT realises an additional left-moving ${\rm U}(1)$ current algebra on the worldsheet. In the bulk, the total gauge group is therefore a maximal rank subgroup of $E_8\times E_8 \times \mathrm{U}(1)^5$ (or of $\mathrm{Spin}(32)/\mathbb Z_2 \times \mathrm{U}(1)^5$).  The extra ${\rm U}(1)$ factor suggests that the dynamics is captured by a CFT whose target is at an orbifold point.

\subsubsection{Non-compact vs.~compact K3}\label{sec:compvsnonc}

The K3 surface that appears in the target space of the $\sigma$-model on the string worldsheet is non-compact since the base of its elliptic fibration is the complex plane $\mathbb{C}_{|\mathbf{z}_0| e^{\ii{\bf b}_3}}$. On the dual heterotic  side one might have expected instead a compactification on a \emph{compact} K3 that is an elliptic fibration over $\mathbb{P}^1$ as opposed to the complex plane $\mathbb{C}_{|\mathbf{z}_0| e^{{\ii \bf b}_3}}$. However, our analysis applies to the string at a point in the moduli space corresponding to a type II limit realised as a Tyurin degeneration. This maps to the tensionless limit for the string in Planck units, cf.~\eqref{eq:stringtension2}. 
In the dual heterotic frame, the tensionless limit amounts to taking
\begin{equation}
    S\equiv \frac{M_{\rm pl}^2}{M_{\rm het}^2}\to \infty\,\label{Shet}
\end{equation}
while keeping all other moduli constant. In particular, the volumes of curves on the heterotic K3$_{\rm het}$ measured in units of $M_{\rm het}$ form part of the scalar components of the hypermultiplets. Therefore, in the tensionless limit for the heterotic string these have to stay constant since we are considering a limit in the vector multiplet sector only. This means that in Planck units the volume of a curve $C_{\rm K3}$ on K3$_{\rm het}$ behaves in the limit \eqref{Shet} as 
\begin{equation}
    \text{vol}(C_{\rm K3})\, M_{\rm pl}^2 =\underbrace{\text{vol}(C_{\rm K3})M_{\rm het}^2}_{=\text{const}} S \to \infty\,. 
\end{equation}
Thus in Planck units the volume of all  curves on K3$_{\rm het}$ diverges in the tensionless limit for the heterotic string, including in particular the base $\mathbb{P}^1$ of the elliptically fibred K3$_{\rm het}$.
 This explains why the analysis of the $\sigma$-model in the Tyurin degeneration only detects a non-compact base of the target space K3$_{\rm het}$.
 By the same reasoning, the volume of the torus fibre of K3$_{\rm het}$ in Planck units must diverge in the emergent string limit. This perfectly matches the divergence of the moduli space metric, \eqref{eq:Sforbi}, for the interacting zero modes of the $\sigma$-model, including the zero modes $(\mathbf{b}^0,\mathbf{c}^0)$ associated with the elliptic fibre. 

 By contrast, the volume of the additional torus factor is part of a vector multiplet, whose saxionic part is of the form $ {\rm vol}(T^2_{(\mathbf{b}^1,\mathbf{b}^2)}) M^2_{\rm Pl}$ and stays constant in the emergent string limit, in which only $S \to \infty$.
 Again, this expectation is reflected in the fact that the moduli space metric \eqref{eq:Sforba} for this vector multiplet remains constant even in the degeneration limit. 
 
Away from the string Tyurin degeneration and hence the string limit $S \to \infty$ on the dual heterotic side, the target space factor K3$_{\rm het}$ is compact also in Planck units.
 The $\sigma$-models arising from the worldsheet theory of the strings  in type II degenerations of Calabi--Yau threefolds hence indeed describe the tensionless limit of a heterotic string whose target space contains a compact K3$_{\rm het}$.

\subsubsection{BPS invariants and modularity}\label{subsec:modular}

Having established that the EFT string localised on the divisor $Z$ of a Tyurin degeneration is a critical heterotic string, we come back to the interpretation of the BPS invariants associated with special Lagrangian 3-cycles dual to elements in $ H^3(V, \mathbb Z) \cap {\rm Gr}_2(\Delta)$.  
 As explained in Section \ref{subsubsec-BPStower}, these are the 3-cycles of asymptotically vanishing volume in the infinite distance limit. To each such special Lagrangian 3-cycle $\Gamma_0$ one can associate a 2-cycle  $C_0$ on $Z$ whose class lies in the transcendental lattice $\Lambda_{\rm trans}$ of the K3 surface $Z$.
  The heterotic string interpretation now suggests that the BPS indices $\Omega_{\rm BPS}(\Gamma_0)$ count BPS states associated with the winding and Kaluza-Klein states  of the heterotic string with respect to the torus factors on the dual heterotic target space ${\cal M}$.
 To see this, recall that in a type II$_b$ degeneration, $\Lambda_{\rm trans}$ is a sublattice of signature $(2,b)$ of the K3 lattice  $\Gamma_{(3,19)}$,
 \begin{equation}
     \Lambda_{\rm trans} \subset \Gamma_{(3,19)} = U^{\oplus 3} \oplus  \Gamma_{16}   \,,
 \end{equation}
 where $U$ is the hyperbolic lattice of signature $(1,1)$ and $\Gamma_{16}$ collectively refers to the $E_8 \times E_8$ or $\mathrm{Spin}(32)/\mathbb Z_2$ lattice.
  For example, for $b\geq 2$, $\Lambda_{\rm trans}$ can be embedded into a lattice containing two copies of $U$. 
  The projection of the curve $C_0$ to each such hyperbolic factor then counts the wrapping and winding number of a heterotic string state with respect to the two one-cycles in the torus factor of the heterotic target space ${\cal M}$, while the projection on the $\Gamma_{16}$  factor accounts for the perturbative $E_8 \times E_8$ or $\mathrm{Spin(32)}/\mathbb Z_2$ charge. This generalises also to $b=1$ or $b=0$: The only difference is that the momentum and winding along one or both of the torus 1-cycles are not independent, along the lines explained in Section \ref{sec:rksmall}. 
 As first observed in \cite{Harvey:1995fq} by computing the threshold corrections induced by the running of these states in the supergravity, the BPS indices of the winding or KK states are counted by a generating function with certain modular properties.
 In absence of space-time filling NS5-branes in the heterotic target space, the generating function is a meromorphic modular form. In the presence of NS5-branes, 
 experience from similar setups, e.g. \cite{Troost:2010ud,Gaiotto:2019gef,Dabholkar:2020fde}, suggests the appearance of modular anomalies which reflect incomplete cancellations 
 in the BPS index. These are due to the backreaction of the NS5-branes on the heterotic $\sigma$-models. In the worst case this is expected to give rise to mock modular forms, which includes the quasi-modular forms \cite{Klemm:2012sx,Huang:2015sta} that appear for the elliptic genus of the 5d/6d heterotic string, see \cite{Lee:2018urn,Lee:2018spm,Lee:2020blx} and references therein.

As will be discussed in Section \ref{subsec:comparisonWS}, the presence of NS5-branes in the heterotic target can be read off from the fact that the space ${\rm Gr}_3$ associated with the type II$_b$ degeneration is non-empty.
 In combination these considerations lead to the following conjecture:

\begin{conjecture}\label{conj-modular}
Consider a type II$_{b}$ limit associated with a Tyurin degeneration of the form \eqref{eq:V0split} and a special Lagrangian 3-cycle $\Gamma_0$ dual to an element in $ H^3(V, \mathbb Z) \cap {\rm Gr}_2(\Delta)$. 
Then there  exists a meromorphic mock-modular form (in general with respect to a subgroup $G \subseteq SL(2,\mathbb Z)$)
\begin{equation}
\theta(q) = \sum_{n \in {\cal I}}  c(n) q^n\,,
\end{equation}
with ${\cal I}$ a suitable subset of $\mathbb Q$, whose coefficients count the 4d $\cN =2$ BPS invariants associated with $\Gamma_0$ according to the identification
\begin{equation}
    \Omega_{\rm BPS}(\Gamma_0) = c\left(\frac{1}{2} C_0 \cdot_Z C_0\right) \,,
\end{equation}
where $C_0=\partial_* \iota_*\Gamma_0 \in \Lambda_{\rm trans}$ denotes the curve on $Z$ over which the asymptotic form of $\Gamma_0$ is fibered close to the Tyurin degeneration (cf. the discussion below (\ref{eq:isom})). 
If ${\rm Gr}_3 = \emptyset$, $\theta(q)$ is in fact modular.
\end{conjecture}
In particular this guarantees the existence of a BPS tower for 
$\Gamma_0$ with $C_0 \cdot_Z C_0 \geq 0$.

Note that a suitable refinement of the BPS index by fugacities associated with the abelian gauge charges should lead to quasi-Jacobian forms or generalisations thereof, whose elliptic index will be encoded in the transcendental lattice $\Lambda_{\rm trans}$ of $Z$.
 It would be very interesting to find direct evidence for this behaviour by studying the BPS indices with the special Lagrangian 3-cycles in the Type IIB geometry. A good starting point could be the explicit geometries of \cite{Doran:2024kcb} that admit Tyurin degenerations and for which the BPS invariants on the mirror Type IIA side have been computed in terms of modular forms.

\section{Mirror Symmetry to Emergent Strings in Type IIA}
\label{sec:mirror_symmetry}
In the previous sections, we kept our discussion strictly to Calabi--Yau threefold compactifications of \emph{Type IIB} string theory. At no stage did we invoke mirror symmetry as a physical input to use known results about emergent strings in the vector multiplet moduli space of \emph{Type IIA} compactifications on the mirror. Our results thus apply to \emph{general} Tyurin degenerations; in particular, they are not restricted to the large complex structure regime of $V$, which, under mirror symmetry, maps to the large volume regime of Type IIA on the mirror manifold $\widehat{V}$. 

It is nonetheless interesting to relate the emergent EFT strings in Type IIB complex structure degenerations and those arising in the mirror K\"ahler moduli space of Type IIA string theory as discussed in~\cite{Lee:2019oct}. The natural expectation, fueled in part by \cite{doran2016mirrorsymmetrytyurindegenerations}, is that they are precisely mapped to each other by mirror symmetry. To demonstrate this explicitly, we would have to know the exact geometric action of mirror symmetry, which is not available in general. Nevertheless, we can compare the worldsheet theories of the emergent strings arising in Type IIA string theory to the worldsheet theories discussed in Section \ref{sec:Worldsheet}. 

From the analysis in \cite{Lee:2019oct} we recall that emergent strings in Type IIA arise in the large base limit for K3-fibred Calabi--Yau threefolds and correspond to NS5-branes wrapping the generic K3-fibre of the threefold. The worldsheet theory on these strings can then be obtained by dimensionally reducing the NS5-brane worldvolume theory on the K3. We discuss the details of this in Section~\ref{subsec:IIA_mirror}. In Section~\ref{subsec:comparisonWS}, we then compare the worldsheet theories obtained from the NS5-branes in Type IIA string theory to the worldsheet theories discussed in Section \ref{sec:Worldsheet}. Finally, in Section~\ref{subsec:directomirror} we consider special cases where the action of mirror symmetry is known explicitly and demonstrate that, indeed, it maps the NS5-brane strings of Type IIA into the Type IIB EFT string geometries that we discussed in Section \ref{sec:Worldsheet}. 

\subsection{Worldsheet Theory of NS5-Brane String in Type IIA}
\label{subsec:IIA_mirror}

\begin{figure}[t]
    \centering
    \includegraphics[width=0.5\linewidth]{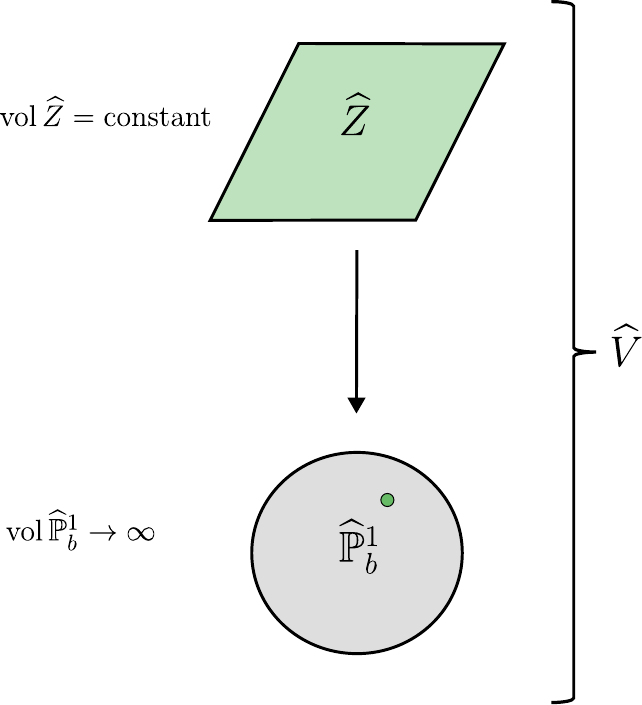}
    \caption{A geometric depiction of the set-up described around equation \eqref{eq:hatVfibration}. The Type IIA NS5-brane (indicated in green) is located at a generic point in the base $\widehat{\mathbb{P}}^1_b$ and wraps the generic K3-fibre $\widehat{Z}$. The emergent string is obtained in the limit where the volume of the base becomes large, while the volume of the generic fibre stays constant.}
    \label{fig:NS5}
\end{figure}

Let us consider Type IIA string theory on a Calabi--Yau threefold $\widehat{V}$ that admits a K3-fibration over $\widehat{\mathbb P}^1_b$ with generic K3-fibre class $[\widehat{Z}]\in H_4(\widehat{V})$,
\begin{equation}\begin{aligned} \label{eq:hatVfibration}
\widehat Z \ \stackrel{\widehat\iota}{\hookrightarrow} & \  \ \widehat V \cr 
&\ \ \downarrow\cr 
& \ \  \widehat{\mathbb P}^1_b \,.
\end{aligned}\end{equation}
The limit in the K\"ahler moduli space in which  $ \widehat{\mathbb P}^1_b$ becomes large while the volume of the generic fibre remains constant corresponds to an emergent string limit. The emergent string in this case is given by the NS5-brane wrapping the generic K3-fibre and extended along two of the non-compact directions \cite{Lee:2019oct}. The situation is depicted in Figure \ref{fig:NS5}.

We are now interested in the worldvolume theory of an NS5-brane wrapping $\widehat{Z}$. By duality to M-theory on $\widehat{V}\times S^1$, the resulting string can be viewed as a special case of an MSW string \cite{Maldacena:1997de} obtained by wrapping an M5-brane on $\widehat{Z}\subset \widehat{V}$ and not on $S^1$. The bosonic massless modes on the worldvolume theory of the NS5-brane are given by a self-dual 2-form $\widehat{B}_2$, four non-compact real scalars and one compact scalar; together with their fermionic partners they form a matter multiplet of 6d $\cN=(0,2)$ supersymmetry. In 10d, the non-compact real scalars describe the extended directions normal to the NS5-branes. In the M-theory dual, the compact real scalar describes the position of the M5-brane along the additional $S^1$. 

For the string obtained by wrapping the Type IIA NS5-brane on $\widehat{Z}$, the massless spectrum on the worldsheet is obtained by dimensional reduction. The space transverse to $\widehat{Z}$ is topologically $ \widehat{\mathbb P}^1_b\times \mathbb{C}^*$, where $ \widehat{\mathbb P}^1_b$ is the base of the K3-fibration for $\widehat{V}$. Thus, two of the non-compact, real scalars on the worldvolume theory of the NS5-branes become compact and combine into the complex coordinate $\widehat{\mathbf{v}}\in \widehat{\mathbb{P}}^1_b$. The other two real scalars form a complex coordinate $\widehat u\in\mathbb{C}^*$. In addition, the compact scalar on the worldvolume theory of the NS5-branes remains in the spectrum after dimensional reduction. In the following, we refer to this scalar as $\widehat{\mathbf{b}}^1$. On the string worldsheet, these scalar fields split into left- and right-moving scalars, with the latter having fermionic superpartners to form multiplets of the $\cN = (0,4)$ super-algebra on the worldsheet. In addition to the scalar modes on the 2d worldsheet arising from the scalar modes already present in the 6d $\cN=(0,2)$ parent theory, the self-dual 2-form $\widehat{B}_2$ yields further scalar modes when expanded in 2-forms on $\widehat{Z}$. These scalars are compact. As for the MSW-string discussed in \cite{Maldacena:1997de}, $\widehat{B}_2$ gives rise to $b_2^+(\widehat{Z})$ right-moving and $b^-_2(\widehat{Z})$ left-moving compact scalar zero modes. Since by assumption $\widehat{Z}$ is a K3-surface we hence end up with $b_2^-(\widehat{Z})=19$ left-moving scalar modes and $b_2^+(\widehat{Z})=3$ right-moving scalar modes. These latter scalar zero modes are accompanied by fermionic zero modes and form appropriate $\cN =(0,4)$ supermultiplets. 

The massless spectrum of the worldsheet theory thus has left- and right-moving central charges 
\begin{equation}\label{centralcharges}
    c_L = 24\,,\qquad c_R=12\,,
\end{equation}
which agrees with the central charges of the critical heterotic string in light-cone gauge. To identify the actual heterotic dual we must consider the interactions for the scalar fields. As in Section~\ref{sec:Worldsheet}, we detect such interactions from the kinetic term of the scalar fields in the worldsheet action, which is of the form 
\begin{equation}
    S_{\rm WS,kin} = \frac12 \int {\rm d}^2 \sigma \; g_{\mu \nu}(\widehat \phi^k)\, \partial_\sigma \widehat\phi^\mu \partial^\sigma \widehat\phi^\nu \,,
\end{equation}
where $\widehat\phi^\mu$ accounts for all scalar fields in the theory. As in Section~\ref{subsec:interactions}, the interactions are encoded in the field-dependence of $g_{\mu \nu}$. The metric for the scalar fields $\widehat{\mathbf{b}}^1$ and $\widehat u$ is constant, so that these fields can be treated as free modes. However, due to the nontrivial fibration, the field space metric for the modes arising from $\widehat{B}_2$ can have a nontrivial dependence on $\widehat{\mathbf{v}}$ implying that these fields are interacting. 

Let us fix a fibre $\widehat{Z}_0$ over a point $\widehat{\mathbf{v}}\in  \widehat{\mathbb P}^1_b$. At this point we can consider the scalar fields on the worldsheet obtained by expanding $\widehat{B}_2$ in a basis of $H^2(\widehat{Z}_0)$ as
\begin{equation}
    \widehat{B}_2 = \widehat{b}_2^\a \, \widehat{\omega}_\a \,,\qquad \widehat{\omega}_\a\in H^2(\widehat{Z}_0)\,. 
\end{equation}
If $\widehat{Z}$ is non-trivially fibred over $ \widehat{\mathbb P}^1_b$ the basis of 2-forms chosen on $\widehat{Z}_0$ undergoes non-trivial monodromies around points in $ \widehat{\mathbb P}^1_b$. Thus, as we change $\widehat{\mathbf{v}}$, the basis of 2-forms in which we expand $\widehat{B}_2$ changes. As a consequence, the kinetic term for those $\widehat{b}_2^i$ whose associated $\widehat \omega_i$ does not correspond to a  
harmonic 2-form globally defined on $\widehat V$
 depends on $\widehat{\mathbf{v}}$. In contrast, if a 2-form in $\widehat \omega_a\in H^2(\widehat{Z})$ \textit{is} globally defined on $\widehat V$, the kinetic term for the corresponding $\widehat b^a_2$ is independent of $\widehat{\mathbf{v}}$ and in fact completely field-independent. The $\widehat{b}_2^a$ are hence free fields on the worldsheet. In case $\widehat{\omega}_a$ is globally defined on $\widehat{V}$ it is the pull-back of a 2-form 
  in $H^{1,1}(\widehat{V})$
 under the inclusion map $\widehat\iota: \widehat{Z} \hookrightarrow \widehat{V}$. The lattice $\widehat\iota^*(H^{1,1}(\widehat{V}))$ is known as the polarization lattice, $\widehat{\Lambda}_{\rm pol}$, of the K3-fibre $\widehat{Z}$ with signature\footnote{Note that since   $\widehat{Z}$ is nontrivially fibred over base $ \widehat{\mathbb P}^1_b$, the case $\widehat{\rho}=19$ is excluded.} 
\begin{equation}
\text{sgn}(\widehat{\Lambda}_{\rm pol})=(1,\widehat \rho), \qquad  0\leq \widehat\rho <19   \,, \qquad \widehat{\Lambda}_{\rm pol} = \widehat\iota^*(H^{1,1}(\widehat{V}))   \,.  
\end{equation}
Accordingly, there are one right-moving and $\widehat \rho$ left-moving free scalar modes on the string worldsheet. 

For $1\leq \widehat{\rho}$, the free right-moving scalar together with one of the free left-moving scalars forms a free compact real scalar on the worldsheet of the string. We denote this real scalar field by $\widehat{\mathbf{b}}^2$. It realises a global left-moving and right-moving ${\rm U}(1)$ algebra on the string, leading to a massless $U(1)^2$ gauge theory in the bulk associated with an $S^1$ target space. Together with the scalar field $\widehat{\mathbf{b}}^1$ this constitutes a target space factor $T^2_{(\widehat{\mathbf{b}}^1,\widehat{\mathbf{b}}^2)}$. 
 The other free compact left-moving modes realise a rank $\widehat{\rho}-1$ current algebra on the string worldsheet, leading to a gauge theory in the bulk with gauge group of rank ${\widehat \rho-1}$.\footnote{If $\widehat{\rho}=0$, the $S^1$ radius is fixed to the self-dual radius and the left-moving ${\rm U}(1)$ is broken similar to our discussion in Section \ref{sec:rksmall}.}

On the other hand, the interacting modes on the string worldsheet are associated with the orthogonal complement of $\widehat{\Lambda}_{\rm pol}\subset \Gamma_{3,19}$ which corresponds to the transcendental lattice $\widehat{\Lambda}_{\rm trans}$ of $\widehat{Z}$ and has signature
\begin{equation}
\text{sgn}(\widehat{\Lambda}_{\rm trans})=(2,19-\widehat\rho) \,.
\end{equation} 
For $\widehat{\rho}<18$, the transcendental lattice $\widehat \Lambda_{\rm trans}$ contains two left-moving interacting fields which form real scalar fields on the worldsheet once combined with the two right-moving interacting fields. We denote the resulting compact real scalar fields by $(\widehat{\mathbf{b}}^0, \widehat{{\mathbf{c}}}^0)$. They can be viewed as the coordinates of a torus in the target space of the sigma-model on the string worldsheet that is nontrivially fibred over $ \widehat{\mathbb P}^1_b$.\footnote{If $\widehat{\rho}=18$ we are in similar situation as discussed in Section~\ref{sec:rk21}.} The remaining left-moving interacting scalar fields do not give rise to a massless gauge theory in the bulk. 

To summarise, the worldsheet theory of the Type IIA NS5-brane on $\widehat{Z}$ is a sigma-model with central charges as in \eqref{centralcharges} with target space
\begin{equation}\label{eq:hatM}
    \widehat{\cM} = T^2_{(\widehat{\mathbf{b}}^1,\widehat{\mathbf{b}}^2)} \times \left(T^2_{(\widehat{\mathbf{b}}^0, \widehat{\mathbf{c}}^0)} \to  \widehat{\mathbb P}^1_{\hat{\mathbf{v}}}\right) \times \mathbb{C}^*_{\widehat u}\,. 
\end{equation}
To ensure $\cN=(0,4)$ worldsheet supersymmetry, the surface given by the $T^2$ fibration over $ \widehat{\mathbb P}^1_{\widehat{\mathbf{v}}}$ has to be a K3 surface. We hence obtain a heterotic sigma model with target space K3$_{\rm het}\times T^2$ with a gauge group $G_{\widehat \rho -1} \times U(1)^2_{\rm L} \times U(1)^2_{\rm R}$ of total rank $\widehat{\rho}+3$, where the two Abelian factors arise from $T^2_{(\widehat{\mathbf{b}}^1,\widehat{\mathbf{b}}^2)}$. In the vector multiplet moduli space of the 4d $\cN=2$ effective action obtained by compactification of Type IIA string theory on $\widehat{V}$, the weak coupling limit for the NS5-brane string corresponds to the limit where the volume of $ \widehat{\mathbb P}^1_b$ is taken to infinity with the volume of $\widehat{Z}$ remaining constant. Therefore, in the weak coupling limit, $ \widehat{\mathbb P}^1_b$ decompactifies and can effectively be replaced by $\mathbb{C}$. This is the Type IIA version of our discussion in Section \ref{sec:compvsnonc}.

\subsection{Comparison of Worldsheet Theories}
\label{subsec:comparisonWS}
We can now compare the worldsheet theories of the NS5-brane string of Type IIA discussed in the previous section to the strings of Type IIB associated with Tyurin degenerations. In the following we denote by $\widehat{\mathtt{S}}_{\widehat{Z}}$ the string obtained by wrapping a Type IIA NS5-brane on the K3-fibre $\widehat{Z}$ of a Calabi--Yau threefold $\widehat{V}$. Similarly, we refer to the string becoming light at a Tyurin degeneration of a Calabi--Yau threefold $V$ as $\mathtt{S}_{Z}$, where $Z$ is the K3 surface arising at the degeneration of $V$. As in Section~\ref{sec:Worldsheet}, this comparison can be done at different levels of detail for the worldsheet theory. In the remainder of this section we will show the following statements:
\begin{enumerate}
    \item The massless spectrum on the worldsheet of $\widehat{\mathtt{S}}_{\widehat Z}$ is identical to the spectrum of fields on $\mathtt{S}_{Z}$. This is simply ensured by $\widehat{Z}$ and $Z$ being K3 surfaces. 
    \item The number of free fields on $\widehat{\mathtt{S}}_{\widehat Z}$ equals that on $\mathtt{S}_{Z}$ if the transcendental lattice of $Z$ corresponds to the quantum Picard lattice of $\widehat{Z}$. In other words, the number of free fields is the same if $Z$ and $\widehat{Z}$ are mirror to each other. 
    \item In setups with a perturbative heterotic dual string (i.e.~a heterotic string compactification without spacetime filling NS5-branes), for the interactions on the string worldsheets of $\widehat{\mathtt{S}}_{\widehat Z}$ and $\mathtt{S}_Z$ to agree, a necessary condition is $h^{2,1}(\widehat V)=h^{1,1}(V)$ and vice versa, which is ensured if $V$ and $\widehat V$ are mirror to each other. 
\end{enumerate}
Thus, the string $\widehat{\mathtt{S}}_{\widehat {Z}}$ becoming light in the large-base limit for a K3-fibred Calabi--Yau threefold $\widehat V$ in Type IIA  gives rise to the same worldsheet theory as the string $\mathtt{S}_{Z}$ emerging in Type IIB string theory on at a Tyurin degeneration of the Calabi--Yau threefold $V$ if $V$ and $\widehat{V}$ are mirror to each other. 

\paragraph{1. Worldsheet Spectrum.} Let us first compare the spectrum of the worldsheet theories on $\widehat{\mathtt{S}}_{\widehat Z}$ and $\mathtt{S}_Z$. The worldsheet theory of either string is a $\sigma$-model with $c_L=24$ and $c_R=12$. The respective target spaces are given by $\cM$ in \eqref{eq:targetspace} and $\widehat{\cM}$ in \eqref{eq:hatM}. As discussed at the end of the previous section and in Section \ref{sec:compvsnonc}, in the tensionless limit for the string, the base of the K3 is in both cases non-compact. The $\sigma$-models of the strings obtained, respectively, in Type IIA and Type IIB hence have the same target space upon identifying 
\begin{equation}
    (\mathbf{b}^{0},\mathbf{b}^1, \mathbf{b}^2,\mathbf{c}^0,u)\quad  \longleftrightarrow \quad  (\widehat{\mathbf{b}}^0,\widehat{\mathbf{b}}^1, \widehat{\mathbf{b}}^2, \widehat{\mathbf{c}}^0,\widehat u)\,. 
\end{equation}
See also Table \ref{tab:scalars} for an overview of the various fields involved in the identification. Thus, the fact that $Z$ and $\widehat Z$ are K3 surfaces ensures that the spectrum of massless fields on the worldsheet theory on $\widehat{\mathtt{S}}_{\widehat Z}$ is identical to that of $\mathtt{S}_Z$.

\paragraph{2. Free Fields and Bulk Gauge Theory.} For the string $\mathtt{S}_{Z}$, the transcendental lattice of $Z$ gives rise to the free modes on the worldsheet while for $\widehat{\mathtt{S}}_{\widehat Z}$ the polarization lattice $\widehat{\Lambda}_{\rm pol}$ gives free modes. Conversely, while the polarization lattice $\Lambda_{\rm pol}$ of $Z$ gives interacting modes on the worldsheet of $\mathtt{S}_{Z}$, $\widehat{\Lambda}_{\rm trans}$ gives the interacting modes on $\widehat{ \mathtt{S}}_{\widehat Z}$. Thus, from the perspective of the worldsheet theory the roles of polarization and transcendental lattice are exchanged. 

\begin{table}[t]
\begin{center}
\begin{tabular}{c@{\quad}c}
     \textbf{IIA NS5-brane on K3-fibre} & \textbf{IIB EFT string in Tyurin degeneration}\\
     \begin{tabular}{|c|c|c|}
     \hline&&\\[-1em]
          Field & Origin & Free/interacting \\ &&\\[-1em]\hline&&\\[-1em]
          $\widehat{\mathbf{b}}^0,\widehat{\mathbf{c}}^0$ & $\widehat{\Lambda}_{\mathrm{trans}}$ & interacting\\ &&\\[-1em]\hline&&\\[-1em]
          $\widehat{\mathbf{b}}^1$ & M-theory circle & free \\&&\\[-1em] \hline&&\\[-1em]
          $\widehat{\mathbf{b}}^2$ & $\widehat{\Lambda}_{\mathrm{pol}}$ & free \\&&\\[-1em] \hline
     \end{tabular}

     &

     \begin{tabular}{|c|c|c|}
     \hline&&\\[-1em]
          Field & Origin & Free/interacting \\ &&\\[-1em]\hline&&\\[-1em]
          $\mathbf{b}^0,\mathbf{c}^0$ & reduction $B_2,C_2$ & interacting\\&&\\[-1em] \hline&&\\[-1em]
          $\mathbf{b}^1,\mathbf{b}^2$ & $\Lambda_{\mathrm{trans}}$ & free \\&&\\[-1em] \hline
     \end{tabular}
\end{tabular}
\end{center}
\caption{An overview of the various worldsheet scalar fields obtained (left) from wrapping a Type IIA NS5-brane on the K3-fibre of a Calabi--Yau threefold $\widehat{V}$, and (right) from a Type IIB EFT string probing a Tyurin degeneration of a Calabi--Yau threefold $V$.}
\label{tab:scalars}
\end{table}

As far as the gauge sector is concerned, the Type IIA NS5-brane string gives rise to a gauge group in the bulk of rank $\widehat{\rho} +3$, while the strings discussed in Section~\ref{sec:Worldsheet} give rise to a rank $21-\rho$ gauge theory. Thus, the string $\mathtt{S}_{Z}$ gives rise to a bulk gauge theory of the same rank as the string $\widehat{\mathtt{S}}_{\widehat Z}$ if 
\begin{equation}\label{eq:sumrhos}
    \rho + \widehat\rho = 18\,. 
\end{equation}
This is achieved if the transcendental lattice of $Z$ is related to $\widehat{\Lambda}_{\rm pol}$ via 
\begin{equation}\label{eq:Lambdamirror}
    \Lambda_{\rm trans} = \widehat{\Lambda}_{\rm pol} \oplus U \,,
\end{equation}
where $U$ is the hyperbolic plane. Hence, the transcendental lattice of $Z$ is indeed exchanged with the polarization lattice of $\widehat{Z}$ up to the additional factor of $U$ which accounts for the fact that the free real scalar $\mathbf{b}^1$ on the worldsheet of $\mathtt{S}_{Z}$ arises from $H^2(Z)$ whereas $\widehat{\mathbf{b}}^1$ on the worldsheet of $\widehat{\mathtt{S}}_{\widehat Z}$ is already present as a real scalar on the worldvolume theory of the NS5-brane prior to compactification. The relation \eqref{eq:Lambdamirror} identifies the transcendental lattice of $Z$ with the quantum Picard lattice of $\widehat{Z}$. Since these two lattices are exchanged under K3-mirror symmetry, for two strings $\mathtt{S}_Z$ and $\widehat{\mathtt{S}}_{\widehat{Z}}$ to have the same amount of free fields on their respective worldsheets (and thus to lead to the same bulk gauge theory), the two K3-surfaces $Z$ and $\widehat{Z}$ must be the mirror of each other. 

\paragraph{3. Worldsheet Interactions and Bundle Moduli Space.}
For the worldsheet theories on $\widehat{\mathtt{S}}_{\widehat Z}$ and $\mathtt{S}_Z$ to be identical, the interactions of the non-free  fields on the worldsheets must be identical. In either case, the interactions are determined by the embedding of the respective K3 surfaces in the full Calabi--Yau threefolds. For example, the details of the interactions of $\widehat{\mathtt{S}}_{\widehat Z}$ depend on the exact twist of the fibration of $\widehat{Z}$ over the base of $\widehat{V}$. The interactions are hence determined by global data of the Calabi--Yau threefolds and cannot just be inferred from the local geometry in the vicinity of $\widehat Z$ and $Z$. 

Similarly, in the respective heterotic dual compactifications, the bundles on the target space of the $\sigma$-model must be specified. We already mentioned that supersymmetry requires the target space to be the product K3$_{\rm het}\times T^2$, and it remains to specify a gauge background for the perturbative heterotic gauge group. In the following we do not attempt to translate the details of the embedding $\widehat{Z}\hookrightarrow \widehat V$ into the worldsheet interactions for the generic case. Instead, we focus on the simple case with a \emph{perturbative} heterotic dual, i.e., a heterotic compactification on K3$_{\rm het}\times T^2$ without spacetime filling NS5-branes. In particular, this means that all massless vector multiplets and hypermultiplets must arise from excitations of the perturbative heterotic string. In this case, we have
\begin{equation}
    n_{\rm V}^{\rm pert} = \text{rk}(G) -1\,,\qquad n_{\rm H}^{\rm pert} = n_{\rm H, gen.}+ n_{\rm H,bundle}\,,
\end{equation}
where $\mathrm{rk}\,G$ denotes the rank of the unbroken part of the perturbative heterotic gauge group, $n_{\rm H, gen.}$ is the number of the generic massless hypermultiplets arising from the deformation moduli of the K3$_{\rm het}$ and $n_{\rm H,bundle}$ counts the model-dependent massless hypermultiplets corresponding to deformations of the gauge bundle.

Let us first discuss how the condition of a \textit{perturbative} heterotic dual translates to the Type IIA and Type IIB perspectives. From the Type IIA perspective, we are imposing the condition
\begin{equation}\begin{aligned}
\label{eq:nVIIA}
    n_{\mathrm{V}}^{\rm IIA}&\stackrel{!}{=}n_{\rm V} ^{\rm pert}\\
    &= \mathrm{rk}\,G-1 \\
    &= 1+\mathrm{rk}\,\widehat{\Lambda}_{\mathrm{pol}}\,.
\end{aligned}\end{equation}
At the same time, the number of vector multiplets in Type IIA compactified on $\widehat{V}$ is given by $h^{1,1}(\widehat{V})$. Hence, we recover the result that there are no additional fibral divisors localised over points in the base of the K3-fibred threefold $\widehat V$. 

From the Type IIB perspective, we are imposing the condition
\begin{equation}\begin{aligned}
    \label{eq:nVIIB}
    n_{\mathrm{V}}^{\rm IIB}&\stackrel{!}{=}n_{\rm V} ^{\rm pert}\\
    &= \mathrm{rk}\,G-1 \\
    &= -1+\mathrm{rk}\,\Lambda_{\mathrm{trans}}\\
    &=h^{2,1}(V)-\frac{1}{2}\mathrm{dim}\,\mathrm{Gr}_3\,.
\end{aligned}\end{equation}
The last line uses 
\eqref{eq:H3decomTypeII} along with \eqref{eq:defLambdatrans}.
Since the number of vector multiplets in Type IIB compactified on $V$ is given by $h^{2,1}(V)$, $\mathtt{S}_Z$ only gives rise to a perturbative heterotic string if $\mathrm{Gr}_3=\emptyset$. Conversely, whenever $\mathrm{Gr}_3$ is non-empty, the dual heterotic theory contains additional spacetime-filling NS5-branes. 

Let us now return to the conditions under which $\mathtt{S}_{Z}$ and $\widehat{\mathtt{S}}_{\widehat{Z}}$ give rise to the same perturbative heterotic string. First, the perturbative gauge groups must agree. For the perturbative heterotic string by \eqref{eq:nVIIA} and \eqref{eq:nVIIB} this is the case if $h^{2,1}(V)=h^{1,1}(\widehat{V})$. Next, we require that the dimension of the moduli space of the heterotic gauge bundle is the same. This is a necessary condition for the gauge bundles of $\mathtt{S}_Z$ and $\widehat{\mathtt{S}}_{\widehat Z}$ (and thus the interactions of the worldsheet theory) to agree. Since we are focusing on the perturbative heterotic string for which all massless hypermultiplets are perturbative string excitations, this imposes the condition $h^{2,1}(\widehat V) = h^{1,1}(V)$. Together, both conditions are satisfied if $\widehat V$ and $V$ are mirror to each other.

To summarise, for the worldsheet theory of the strings $\mathtt{S}_Z$ and $\widehat{\mathtt{S}}_{\widehat Z}$ to be the same, including interactions, the manifolds $V$ and $\widehat{V}$ on which Type IIB and, respectively, IIA string theory are compactified have to be mirror to each other. Conversely, consider a Calabi--Yau threefold $V$ that allows for a Tyurin degeneration giving rise to a string $\mathtt{S}_Z$ upon compactifying Type IIB string theory on it. Then the mirror $\widehat V$ has to be K3-fibred with generic fibre $\widehat Z$ such that Type IIA compactified on $\widehat V$ has an emergent string with the same worldsheet theory as $\mathtt{S}_Z$. This is consistent with the expectation of \cite{doran2016mirrorsymmetrytyurindegenerations}: A Calabi--Yau threefold $V$ which allows for a Tyurin degeneration of the form $V \to X_1 \cup_Z X_2$ such that $X_1$ and $X_2$ induce a lattice polarization on $Z$ of rank $\rho+1$ is mirror dual to a threefold $\widehat{V}$ allowing for a fibration by a K3-surface $\widehat{Z}$ with polarization lattice of rank $19-\rho$.

\subsection{Direct Application of Mirror Symmetry to NS5-Brane String}\label{subsec:directomirror}
In the previous section, we 
 have compared 
 the string $\widehat{\mathtt{S}}_{\widehat Z}$ arising from a wrapped NS5-brane in Type IIA string theory on $\widehat V$
 and the Type IIB EFT string $\mathtt{S}_Z$ arising at a Tyurin degeneration of $V$. We have argued that both strings
   can have the same worldsheet theory provided the Calabi--Yau threefolds $V$ and $\widehat{V}$ are mirror dual to each other. This suggests that the strings $\mathtt{S}_{Z}$ and $\widehat{\mathtt{S}}_{\widehat Z}$ are exchanged under mirror symmetry. 
    Our arguments so far are purely based on the identification of the respective worldsheet theories. It remains to demonstrate how, explicitly, mirror symmetry maps the NS5-brane string $\widehat{\mathtt{S}}_{\widehat Z}$ in Type IIA string theory to the EFT string geometry $\cV$ associated with $\mathtt{S}_{Z}$. That this should be possible is also suggested by the Doran--Harder--Thompson conjecture \cite{doran2016mirrorsymmetrytyurindegenerations}, according to which mirror symmetry maps a Calabi--Yau threefold with a Tyurin degeneration to a mirror threefold which is K3-fibred. For recent explicit studies of such mirror pairs, we refer to \cite{Doran:2024kcb}.

To achieve this requires knowledge of the detailed action of mirror symmetry on the geometry of $\widehat V$, which is not available in general. However, the SYZ conjecture \cite{Strominger:1996it} suggests that mirror symmetry can be viewed as three T-dualities along a special Lagrangian $T^3$ fibration of $\widehat V$. Thus, if we are able to identify the relevant $T^3\subset \widehat V$, we can apply three T-dualities to infer the geometric action of mirror symmetry. 

Consider again the K3-fibred Calabi--Yau threefold $\widehat{V}$ with generic fibre $\widehat Z$, as defined in \eqref{eq:hatVfibration}. To identify the SYZ special Lagrangian $T^3$-fibre, we first notice that $\widehat{Z}$ itself allows for a fibration by a calibrated curve $\widehat \cE$ that is topologically a $T^2$ and satisfies $J_{\widehat{Z}}|_{\widehat \cE}=0$. Since $\widehat \cE$ is not a holomorphic curve, it is not defined globally on $\widehat{V}$, but is subject to monodromies around the points in $\widehat{\mathbb{P}}^1_b$ over which $\widehat{Z}$ degenerates. Still, if there exists a non-contractible path $\widehat \gamma \subset \widehat{\mathbb{P}}^1_b$ along which $\widehat \cE$ does not undergo a monodromy, the restriction of the K3 fibration to $\widehat \gamma$ gives a 3-cycle. However, since $h^1(\widehat{\mathbb{P}}^1_b)=0$ no such $\widehat \gamma$ exists on $\widehat{\mathbb{P}}^1_b$, but there can exist multi-covers of $\widehat{\mathbb{P}}^1_b$ that are branched over the degeneration points of $\widehat{Z}$ in $\widehat{\mathbb{P}}^1_b$ with a different topology. In the following, we discuss one such instance. \\

\begin{figure}[t]
    \centering
    \begin{subfigure}[t]{0.4\textwidth}
        \centering
        \includegraphics[scale=1.5]{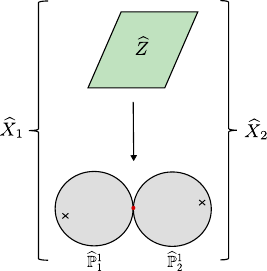}
        \caption{}
        \label{fig:Tyurin-degeneration_IIA}
    \end{subfigure}
    \hspace{2cm}
    \begin{subfigure}[t]{0.4\textwidth}
        \centering
        \includegraphics[scale=1.5]{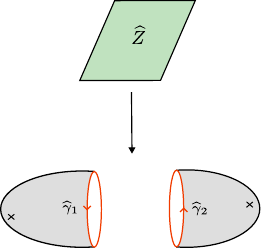}
        \caption{}
        \label{fig:Tyurin-degeneration_IIA-split}
    \end{subfigure}
    \caption{(a) The Tyurin degeneration \eqref{TyurinhetV} in which the base of the K3-fibration splits into two components $\widehat{\mathbb{P}}^1_1$ and $\widehat{\mathbb{P}}^1_2$ intersecting over a single point. (b) The same situation as in Figure (a), but with the intersection point removed, giving rise to two asymptotically cylindrical geometries the 1-cycles $\widehat{\gamma}_1$ and $\widehat{\gamma}_2$ at the boundary.}
\end{figure}

To this end, suppose that $\widehat V$ admits a limit in its complex structure moduli space in which it degenerates into a union of two threefolds, $\widehat{X}_1$ and $\widehat{X}_2$, which are K3-fibred with generic fibre $\widehat{Z}$. Importantly, over $\widehat{X}_1$ and $\widehat{X}_2$ the fibre $\widehat{Z}$ is twisted half as much as it would have to be if $\widehat{X}_{1/2}$ were Calabi--Yau threefolds. We hence refer to $\widehat X_{1,2}$ as ``half'' Calabi--Yau threefolds. This degeneration is a special case of a Tyurin degeneration 
\begin{equation}\label{TyurinhetV}
    \widehat{V} \to \widehat{V}_0=\widehat{X}_1 \cup_{\widehat Z} \widehat X_2\,. 
\end{equation}
See Figure \ref{fig:Tyurin-degeneration_IIA} for a geometric depiction.
To avoid confusion, let us stress that for $\widehat{V}$ we are not interested in the physics of the Tyurin degeneration since we consider the vector multiplet moduli space of Type IIA compactified on $\widehat{V}$. In the sequel, we merely need to assume that there exists such a Tyurin degeneration for $\widehat{V}$. For a large class of examples with this property, see \cite{Braun:2017ryx,Doran:2024kcb}. Notice that for the Tyurin degeneration depicted in~\ref{fig:Tyurin-degeneration_IIA} this requires that the degeneration points of the K3-fibration $\widehat V$ can be separated into two sets such that the net monodromy of the fibre associated with each subset is trivial.

\begin{figure}[t]
    \centering
    \begin{subfigure}[t]{0.4\textwidth}
        \centering
        \includegraphics[scale=0.7]{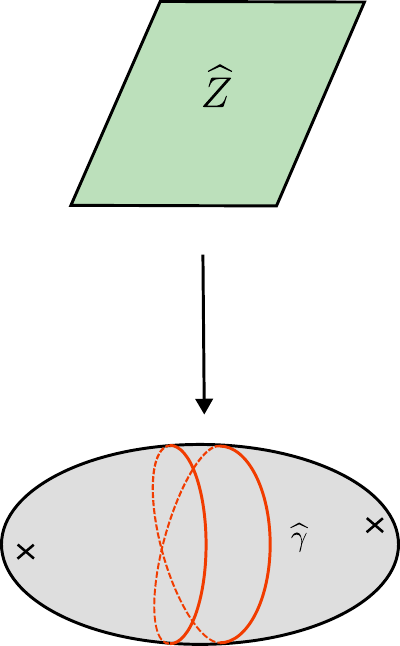}
        \caption{}
        \label{fig:double_cover}
    \end{subfigure}
    ~
    \begin{subfigure}[t]{0.4\textwidth}
        \centering
        \includegraphics[scale=1.3]{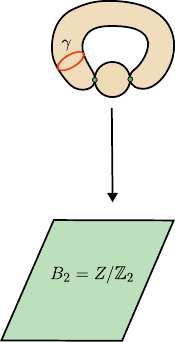}
        \caption{}
        \label{fig:Sen}
    \end{subfigure}
    \caption{The geometry of $\widehat V$ (a) and its mirror $V$ (b). Indicated is the one-cycle $\widehat \gamma$ that lives on the double cover of the base $\widehat{\mathbb{P}}^1_b$ of $\widehat V$ and its mirror $\gamma$ on the genus one-fibre of $V$. Furthermore the K3-fibre $\widehat Z$ and its mirror $Z$ realised as a bi-section of the genus-one fibration of $V$ are indicated in green.}
\end{figure}

Since $\widehat{X}_{1}$ is a half Calabi--Yau threefold, we can remove a point in the base, $\widehat{z}_1\in \widehat{\mathbb{P}}^1_1$, of $\widehat{X}_1$ over which the fibre $\widehat{Z}_1$ is smooth and glue in one boundary of a cylinder at $\widehat{z}_1$. Let us denote the  boundary of the cylinder by $\widehat \gamma_1$. Similarly, we can replace a point in the base of $\widehat{X}_2$ by a cylinder with boundary $\widehat \gamma_2$. Since both $\widehat{X}_1$ and $\widehat{X}_2$ thus have asymptotically cylindrical regions, the monodromy generated by transporting $\widehat{Z}_{1/2}$ around $\widehat\gamma_{1/2}$ is trivial. See also Figure \ref{fig:Tyurin-degeneration_IIA-split}. We obtain $\widehat{V}_0$ by gluing the asymptotic regions of $\widehat{X}_1$ and $\widehat{X}_2$. Let $\widehat i_k :\widehat X_k \hookrightarrow \widehat V_0$, $k=1,2$ be the embeddings of $\widehat X_k$ in $\widehat V_0$. On $\widehat V_0$ we can then consider the one-cycle
\begin{equation}
    \widehat \gamma = \widehat i_{1,\ast}(\widehat \gamma_1) - \widehat i_{2,\ast} (\widehat \gamma_2)\,,
\end{equation}
where the minus sign takes care of the orientation of $\widehat \gamma_{1,2}$ such  that the cycle $\widehat \gamma$ is nontrivial. Since $\widehat{\mathbb{P}}^1_b$ has no one-cycles, $\widehat \gamma$ cannot be viewed as a one-cycle on $\widehat{\mathbb{P}}^1_b$ identified as a section of the K3-fibration. Instead, in the simplest case, $\widehat \gamma$ is a one-cycle on a bi-section $\widehat B$ of the K3 fibration that is a double cover of $\widehat{\mathbb{P}}^1_b$. Since this bi-section has $h^1(\widehat B)>0$, it is a higher-genus curve. For this to be possible, the bi-section must be irreducible, which can occur due to branching over the points in $\widehat{\mathbb{P}}^1_b$ over which $\widehat Z$ degenerates. The simplest case would be that the bi-section is topologically a $T^2$ with one of the one-cycles of the $T^2$ given by the path $\widehat \gamma$. For definiteness, we will make this additional assumption from now on.
 Since the monodromy around $\widehat \gamma_{1,2}$ is  trivial, also the monodromy around $\widehat \gamma$ is trivial such that $\widehat{\mathcal{E}}\times \widehat \gamma$ gives a well-defined 3-cycle of $T^3$ topology. The situation is depicted schematically in Figure \ref{fig:double_cover}. 

The action of mirror symmetry can now be split into performing one T-duality along $\widehat \gamma$ and two T-dualities along $\widehat \cE$. By applying the SYZ picture to $\widehat Z$, the latter two T-dualities amount to mirror symmetry for the K3 surfaces. In addition, since $\widehat B$ is a genus-one curve, T-duality along $\widehat \gamma$ maps this torus to its dual, $B$. Hence the mirror $V$ has a similar fibration structure as $\widehat V$, i.e.~it is a K3-fibration with generic fibre $Z$. This is consistent with the Doran--Harder--Thomson conjecture~\cite{doran2016mirrorsymmetrytyurindegenerations} that a Calabi--Yau threefold with a Tyurin degeneration is mirror to a K3-fibration, which we are applying to the Tyurin degeneration of $\widehat V$. The base of this fibration is a genus-zero curve, $\mathbb{P}^1_b$, which has a double cover $B$ with the topology of a torus. 
We denote by $\gamma\in H_1(B)$ the T-dual of $\widehat \gamma$.

Consider now the large base limit $\text{vol}(\widehat{\mathbb{P}}^1_b)\to \infty$ for the threefold $\widehat V$. In this limit $\widehat \gamma\in H_1(\widehat B)$ grows to infinite length. Hence, in the mirror dual of the large base limit, the dual cycle $\gamma\in H_1(B)$ pinches. Let us further assume that the normal bundle of $B$ inside $V$ is trivial, $\cN_{B|V}=\mathcal{O}(0)\oplus \mathcal{O}(0)$. Then, apart from the fibration with generic fiber $Z$, the threefold $V$ allows for a second, incompactible genus-one fibration with generic fiber $B$.  An example of such a geometry can be found in~\cite{FierroCota:2023bsp}. In this case, the mirror dual of the large base limit for $\widehat V$ corresponds to the limit in which the generic fibre of the genus-one fibration of $V$ degenerates.\\

With this preparation, we can now study the action of mirror symmetry on the NS5-brane string $\widehat{\mathtt{S}}_{\widehat{Z}}$ by applying T-duality to the cycles $\widehat \cE$ and $\widehat \gamma$. Recall that the action of T-duality for NS5-branes differs depending on whether the $S^1$ along which we T-dualise is parallel or normal to the NS5-brane: If the $S^1$ is parallel to the NS5-brane worldvolume, the NS5-brane is not affected by it. For an $S^1$ normal to the NS5-brane, T-duality instead turns the NS5-brane into a KK-monopole for which the T-dual $S^1$ shrinks at the location of the (former) NS5-brane. 

Given the split of mirror symmetry for $\widehat{V}$ into mirror symmetry for $\widehat{Z}$ and a T-duality along $\widehat \gamma$, the NS5-brane wrapping $\widehat{Z}$ is not affected by the application of K3-mirror symmetry to $\widehat{Z}$. Instead, T-duality along $\widehat \gamma$ turns the NS5-brane into a KK-monopole. Four directions of the worldvolume theory of the monopole are hence identified with $Z$. Moreover, the cycle $\gamma$ dual to $\widehat \gamma$ takes over the role of the KK-circle and shrinks at the location of the monopole. The monopole geometry is illustrated in Figure \ref{fig:Sen}. At the centre of the monopole, the genus-one fibre of $V$ is hence degenerate. Since $\widehat \g$ is a one-cycle on the double-cover of $\widehat{\mathbb{P}}^1_b$, the worldvolume of the monopole is a double-cover of the base $B_2$ of $V$ viewed as a genus-one fibration. Moreover, at the location of the KK-monopole the discriminant of the genus-one curve $B$ has a double zero. Altogether, we conclude that at the centre of the monopole we obtain a degenerate Calabi--Yau threefold $V_0$ that is a genus-one fibration with a bi-section identified as $Z$. The degeneration amounts to 
an $I_2$ singularity in the generic genus-one fibre. In F-theory language, the centre of the KK-monopole hence realises a Sen-limit \cite{Clingher:2012rg}.\footnote{See \cite{Lee:2021qkx,Lee:2021usk,Alvarez-Garcia:2023qqj,Alvarez-Garcia:2023gdd} for the description of type II limits of elliptic K3 and Calabi-Yau threefolds as Sen-limits.} After resolving the $I_2$ singularity, the fibre $V_0$ over the centre of the monopole splits into the union of two Fano threefolds 
\begin{equation}
    V_0 = X_1 \cup_Z X_2\,,
\end{equation}
intersecting over $Z$.  
 Thus, the mirror dual of the NS5-brane string gives a geometry that realises a degeneration for the Calabi--Yau threefold $V_0$ of Tyurin type. Moreover, the total geometry of the KK-monopole is an example of the geometry $\cV$ associated with the EFT strings discussed in Section~\ref{sec:Worldsheet}. \\

To summarise, we have 
constructed the mirror dual for a configuration of emergent strings 
 obtained by wrapping an NS5-brane on the K3-fibre $\widehat Z$ of a Calabi--Yau threefold $\widehat{V}$,  under the assumption  that $\widehat{V}$ allows for a Tyurin degeneration into two half-Calabi--Yau threefolds which are themselves fibred by $\widehat Z$. For such setups, the mirror dual of the emergent strings corresponds to the Type IIB EFT strings discussed in Section~\ref{sec:Worldsheet} with the special property that the Tyurin degeneration on the Type IIB side is a Sen-limit for a genus-one fibred Calabi--Yau threefold $V$ and the double cover of the base $B_2$ of this fibration is the mirror $Z$ of $\widehat{Z}$.

\section{Beyond Tyurin Degenerations}\label{sec:beyondTyurin}
We have focused, so far, on a special class of type II limits in the complex structure moduli space of Calabi--Yau threefolds corresponding to Tyurin degenerations. In these degenerations, the generic Calabi--Yau threefold degenerates into two Fano threefolds intersecting over a K3 surface. A natural question is whether there are more general kinds of type II degenerations. By similar reasoning as in Section~\ref{sec:mirror_symmetry}, we expect Calabi--Yau threefolds with an Abelian surface fibration to be mirror to threefolds that allow for a type II degeneration in which the threefold splits into two components intersecting over the mirror surface, which is again an Abelian surface. More generally, one might ask whether a type II limit could correspond to a degeneration of the form
\begin{equation}\label{eq:generalTypeII}
    V\to V_0 = \bigcup_{i=1}^n X_i\,,\qquad M_{i,j}=X_i \cap X_j\,, 
\end{equation}
where the manifolds $M_{i,j}$ arising at the intersection of two components $X_i$ and $X_j$ are not necessarily K3 or Abelian surfaces. While we are not aware of an example of a type II degeneration of this kind, we do not know of a geometric argument excluding it either.
 The goal of this section is to study more general type II degenerations and their interplay with the Emergent String Conjecture. In Section \ref{sec:abelian-surface} we discuss type II limits with Abelian surface defects. In Section \ref{sec:mutlipleII} we use the Emergent String Conjecture to constrain the possible geometries as in \eqref{eq:generalTypeII} that can occur in type II limits.

\subsection{Type II Degenerations with Abelian Surfaces}  \label{sec:abelian-surface}
Consider a type II singularity in the complex structure moduli space of a Calabi--Yau threefold $V$ such that at the singularity $V$ degenerates as 
\begin{equation}\label{Abeliandeg}
    V\to V_0 = X_1 \cup_{A} X_2 \,,
\end{equation}
where $A$ is an Abelian surface. As before, we can analyse the physical theory obtained by realising this type II limit at spatial infinity. Since this analysis parallels the discussion in Sections~\ref{sec:BPSstates} and \ref{sec:Worldsheet}, we will be rather brief and only highlight the aspects that differ if we replace the K3 surface $Z$ appearing in Tyurin degenerations by an Abelian surface $A$ in $V_0$. 

Since the algebraic properties of the limiting mixed Hodge structure are insensitive to the geometric details of the degeneration, all aspects of the low-energy theory derived from this and discussed at the end of Section~\ref{sec:EFTstrings} are completely unchanged. In particular, this concerns the mass of BPS particles with charge in $\text{Gr}_2$ and the tension of the EFT string realising the type II limit. By contrast, our arguments in Section~\ref{sec:BPSstates} establishing the existence of a tower of BPS states that become massless in the type II limit make use of the precise geometry of $V_0$, and so do the arguments establishing the EFT string associated with the degeneration as a critical heterotic string. These arguments must be adapted to the case of the Abelian surface to infer the details of the quantum gravity theory in the degeneration limit~\eqref{Abeliandeg}. 

\paragraph{Worldsheet Spectrum of EFT string.} 
To determine the spectrum on the worldsheet of the EFT string realising the type II degeneration of the form \eqref{Abeliandeg}, we proceed as in Section~\ref{sec:Worldsheet} and study the modes that localise on $V_0$. First, the zero modes  $(\mathbf{z}_0, \mathbf{b}
^0, \mathbf{c}^0, \Phi)$ corresponding to the directions transverse to $A$ in $\cV$ as well as those obtained from $B_6$ and $C_6$ are universally present and hence give rise to five left- and five right-moving scalars on the worldsheet. Additional zero modes arise from reducing $C_4$ over elements in $H^2(A)$. For $A$ an Abelian surface, $H^2(A, \mathbb{Z})$ can be identified with the lattice $\Gamma_{3,3}$. In total this gives rise to three left- and three right-moving scalars, making a total of eight left- and eight right-moving scalars. As before the string solution is a BPS solution leading to an $\cN =(0,4)$ superalgebra that is manifest on the string. Therefore, as in Section~\ref{sec:Worldsheet}, the right-moving scalar fields are accompanied by right-moving fermions to form full $\cN=(0,4)$ multiplets. However, unlike for the case discussed in Section~\ref{sec:Worldsheet}, for the degeneration \eqref{Abeliandeg} there are additional left-moving fermions arising from reducing the 10d gravitinos over one-cycles localised on $V_0$. The number of the left-moving fermions arising in this way is counted by $b_1(A)$, which for an Abelian surface gives $b_1(A)=4$ fermionic zero modes per gravitino, hence eight left-moving fermions in total. The spectrum of the EFT string associated with the degeneration~\eqref{Abeliandeg} thus gives rise to a 2d worldsheet theory with central charges 
\begin{equation}
    c_L=c_R=12\,. 
\end{equation}
Notice that, including the additional left-moving fermions, the worldsheet degrees of freedom assemble in complete multiplets of an accidental $\cN=(4,4)$ supersymmetry on the worldsheet. The worldsheet theory can thus be identified with that of a critical Type II string on some background breaking only half of the super-charges of the 10d Type II string theory. To identify the details of the Type II background, we would have to study the worldsheet interactions. We will not perform this analysis here; for our purposes it suffices that from the worldsheet spectrum we can conclude that the resulting EFT string is a critical string as required by the Emergent String Conjecture.  

\paragraph{Tower of BPS states.} 

The excitation spectrum of the critical Type II string emergent in the limit (\ref{Abeliandeg}) must be accompanied by a tower of particle states whose mass scales like the square root of the string tension. These correspond to KK modes of the 10d string excitations. As in Section~\ref{sec:BPSstates}, such a tower can arise from D3-branes wrapping special Lagrangian 3-cycles dual to elements in $\text{Gr}_2$. Existence of a tower requires a 3-cycle $\Gamma_0$ of the aforementioned type that allows for multi-wrapping. As discussed in Section \ref{subsubsec-BPStower}, the 3-cycle $\Gamma_0$ can be written as an $S^1$ fibration over curves in $A$. For an Abelian surface, all curves $C_0$ of this kind are multi-covers of tori implying that $b_1(\Gamma_0)\geq 3$ such that, indeed, the 3-cycles $\Gamma_0$ can be multi-wrapped. Also in this case, we can apply Conjecture~\ref{conj-modular} to relate the BPS indices for $n\Gamma_0$ to meromorphic modular forms. However, since the string arising in the type II limit has enhanced supersymmetry, all BPS indices vanish due to an exact cancellation between vector- and hypermultiplets. Even though the BPS indices vanish, there are still towers of BPS particles becoming massless at the same rate as the (square-root of the) tension of the emergent string. \\

We thus conclude that also type II limits in which the Calabi--Yau threefold splits into two components intersecting over an Abelian surface give rise to an emergent critical string that is accompanied by a tower of BPS particles, in accordance with the Emergent String Conjecture.

\subsection{Multi-Component Type II Degenerations} \label{sec:mutlipleII}
We now turn to more general degenerate geometries that could in principle occur at type II singularities in the complex structure moduli space of Calabi--Yau threefolds. Whereas in the rest of this paper, we used the geometry of Tyurin degenerations to study the physics of emergent string limits of Type IIB compactifications on Calabi--Yau threefolds, we now reverse the logic and use the Emergent String Conjecture to make predictions about the possible geometries arising at type II degenerations. More precisely, we will show that in the specific context of Type IIB compactifications on Calabi--Yau threefolds, the Emergent String Conjecture implies that in type II degenerations of Calabi--Yau threefolds as in \eqref{eq:generalTypeII} the manifolds $M_{ij}$ must be either all K3 surfaces polarised by the same lattice $\Lambda_{\rm pol}$ or Abelian surfaces sharing the same complex structure.\\

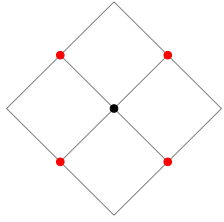
\begin{figure}[t]
    \centering 
\begin{equation*}\begin{aligned}
     \begin{tikzpicture}[scale=1,cm={cos(45),sin(45),-sin(45),cos(45),(15,0)}]
  \draw[step = 1, gray, ultra thin] (0, 0) grid (2, 2);

  \draw[fill,black] (1, 1) circle[radius=0.05];
  \draw[fill,red] (2,1) circle[radius=0.05];
  \draw[fill,red] (1,2) circle[radius=0.05];
    \draw[fill,red] (0,1) circle[radius=0.05];
  \draw[fill,red] (1,0) circle[radius=0.05];
\end{tikzpicture}
\end{aligned}\end{equation*} 
\caption{Hodge-Deligne diamond for a type II degeneration of a K3 surface, with the populated horizontal rows representing the spaces ${\rm Gr}_1$, ${\rm Gr}_2$, ${\rm Gr}_3$. \label{fig:K3-II}}
\end{figure}

As a warm-up, let us consider type II degenerations of K3 surfaces. According to the Kulikov classification \cite{Kulikov1977,Persson1977,Kulikov1981,PerssonPinkham1981}, for a K3 surface $S$ a type II limit in the complex structure moduli space corresponds (up to birational transformation and base change) to a degeneration
\begin{equation}\label{eq:K3degen}
    S\to S_0 = \bigcup_{a=1}^n S_a\,, \qquad C_{a,b}=S_a \cap S_b\,,
\end{equation}
where the end-components $S_1$ and $S_n$ are rational surfaces. The components $S_a$ intersect in such a way that only $C_{a, a+1}$ are non-vanishing. The $C_{a,a+1}$ are elliptic curves sharing the same complex structure. For K3 surfaces this condition on type II degenerations is already manifest at the level of the mixed Hodge structure. To see this, consider the Hodge--Deligne diamond for a K3 surface, which is defined in complete analogy to the Hodge--Deligne diamond for threefolds. For type II degenerations of K3 surfaces the Hodge--Deligne diamond is depicted in Figure \ref{fig:K3-II}.
 Here the spaces represented by red dots are one-dimensional. The graded space $\text{Gr}_1$ can hence be identified with the pure Hodge structure on the middle cohomology of a torus. In analogy to the Tyurin degenerations, this torus can be identified with the curves $C_{a,a+1}$ arising at the intersections of the irreducible components of $S_0$. The space $\text{Gr}_1$ being isomorphic to the Hodge structure of a single torus further explains why all curves $C_{a, a+1}$ have the same complex structure. In the case of K3 surfaces, the limiting mixed Hodge structure associated with a type II limit fixes the possible kind of degenerate geometries of the form \eqref{eq:K3degen} that can arise in these limits.  \\

We now return to general type II limits for Calabi--Yau threefolds. Based on our discussion at the end of Section~\ref{sec:EFTstrings}, from the perspective of low-energy theory general type II limits are expected to correspond to emergent string limits.  The Hodge--Deligne diamond for such limits is depicted in Figure~\ref{fig:II}. As reviewed in Section~\ref{subsec:Algprops}, the Calabi--Yau condition for the underlying threefold $V$ constrains $i^{2,0}=i^{0,2}=1$. Therefore, in a generic type II limit we obtain the mixed Hodge structure for a K\"ahler surface with a unique holomorphic $(2,0)$-form, which can either be a K3 or an Abelian surface. If we consider a semi-stable degeneration of the kind in \eqref{eq:generalTypeII} this implies that at least one of the $M_{i,j}$ is a K3 or Abelian surface with a pure Hodge structure contained in $\text{Gr}_2$. If there are multiple $M_{i,j}$ of this kind, they all have to be describable by the same Hodge structure --- otherwise $i^{2,0}$ would be larger than $1$ because each surface would contribute an independent $(2,0)$ form.
 Therefore these $M_{i,j}$ are either all Abelian surfaces with the same complex structure or all K3 surfaces that are polarised by the same lattice $\Lambda_{\rm pol}$. However, unlike for type II Kulikov models, from the algebraic perspective this does not yet exclude the existence of further non-zero $M_{i,j}$ corresponding to Fano surfaces with $h^{2,0}=h^{0,2}=0$. The reason is that the Hodge theory does not constrain the dimension of $I^{1,1}$, i.e.~the dimension of the central red dot in Figure~\ref{fig:II}, which is the free parameter $b = {i}^{1,1}$ characterising the II$_b$ degeneration.

We now argue that such a geometry would be in tension with the Emergent String Conjecture. Assume that we can realise a type II degeneration as in \eqref{eq:generalTypeII} for which at least one of the $M_{i,j}$ is a Fano surface, denoted as $M_{i_1,j_1}$, that arises in addition to the Abelian or K3 surface $M_{i_0,j_0}$.
The arguments about the tension of the EFT string presented in Section~\ref{sec:EFTstrings} continue to apply in such a setting since they only depend on the growth of the Hodge norms and the properties of the K\"ahler potential in the type II limit. Both are determined entirely by the algebraic properties of the degeneration and hence are not sensitive to the details of the geometry. Moreover, our arguments in Section~\ref{sec:BPSstates} about the existence of towers of BPS states were based on the \emph{local} geometry of $V_0$ near the K3 surface. By zooming in to the vicinity of $M_{i_0,j_0}$ inside $V_0$ we can establish the existence of 3-cycles that can be multi-wrapped, implying the existence of towers of BPS particles. 
 Similar to the string tension, the mass of the tower of BPS states (once established in the local geometry) only depends on the algebraic properties. Hence, irrespective of the details of the global geometry we find an EFT string whose tension goes to zero in the type II limit at the same rate as the lightest tower of states.\footnote{Notice that we did not exclude the possibility that in the presence of an additional Fano surface $M_{a_1,b_1}$ arising at the type II degeneration there is a tower of states with exponentially small mass in $s^k$. However, if this was the case, this by itself would be in contradiction with the Emergent String Conjecture. The reason is the following: Since these states correspond to elements in $\text{Gr}_2$, they are exclusively charged under U(1)s that become weakly coupled in the type II limit. Therefore, these are weakly coupled states which are subject to the constraints of the Emergent String Conjecture and hence either have to be KK states or excitations of a critical string. The latter case is excluded because there is no string with tension that vanishes exponentially fast in $s^k$. On the other hand, from dimensional reduction it is clear that the mass of a KK-tower decays exponentially in the moduli space distance and hence polynomially in $s^k$. Therefore a tower of weakly coupled states with mass decaying exponentially fast in $s^k$ is by itself in contradiction with the Emergent String Conjecture.}

 The details of the global geometry, however, do become relevant since the Emergent String Conjecture requires the emergent string to be a \textit{critical} string~\cite{Lee:2019oct}, which is not ensured by the algebraic properties of the type II degenerations alone. 

To infer the constraints on the geometry from demanding criticality of the emergent string, let us consider the worldsheet theory on the EFT string arising in a hypothetical type II degeneration in which an additional Fano manifold $M_{i_1,j_1}$ arises. As in Section~\ref{sec:Worldsheet}, we can determine the worldsheet spectrum of the string by reducing $C_4$ over 2-forms localised on the central fibre $V_0$,
but we now have to take into account all localised 2-forms including those receiving contributions from $M_{i_1,j_1}$. The rationale behind this is that one could blow down all components of the geometry except its end components $X_1$ and $X_2$, making all  $M_{i,j}$ coincide at the intersection $X_1 \cap X_2$. 
 Such a degeneration is not semi-stable, and to remedy this, one can blow the intersection locus back up.
 The blow-up is therefore an auxiliary procedure
 that artificially splits the localised degrees of freedom into various components. In counting the string modes we must therefore treat all components $M_{i,j}$ as contributing to one physical string that becomes tensionless in the degenerating geometry, rather than to several individual tensionless strings, one per $M_{i,j}$.

Since at least one of the $M_{i_0,j_0}$ has to be a K3 surface, we get $\text{dim}\,H^2(M_{i_0,j_0}) =22$ scalar fields, out of which three are right-moving and 19 are left-moving, or in the case of $M_{i_0,j_0}$ being an Abelian surface $\text{dim}\,H^2(M_{a_0,b_0})=6$ scalar fields (three left- and three right-moving) and $2b_1(M_{i_0,j_0})=8$ left-moving fermions. If there was another, inequivalent surface $M_{i_1,j_1}$  additional localised zero modes from $H^2(M_{i_1,j_1})$ would contribute to the worldsheet spectrum of the string. Furthermore, also upon reduction of $B_6$ and $C_6$ along  $H^4(M_{i_1,j_1})\neq 0$ we would obtain additional massless worldsheet degrees of freedom. Hence, also taking into account the additional zero modes discussed in Section~\ref{sec:Worldsheet} gives a string worldsheet with right-moving central charge
\begin{equation}
   c_R >12\,.
\end{equation}
Since the string worldsheet preserves $\cN =(0,4)$ supersymmetry, this string cannot be a critical string.

To summarise, the only possibility for $V_0$ to split in multiple components is if all $M_{i,j}$ in \eqref{eq:generalTypeII} correspond to the same manifold, i.e.~are either all K3 surfaces with fixed polarization lattice and complex structure or are all Abelian surfaces with the same complex structure. The reason is that in this case, the zero modes arising from the  additional components $M_{i_1,j_1}$ are just copies of the original zero modes associated with $M_{i_0,j_0}$ and do not introduce new, independent degrees of freedom.

We thus conclude that the Emergent String Conjecture constrains the possible geometries arising in type II degeneration limits in the complex structure moduli space of Calabi--Yau threefolds in the following way:

\begin{conjecture}\label{conj1}
    Let $V_u$ be a one-parameter family of Calabi--Yau threefolds describable by a fourfold $\cV \to \mathbf{D}$ over the disk $\mathbf{D}=\{u\in \mathbb{C}\big| |u|<1\}$. Suppose the central fibre $V_0$ realises a type II$_b$ degeneration in the complex structure moduli space of $V$. Then upon birational transformations of $\cV$ and base change $u\mapsto u^k$, the central fibre $V_0$ can be brought into the semi-stable form 
    \begin{equation}
        V_0 = \bigcup_{i=1}^n X_i\,,
    \end{equation}
    for which the non-vanishing double surfaces $M_{i,j}=X_i \cap X_j$ are either all Abelian with the same complex structure or all K3 surfaces that are polarised by the same lattice $\Lambda_{\rm pol}$ of rank $(1,19-b)$. 
    In particular, the parameter $b$ characterzing the II$_b$ degeneration of the family is bounded as 
    \begin{equation}
    0 \leq b \leq 19 \,. 
    \end{equation}

\end{conjecture}

Conjecture~\ref{conj1} can be viewed as a corollary of the Emergent String Conjecture in Type IIB compactifications on Calabi--Yau threefolds. The authors are not aware of a similar statement in the mathematics literature. It would be extremely interesting to independently confirm Conjecture~\ref{conj1} from a purely geometric point of view. 
\section{Discussion and Conclusions}\label{sec:conclusions}
In this work, we have successfully tested the Emergent String Conjecture~\cite{Lee:2019oct} in the vector multiplet moduli space, $\cM_{\rm V}^{\rm IIB}$, of Calabi--Yau compactifications of Type IIB string theory. This setting is special since it lacks higher-dimensional branes that could give rise to BPS strings coupling to the vector multiplets. To establish the existence of emergent strings, we have therefore concentrated on purely geometric string solutions to the effective 4d $\cN=2$ supergravity action that become tensionless in certain infinite distance limits in $\cM_{\rm V}^{\rm IIB}$. 

In general, emergent string limits can be studied at two different levels: First, one can focus on the low-energy effective action and study the behaviour of its gauge couplings, central charges, higher-derivative couplings, or other IR data in specific limits in the moduli space. In the present setup, the couplings of the low-energy effective theory can be reliably described using the techniques of asymptotic Hodge theory \cite{Grimm:2018ohb,Grimm:2018cpv,Grimm:2019wtx,Gendler:2020dfp,Bastian:2020egp,Calderon-Infante:2020dhm,Grimm:2021ikg,Palti:2021ubp,Grimm:2021vpn,Grimm:2022xmj,Bastian:2023shf}. 
  Taking this algebraic perspective one can also compute the tension of certain BPS string solutions to the 4d $\cN=2$ EFT, as discussed in~\cite{Lanza:2020qmt,Lanza:2021udy}. In this way it has been demonstrated that, from a low-energy perspective, general type II limits in the complex structure moduli space qualify as candidates for emergent string limits; in particular they feature a string becoming tensionless at the same rate as the mass square of a potential tower of BPS particles.

However, to actually demonstrate that these limits are emergent string limits in the sense of~\cite{Lee:2019oct} the existence of \emph{some} tensionless string is not sufficient. Rather, the following two points must be shown:
\begin{enumerate}
     \item The BPS string of~\cite{Lanza:2020qmt,Lanza:2021udy} that becomes tensionless at type II degenerations is a   \emph{critical} string. 
    \item There indeed exists a tower of particle states whose mass scales as the square root of this emergent critical string tension in the type II limit and which can be interpreted as an accompanying KK tower. 
   
\end{enumerate}
This cannot be inferred from the data of the IR theory, but requires non-trivial information about the full UV theory of quantum gravity. 
  In fact, that the low-energy effective action is lacking the information to establish the criticality of the string appears to be a general challenge for testing the Emergent String Conjecture. For example, bottom-up approaches
  from consistency of the low-energy effective gravitational action have so far at best established the existence of a light tower of states that \emph{resembles} the excitation tower of a perturbative, critical string~\cite{Basile:2023blg,Bedroya:2024ubj,Kaufmann:2024gqo}. By contrast, a bottom-up argument that identifies this tower of states as arising from a critical string is yet to be found. Let us stress that we do not claim that deriving the criticality of the string from bottom-up considerations is impossible. Rather, in the setups considered in this work, additional non-trivial UV information was key to argue for the existence of a tower of states corresponding to either BPS states or a critical string.

The UV information that allowed us to address the two points above is encoded in the geometric details of the asymptotic degeneration of the Calabi--Yau threefold.   The main part of our geometric analysis focused on Tyurin degenerations, in which the Calabi--Yau threefold splits into a union of two (quasi-)Fano threefolds intersecting over a K3 surface. To establish the criticality of the string as done in Section \ref{sec:Worldsheet}, the appearance of the K3 surface was crucial:  Its topology fixes the central charge on the worldsheet of the emergent string to the critical value $(c_L,c_R)=(24,12)$ in light-cone gauge. The geometry of the Tyurin degeneration therefore establishes the emergent string as a heterotic string. Furthermore, the concrete geometry at the Tyurin degeneration enabled us in Section \ref{sec:BPSstates} to identify infinite families of special Lagrangian 3-cycles that give rise to a tower of light BPS states when wrapped by D3-branes. Interestingly, this tower is not generated by successive application of the monodromy around the singular divisor in $\cM_{\rm V}^{\rm IIB}$. For this reason, the origin of the tower in the type II limits is very different from those towers identified in~\cite{Grimm:2018ohb,Grimm:2018cpv} using a monodromy argument. Moreover, even though there can exist states whose mass vanishes at a parametrically faster rate than the mass of the critical string excitations, we argued that there is no \emph{tower} of such states. This implies that the leading tower of light BPS states is indeed the one whose mass scales like the mass of string excitations. To our knowledge, at least in the context of the Distance Conjecture, our argument is the first one that identifies the leading tower of light BPS states directly in $\cM_{\rm V}^{\rm IIB}$.

 In fact, the light BPS states and the string are closely related: As we argue in Section~\ref{subsec:modular}, the BPS invariants associated with these states count certain winding and KK-momentum states of the heterotic string. Based on experience in other, higher-dimensional setups we then motivated Conjecture~\ref{conj-modular} stating that the generating function for the BPS invariants of the tower of light BPS states is a meromorphic (at worst mock-)modular form and that the BPS invariants for the special Lagrangian 3-cycles are given by certain invariants of the K3 arising at the Tyurin degeneration. It would be extremely interesting to develop techniques to test Conjecture~\ref{conj-modular} directly in the topological A-model. 

Even though the criticality of the string ensures the existence of an infinite tower of states in its tensionless limit, its excitation spectrum depends on the details of the worldsheet theory. This is particularly relevant in the context of the Weak Gravity Conjecture~\cite{ArkaniHamed:2006dz}, for which string excitations play a significant role in weak coupling limits~\cite{Lee:2018urn,Lee:2018spm,Lee:2019jan,Cota:2020zse,Klaewer:2020lfg,Heidenreich:2021yda,Cota:2022yjw,Cota:2022maf,FierroCota:2023bsp,Heidenreich:2024dmr}.   To determine the spectrum of charged string excitations, we have to know the current algebras realised on the string worldsheet. In Section \ref{sec:Worldsheet} we have determined the global worldsheet algebras by studying the interactions on the string. 
      As we have seen,  the free modes are associated with elements in the transcendental lattice, $\Lambda_{\rm trans}$, of the K3 arising at the Tyurin degenerations. The resulting global algebra on the string worldsheet gives rise precisely to the gauge theory in the bulk that becomes weakly coupled in the emergent string limit, as expected. Based on the study of the kinetic terms for the scalar modes on the string, for $4\leq \text{rk} \,\Lambda_{\rm trans}\leq 20$ we have identified the string as the heterotic string on ${\rm K3}_{\rm het}\times T^2$, with a perturbative heterotic gauge algebra of rank $\text{rk}\,\Lambda_{\rm trans}$. In the other cases, we have indicated a possible realisation of the heterotic duality, but due to left-/right-moving asymmetry, the exact details of these models remain to be understood further. 

Our analysis has, however, not determined the exact interactions on the string worldsheet. From the dual heterotic perspective, it remains to identify the gauge bundle for the perturbative gauge group. This would require completing our analysis of the kinetic terms for the zero modes on the string to a computation of the full worldsheet theory. Such an analysis would enable us to elucidate Type II/heterotic duality beyond the known regimes via the framework developed in this paper.
 Detailed knowledge of the interactions is  also required to study the relation between the emergent strings in Type IIB and those in Type IIA compactifications on Calabi--Yau threefolds identified in~\cite{Lee:2019oct}. As we have shown in Section~\ref{sec:mirror_symmetry}, for a perturbative heterotic dual (without spacetime filling NS5-branes) the emergent strings obtained from wrapped NS5-branes in Type IIA are indeed equivalent to those encountered in Type IIB Tyurin degenerations
if both string theories are compactified on a mirror pair of Calabi--Yau threefolds. 
Moreover, by directly applying mirror symmetry and under certain assumptions detailed  in Section~\ref{subsec:directomirror}, we have mapped the NS5-brane emergent strings to the Type IIB BPS string solutions. It would be exciting to further study the action of mirror symmetry in more general classes of examples. 
 Indeed, based on the Emergent String Conjecture and the findings of this paper, it is tempting to speculate that the mirror geometry, for \emph{all} choices of K3-fibred threefolds, admits a Tyurin degeneration or a generalisation as envisioned in Conjecture~\ref{conj1}, thereby reversing the Doran--Harder--Thompson conjecture \cite{doran2016mirrorsymmetrytyurindegenerations}.
 This particularly applies to situations in which the heterotic dual theory has spacetime filling NS5-branes. 

Another exciting avenue, alluded to already above, is a detailed exploration of the exact worldsheet theories of the string arising in Tyurin degenerations with $\text{rk}(\Lambda_{\rm trans})\in\{2,21\}$. Such geometries realise the minimal and maximal possible value for the parameter $b$ characterising a type II$_b$ degeneration. Since these models do not have a geometric Type IIA mirror dual, the exact worldsheet theory of the emergent heterotic string cannot be inferred even if mirror symmetry is applied. Nonetheless, the framework developed in this paper enables the study of the duality between Type IIB and heterotic string theory in these regimes. This especially applies to type II degenerations arising at the intersection of multiple type II divisors in $\cM_{\rm V}^{\rm IIB}$, which cannot occur in the large volume/large complex structure regime~\cite{Lee:2019oct,Grimm:2019bey}. As we have argued, even in such regimes the duality frame described by the multiple strings becoming light at the same rate is \emph{unique}. The details of the perturbative heterotic dual corresponding to such higher-codimension singularities can be explored using the approach developed in this work. Since they arise outside the applicability of standard heterotic/Type II duality, determining their perturbative description via the string solutions analysed in this paper promises new interesting insights. 
 This may shed novel light on the landscape of perturbative string theories in four dimensions beyond the well-studied geometric compactifications of the heterotic or Type IIA string. 

Since the algebraic/IR perspective suggests that \emph{any} type II limit gives rise to an emergent string limit, the geometry must ensure that the associated string is critical even if the type II limit is not realized as a Tyurin degeneration. In Section~\ref{sec:beyondTyurin}, we have discussed the constraints on the possible geometries arising at type II singularities imposed by the Emergent String Conjecture. The results are summarised in Conjecture~\ref{conj1}:  More general multi-component degenerations are possible at type II singularities as long as all components intersect either all over K3 surfaces polarised by the same lattice or all over Abelian surfaces with identical complex structure. The statement of Conjecture~\ref{conj1} can be viewed as the analogue of the constraints on the possible geometries arising at type II Kulikov degenerations of K3 surfaces~\cite{Kulikov1977,Persson1977,Kulikov1981,PerssonPinkham1981}. We are not aware of a similar criterion derived just from geometric considerations of Calabi--Yau threefolds, and it would be intriguing to see whether Conjecture~\ref{conj1}  holds true from a geometric perspective and, if not, how to reconcile this with the Emergent String Conjecture. Conjecture~\ref{conj1} further gives a  new constraint on the possible type II$_b$ singularities in the complex structure moduli space of Calabi--Yau threefolds: From the pure algebraic perspective, the parameter $b$ is bounded to be $b\leq h^{2,1}-1$. However, from our discussion of the possible geometries arising in type II limits, we have concluded that the Emergent String Conjecture imposes a bound $b\leq 19$, which for $h^{2,1}\geq 20$ is a new, stronger constraint on which limiting mixed Hodge structures can have a geometric realisation on Calabi--Yau threefolds. 
 This clearly illustrates that the algebraic perspective on type II limits 
 must be complemented by geometric insights
 to capture the full physics of emergent string limits in $\cM_{\rm V}^{\rm IIB}$, and quite generally represents 
 another example of the fruitful application of quantum gravity ideas to geometry.

\subsubsection*{Acknowledgements}
We thank Charles Doran, Thomas Grimm, Amir Kashani-Poor, Albrecht Klemm, Seung-Joo Lee, Boris Pioline, and Jörg Teschner for discussions. 
The work of B.F.~is supported by Deutsche Forschungsgemeinschaft (DFG, German Research Foundation) under Germany’s Excellence Strategy EXC 2181/1 - 390900948 (the Heidelberg STRUCTURES Excellence Cluster).
The work of J.M., T.W., and M.W.~is supported in part by Deutsche Forschungsgemeinschaft under Germany’s Excellence Strategy EXC 2121 Quantum Universe 390833306, by Deutsche Forschungsgemeinschaft through a German-Israeli Project Cooperation (DIP) grant “Holography and the Swampland” and by Deutsche Forschungsgemeinschaft through the Collaborative Research Center 1624 “Higher Structures, Moduli Spaces and Integrability.”

\appendix

\section{Limiting mixed Hodge Structures}\label{app:hodgestr}

In this section we review some background material on limiting mixed Hodge structures. 

\subsubsection*{Weight filtrations and graded spaces}
Let us denote by $T_i$ the action of the local monodromy group on $H^3(V,\mathbb{C})$ upon encircling the divisor $\Delta_i$, i.e.~sending $u^i\mapsto e^{2\pi {\rm i}}u^i$. After possibly performing a base change, we may assume that each $T_i$ is unipotent \cite{Landman:1973,schmid}, and write
\begin{equation}
    T_i = e^{N_i}\,,
\end{equation}
where the $N_i$ are nilpotent log-monodromy operators. These operators give rise to a number of filtrations which are crucial in order to uncover the limiting mixed Hodge structures associated with the various singular loci $\Delta_{k_1\cdots k_r}$ introduced in \eqref{eq:intDelta}. To be precise, let us fix a limit $u^{k_1},\ldots, u^{k_r}\to 0$ (keeping the remaining moduli away from the limit), and denote
\begin{equation}
    N = N_{k_1}+\cdots +N_{k_r}\,.
\end{equation}
Then the so-called monodromy weight filtration $W_\ell(N)$ associated with $N$ is the unique increasing filtration of rational vector spaces
\begin{equation}
    W_{-1}:=0\subset W_0\subset \cdots \subset W_{6}=H^3(V,\mathbb{C})\,,
\end{equation}
such that the following two conditions are satisfied\footnote{Although we will not make use of it directly, let us record the explicit formula of the weight filtration in terms of images and kernels of $N$
\begin{equation}
    W_{\ell+3}(N) =\sum_{j\geq\mathrm{max}{(-1,\ell)}}\mathrm{ker}\,N^{j+1}\cap \mathrm{im}\,N^{j-\ell}\,. 
\end{equation}}
\begin{enumerate}
    \item $N W_\ell\subset W_{\ell-2}$, for all $\ell=0,\ldots, 6$.
    \item $N^j:\mathrm{Gr}_{3+j}\to\mathrm{Gr}_{3-j}$ is a linear isomorphism, where
    \begin{equation}
    \label{eq:Gr}
        \mathrm{Gr}_\ell:=W_\ell/W_{\ell-1}\,,
    \end{equation}
    are the so-called \textit{graded spaces}.
\end{enumerate}
One of the central results of asymptotic Hodge theory is that the grades spaces $\mathrm{Gr}_\ell$ in fact admit a (pure) Hodge structure of weight $\ell$, i.e.~a decomposition
\begin{equation}
    \mathrm{Gr}_{\ell} = \bigoplus_{p+q=\ell}\mathrm{Gr}_\ell^{p,q}\,,\qquad \overline{\mathrm{Gr}_\ell^{p,q}} = \mathrm{Gr}_\ell^{q,p}\,.
\end{equation}
The intuitive picture is as follows. Even though the Hodge structure on $H^3(V,\mathbb{C})$ will degenerate upon approaching the singularity $\Delta_{k_1\cdots k_r}$, the weight filtration induced by the monodromy allows us to separate the elements in $H^3(V,\mathbb{C})$ into the spaces $\mathrm{Gr}_\ell$ which, individually, \textit{do} admit a Hodge structure. However, due to the action of the monodromy these spaces are non-trivially mixed from the point of view of $H^3(V,\mathbb{C})$. For this reason, the resulting structure is referred to as a (limiting) \textit{mixed Hodge structure}.

\subsubsection*{Hodge--Deligne diamonds}
The properties of such limiting mixed Hodge structures are illuminated by using the so-called Deligne splitting. This is simply a decomposition of the middle cohomology $H^3(V,\mathbb{C})$ into
\begin{equation}
    H^3(V,\mathbb{C}) = \bigoplus_{0\leq p,q\leq 3} I^{p,q}(\Delta_{k_1\cdots k_r})\,,
\end{equation}
such that the various $(p,q)$-components satisfy a certain relation under complex conjugation.\footnote{The precise relation is
\begin{equation}
    \overline{I^{p,q}} = I^{q,p}\quad \mathrm{mod}\quad \bigoplus_{r<q,s<p}I^{r,s}\,.
\end{equation}}
Again, let us emphasise that this decomposition can be different for each singularity $\Delta_{k_1\cdots k_r}$, such that one obtains in total $2^{h^{2,1}}-1$ different limiting mixed Hodge structures, within the local patch $u^{1},\ldots, u^{h^{2,1}}=0$ in $\mathcal{M}_{\mathrm{V}}$ we are currently considering. The relation between the weight filtration $W_\ell$ and the Deligne splitting $I^{p,q}$ is given by
\begin{equation}
    W_\ell = \bigoplus_{p+q\leq\ell}I^{p,q}\,.
\end{equation}
Furthermore, the graded spaces can be recovered from the Deligne splitting using\footnote{Strictly speaking, this identification can only be made after rotating to the so-called $\mathbb{R}$-split Deligne splitting, which satisfies $\overline{I^{p,q}}=I^{q,p}$. As explained in \cite{schmid,CKS}, see also \cite{Grimm:2018cpv,Grimm:2021ckh}, such a rotation can always be performed and will not influence our results. }
\begin{equation}
    \mathrm{Gr}_\ell = \bigoplus_{p+q=\ell}I^{p,q}\,.
\end{equation}
There are a number of conditions on the dimensions $i^{p,q}=\mathrm{dim}\,I^{p,q}$ that greatly restrict the number of possible limiting mixed Hodge structures one can write down. 
\begin{enumerate}
    \item First, there is the following relation 
\begin{equation}
\label{eq:ipq_condition1}
    h^{p,3-p} = \sum_{q=0}^3 i^{p,q}\,,
\end{equation}
where we recall that $h^{p,q}$ denote the Hodge numbers of the underlying Calabi--Yau threefold $V$. In particular, since $h^{3,0}=1$, this relation implies that only one of $i^{3,q}$ can be non-zero, leading to the four different possibilities I through IV outlined in Section \ref{subsec:Algprops}.
\item  Second, there are the following symmetry relations 
\begin{equation}
\label{eq:ipq_condition2}
    i^{p,q}=i^{q,p}=i^{3-q,3-p}\,,
\end{equation}
which, together with \eqref{eq:ipq_condition1}, reduce the number of independent $i^{p,q}$ down to only a single number, which we choose to be $i^{2,2}$. 
\item Third, there is one final constraint 
\begin{equation}
\label{eq:ipq_condition3}
    i^{p-1,q-1}\leq i^{p,q}\,,\quad \text{for $p+q<3$}\,.
\end{equation}
\end{enumerate}
In terms of the Hodge--Deligne diamond introduced in Section \ref{subsec:Algprops}, these conditions have the following interpretation. The first condition states that sum of the dimensions along the various diagonals have to add up to $h^{p,3-p}$. The second condition simply implies that the diamond is symmetric under reflections about the vertical and horizontal axes. Finally, the third condition implies that the dimensions cannot increase when moving straight downwards from the central horizontal row. The latter condition arises from the fact that the log-monodromy operator $N$ acts as $N I^{p,q}\subset I^{p-1,q-1}$. Importantly, any collection of $i^{p,q}$ satisfying the conditions \eqref{eq:ipq_condition1}--\eqref{eq:ipq_condition3} can be realised as the Hodge--Deligne diamond associated to a (polarised) mixed Hodge structure. Additionally, if two polarised mixed Hodge structures give rise to the same Hodge--Deligne diamond, then they are related by a change of basis \cite{Kerr2017}. This means that the Hodge--Deligne diamond is precisely the right object to classify (limiting) mixed Hodge structures. By imposing the conditions \eqref{eq:ipq_condition1}--\eqref{eq:ipq_condition3} one can straightforwardly enumerate all possible Hodge--Deligne diamonds. There are precisely $4h^{2,1}$ possibilities, which are for example listed in Table 3.1 of \cite{Grimm:2018cpv}. 

\subsubsection*{Enhancement chains}
So far, we have discussed properties of the individual mixed Hodge structures associated with each $\Delta_{k_1\cdots k_r}$ separately. It turns out, however, that the possible combinations of these various mixed Hodge structures are also greatly restricted. One can approach this in two steps. First, one can ask how a given mixed Hodge structure associated with $\Delta_{k_1\cdots k_r}$ changes upon approaching an intersection with another divisor $\Delta_{k_{r+1}}$ by sending $u^{k_{r+1}}\mapsto 0$, for which we write
\begin{equation}
    I^{p,q}(\Delta_{k_1\cdots k_r})\rightarrow I^{p,q}(\Delta_{k_1\cdots k_{r+1}})\,.
\end{equation}
This is referred to as an \textit{enhancement}. By sequentially sending $u^{k_i}\mapsto 0$ for $i=1,\ldots, h^{2,1}$, one obtains an \textit{enhancement chain}
\begin{equation}
    \mathrm{I}_0\stackrel{u^{k_1}\to 0}{\to} I^{p,q}(\Delta_{k_1})\to\cdots \stackrel{u^{k_{n}}\to 0}{\to} I^{p,q}(\Delta_{k_1\cdots k_{n}})\,,
\end{equation}
where we put $n=h^{2,1}$. Importantly, depending on the order in which we approach the various divisors, the resulting enhancement chains can differ. For example, if $h^{2,1}=2$ there are two possible enhancement chains, namely
\begin{align}
    &\mathrm{I}_0\stackrel{u_1\to 0}{\to} I^{p,q}(\Delta_1)\stackrel{u_2\to 0}{\to} I^{p,q}(\Delta_{12})\,,\\
   & \mathrm{I}_0\stackrel{u_2\to 0}{\to} I^{p,q}(\Delta_2)\stackrel{u_1\to 0}{\to} I^{p,q}(\Delta_{21})\,.
\end{align}
Furthermore, a consistency condition that must be imposed is that $I^{p,q}(\Delta_{12})$ is identical to $I^{p,q}(\Delta_{21})$. In other words, the limiting mixed Hodge structure associated to the intersection of two divisors should not depend on how one approaches the intersection. In general, there will be $h^{2,1}!$ potentially different enhancement chains within the local path in $\mathcal{M}_{\mathrm{V}}$ we are currently considering. 

It is important to stress that not just any combination of singularity types can occur in an enhancement chain. Intuitively, by intersecting with another divisor the singularity can only become more ``severe''. This is quantified by so-called enhancement rules which restrict how the various singularity types can enhance. As a basic example, these rules imply that it is impossible to decrease the principal type, i.e.~an enhancement of the form $\mathrm{III}\rightarrow\mathrm{II}$ is impossible. The full list of enhancement rules is presented in Table 3.3 of \cite{Grimm:2018cpv}.

\section{The Mayer--Vietoris exact Sequence}  \label{App-MV}
Let $V_0$ be a smooth manifold, and $X_1,X_2$ open subsets of $V_0$ such that
\begin{equation}
    V_0 = X_1\cup_Z X_2\,.
\end{equation}
Then there is the long exact sequence
\begin{equation} \label{eq:MV-sequence1}
    \cdots \stackrel{\delta_{k-1}}{\to} H^k(V_0)\stackrel{\alpha_k}{\to}H^k(X_1)\oplus H^k(X_2)\stackrel{\beta_k}{\to} H^k(Z)\stackrel{\delta_k}{\to}H^{k+1}(V_0)\to\cdots\,.
\end{equation}
For such an exact sequence, we have
\begin{equation}
    {\rm im}(\delta_{k-1}) = {\rm ker}(\alpha_k) \,, \qquad {\rm im}(\alpha_k) = {\rm ker}(\beta_k) \,, \qquad   {\rm im}(\beta_{k}) = {\rm ker}(\delta_k) \,.
\end{equation}
The various maps are defined as follows:
\begin{itemize}
    \item[$\alpha_k$:] Write $k:X_1\hookrightarrow V_0$ and $l:X_2\hookrightarrow V_0$ for the inclusions of $X_i$ in $V_0$, then
    \begin{align*}
        \alpha_k:\quad H^k(V_0)&\to H^k(X_1)\oplus H^k(X_2)\,,\\
        \omega&\mapsto (k^*\omega, l^*\omega)\,.
    \end{align*}
    \item[$\beta_k$:]Write $i:Z\hookrightarrow X_1$, and $j:Z\hookrightarrow X_2$ for the inclusion of $Z$ in $X_i$, then 
    \begin{align*}
        \beta_k:\quad H^k(X_1)\oplus H^k(X_2)&\to H^k(Z)\,,\\
        (\omega_1,\omega_2)&\mapsto i^*\omega_1-j^*\omega_2\,.
    \end{align*}
    \item[$\delta_k$:] Fix a partition of unity $(\rho_1,\rho_2)$ subordinate to the cover $\{X_1,X_2\}$ of $V_0$, then
    \begin{align*}
        \delta_k:\quad H^k(Z)&\to H^{k+1}(V_0)\,,\\
        [\alpha]&\mapsto [\eta]\,,
    \end{align*}
    where the $(k+1)$-form $\eta$ is defined by
    \begin{equation}
        \eta=\begin{cases}
            \mathrm{d}(\rho_2\,\alpha)\,,&\text{on $X_1$}\,,\\
            -\mathrm{d}(\rho_1\,\alpha)\,, &\text{on $X_2$}\,.
        \end{cases}
    \end{equation}
    In particular, note that $\eta$ is closed, so that it gives rise to a cohomology class $[\eta]$, but it is not (globally) exact. Alternatively, one may write
    \begin{align*}
        \eta&=\rho_1\eta|_{X_1}+\rho_2\eta|_{X_2}\,,\\&=\rho_1\mathrm{d}(\rho_2\,\alpha)-\rho_2\mathrm{d}(\rho_1\,\alpha) \,,\\
        &=(\rho_1\mathrm{d}\rho_2-\rho_2\mathrm{d}\rho_1)\wedge\alpha\,.
    \end{align*}
    In other words, at the level of differential forms, the map $\delta_k$ simply wedges the $k$-form $\alpha$ with an appropriate ``bump'' 1-form to create a globally well-defined closed $(k+1)$-form.
\end{itemize}

\section{Mayer--Vietoris and Clemens--Schmid exact Sequences and Degenerations of Hodge Structures} \label{app_MV+CS}

In this appendix we review the derivation of the two isomorphisms \eqref{eq:G2isos-1} and \eqref{eq:G2isos-2}. To do so, we will first explain in Section \ref{subsec:MHS_MV} how one associates a mixed Hodge structure to $H^3(V_0,\mathbb{C})$ when $V_0$ is the central fibre of a Tyurin degeneration. The technical tool to do so is the Mayer--Vietoris exact sequence, which was reviewed abstractly in Appendix \ref{App-MV}. Subsequently, in Section \ref{subsec:Clemens--Schmid}, we explain how to relate this ``geometric'' mixed Hodge structure to the ``algebraic'' limiting mixed Hodge structure introduced in Appendix \ref{app:hodgestr} and discussed in Section \ref{subsec:Algprops}. Here the main tool is the so-called Clemens--Schmid exact sequence.

\subsection{Mixed Hodge Structures from Mayer--Vietoris}
\label{subsec:MHS_MV}
Because of the technical nature of the construction, we first discuss a simpler case to illustrate the main idea. 

\subsubsection*{Example: }
\begin{figure}[t]
    \centering
    \includegraphics[scale=0.45]{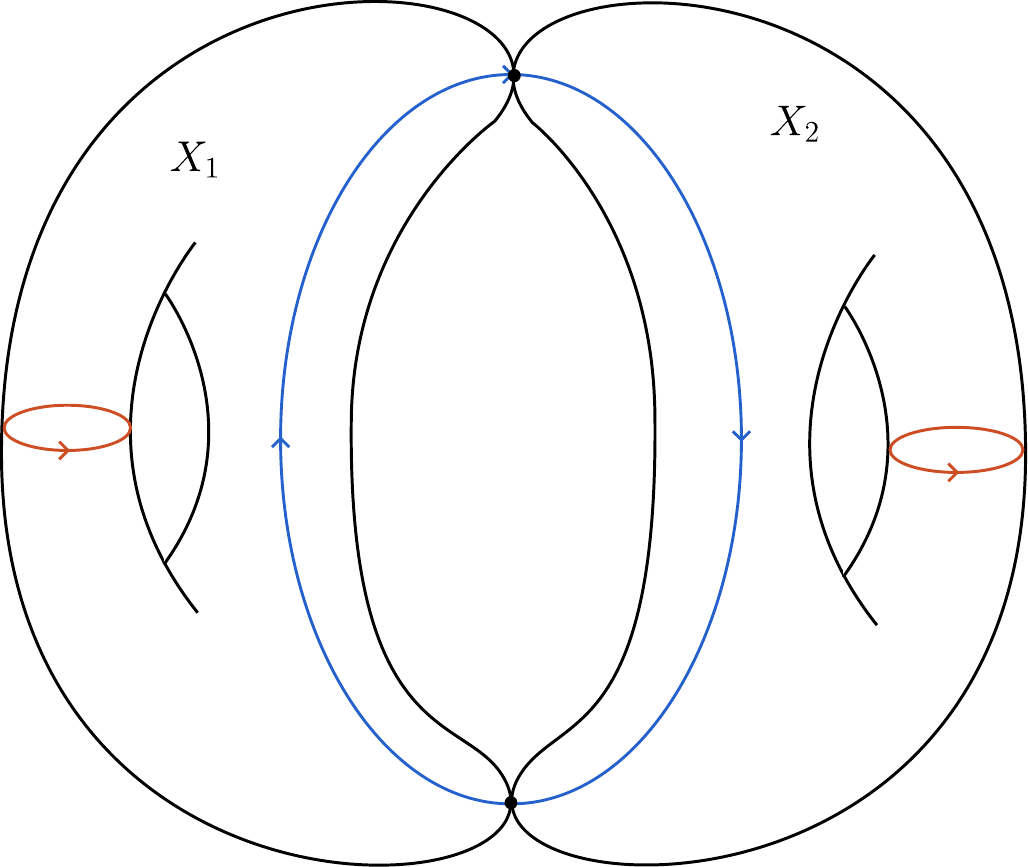}
    \caption{A degeneration of a genus three curve that acquires two nodal singularities. The 1-cycles indicated in red are examples of 1-cycles in $H_1(X_1)\oplus H_1(X_2)$, while the 1-cycle indicated in blue arises from the combination of two 1-chains on $X_1$ and $X_2$ combining at the intersection points to give rise to a 1-cycle on the total space $V_0$.}
    \label{fig:genus3}
\end{figure}

Consider the case where, instead of a Tyurin degeneration, $V_0$ is a genus three curve that acquires two nodal singularities, as illustrated in Figure \ref{fig:genus3}. 
In particular, we have
\begin{equation}
    V_0 = X_1\cup_Z X_2\,,
\end{equation}
where each $X_i$ is a genus one curve, and the intersection $X_1\cap X_2=Z$ is given by two points. Our goal is to characterise $H^1(V_0)$. To do so, let us first discuss the homology $H_1(V_0)$ instead. The picture in Figure \ref{fig:genus3} suggests that there are two kinds of 1-cycles. First, there are the 1-cycles coming from $H_1(X_i)$ separately, some of which are indicated in red. Second, there is the 1-cycle indicated in blue which consists of two 1-chains on $X_1$ and $X_2$ that meet at the intersection points. This intuitive picture is made precise by the Mayer--Vietoris exact sequence for homology, the relevant part of which is given by
\begin{equation}
    0\to H_1(X_1)\oplus H_1(X_2)\to H_1(V_0)\stackrel{\partial_*}{\to} H_0(Z)\to \dots\,.
\end{equation}
Correspondingly, in cohomology we have the exact sequence
\begin{equation}
    0\to H^0(Z)\stackrel{d^*}{\to} H^1(V_0)\stackrel{(k^*,l^*)}{\to} H^1(X_1)\oplus H^1(X_2)\to\cdots\,.
\end{equation}
Again, there are now two kinds of 1-cocycles on $V_0$. First, the ones which restrict to 1-cocycles on $X_1$ or $X_2$. Second, the 1-cocyles which arise from the 0-cocycles on $Z$ via the boundary map $d^*$, which is the dual of the boundary map $\partial_*$. This motivates the following \textit{weight filtration} on $H^1(V_0)$
\begin{equation}
    W_1 = H^1(V_0)\,,\qquad W_0 = \mathrm{im}\,d^*\subset W_1\,.
\end{equation}
As the name suggests, this weight filtration plays a very similar role to the monodromy weight filtration introduced in Appendix \ref{app:hodgestr}. Of particular interest are the corresponding graded spaces, which are given by
\begin{equation}\begin{aligned}
    \mathrm{Gr}_1 &= W_1/W_0\,,\\
    &= H^1(V_0)/\mathrm{im}\,d^*\,,\\
    &\stackrel{\mathrm{(a)}}{=} H^1(V_0)/\mathrm{ker}(k^*,l^*)\,,\\
    &\stackrel{\mathrm{(b)}}{\cong}\mathrm{im}(k^*,l^*)\,,
\end{aligned}\end{equation}
where we used in step (a) exactness of the sequence at $H^1(V_0)$, and in step (b) the first isomorphism theorem. In particular, we see that the graded space $\mathrm{Gr}_1$ can be viewed as those 1-cocycles on $V_0$ which restrict to 1-cocycles on $X_1$ or $X_2$. Similarly, we have
\begin{equation}\begin{aligned}
    \mathrm{Gr}_0 &= W_0\,,\\
    &=\mathrm{im}\,d^*\,,\\
    &\stackrel{\mathrm{(a)}}{\cong}H^0(Z)/\mathrm{ker}\,d^*\,,\\
    &\stackrel{\mathrm{(b)}}{=}H^0(Z)\,,
\end{aligned}\end{equation}
where now we used in step (a) the first isomorphism theorem, and in step (b) exactness of the sequence at $H^0(Z)$, which in particular implies that $d^*$ is injective. In particular, we see that the graded space $\mathrm{Gr}_0$ can be viewed as those 0-cocycles on $Z$ which give rise to 1-cocycles on $V_0$ via the boundary map.

The central statement is now that the spaces $\mathrm{Gr}_1$ and $\mathrm{Gr}_0$ indeed admit a Hodge structure, so that one can indeed assign a mixed Hodge structure to the degenerate variety $V_0$. We will refrain from giving the exact arguments for this statement and instead refer the interested reader to the excellent book \cite{Carlson:2017} and references therein. 

\subsubsection*{Returning to the Tyurin degeneration} With the above example in mind, let us now return to the Tyurin degenerations of Calabi--Yau threefolds. Now we are instead interested in the space $H^3(V_0)$. The relevant part of the Mayer--Vietoris exact sequence is now given by
\begin{equation}
    \cdots\to H^2(X_1)\oplus H^2(X_2)\stackrel{i^*-j^*}{\to} H^2(Z)\stackrel{d^*}{\to} H^3(V_0)\stackrel{(k^*,l^*)}{\to} H^3(X_1)\oplus H^3(X_2)\to 0\,.
\end{equation}
In complete analogy with the previous example, we want to distinguish between the 3-cocycles in $H^3(V_0)$ which restrict to 3-cocycles on $X_1$ or $X_2$, and those 3-cocycles which come from 2-cocycles on $Z$ via the boundary map. As before, this motivates us to define a weight filtration by setting
\begin{equation}
\label{eq:weight-filtration_Tyurin}
    W_3 = H^3(V_0)\,,\qquad W_2 = \mathrm{im}\,d^*\subset H^3(V_0)\,,\qquad W_1=W_0=0\,.
\end{equation}
Then
\begin{equation}\begin{aligned}
    \mathrm{Gr}_3 &= W_3/W_2\,,\\
    &= H^3(V_0)/\mathrm{im}\,d^*\,,\\
    &\stackrel{\mathrm{(a)}}{=}H^3(V_0)/\mathrm{ker}(k^*,l^*)\,,\\
    &\stackrel{\mathrm{(b)}}{\cong}\mathrm{im}(k^*,l^*)\,,\\
    &\stackrel{\mathrm{(c)}}{=}H^3(X_1)\oplus H^3(X_2)\,,
\end{aligned}\end{equation}
where we used in step (a)  exactness of the sequence at $H^3(V_0)$, in step (b) the first isomorphism theorem, and in step (c) exactness of the sequence at $H^3(X_1)\oplus H^3(X_2)$. In words, we see that the space $\mathrm{Gr}_3$ consists precisely of those 3-cocycles on $V_0$ which restrict to 3-cocycles on $X_1$ and $X_2$. Similarly, we have
\begin{equation}\begin{aligned}
    \mathrm{Gr}_2&=W_2/W_1\,,\\
    &=\mathrm{im}\,d^*\,,\\
    &\stackrel{\mathrm{(a)}}{\cong} H^2(Z)/\mathrm{ker}\,d^*\,,\\
    &\stackrel{\mathrm{(b)}}{=}H^2(Z)/\mathrm{im}(i^*-j^*)\,,\\
    &\stackrel{\mathrm{(c)}}{=}H^2(Z)/(\mathrm{im}(i^*)+\mathrm{im}(j^*))\,,
\end{aligned}\end{equation}
where we used in step (a) the first isomorphism theorem, in step (b) exactness of the sequence at $H^2(Z)$, and in step (c) the fact that $\mathrm{im}(i^*-j^*)=\mathrm{im}(i^*)+\mathrm{im}(j^*)$. In words, we see that the space $\mathrm{Gr}_2$ consists of those 3-cocycles on $V_0$ which come from those 2-cocycles on $Z$ that do \textit{not} come from the restriction of 2-cocycles on $X_1$ and $X_2$. 

As in the previous example, the central statement is that this construction again defines a mixed Hodge structure on $H^3(V_0,\mathbb{C})$ of the central fibre $V_0$ of a Tyurin degeneration. Furthermore, from the fact that $W_1=W_0=0$ one immediately reads off that this is a type II$_b$ degeneration. Furthermore, we find that the subindex $b$ precisely counts the number of holomorphic curves in $Z$ which do not come from $X_1$ or $X_2$. 

As an aside, let us mention that the above construction of the weight filtration can be generalised to the case where $V_0$ is instead the union of many varieties $X_1,\ldots, X_r$, by employing a certain spectral sequence that generalises the Mayer--Vietoris exact sequence. See for example Section 4 of \cite{Griffiths:1973} for further details. 

\subsection{The Clemens--Schmid exact Sequence}
\label{subsec:Clemens--Schmid}
Let us briefly recap the state of affairs. In Appendix \ref{app:hodgestr} we have explained that one can associate a limiting mixed Hodge structure with $H^3(V_z)$, where $V_z$ is the generic fibre. In addition to this, in the previous section we have reviewed a construction due to Deligne that allows one to instead associate another mixed Hodge structure with $H^3(\cV)\cong H^3(V_0)$, where $V_0$ is the central fibre. In essence, the Clemens--Schmid sequence gives the relation between these two constructions. In the following, we will follow the exposition in \cite{Morrison:1984}, to which we also refer the reader for more precise statements and proofs. To be explicit, in the current setting the sequence reads\footnote{Here we have already used the fact that $H_5(\cV)=0$, which follows from a Mayer--Vietoris argument, see Lemma 5.2 of \cite{doran2016mirrorsymmetrytyurindegenerations}.}
\begin{equation}
\label{eq:Clemens--Schmid}
    0\to H^3(\cV)\stackrel{\iota^*}{\to}H^3(V_z)\stackrel{N}{\to}H^3(V_z)\to H_3(\cV)\to\cdots\,,
\end{equation}
where $\iota:V_z\hookrightarrow\mathcal{V}$ denotes the inclusion map and $N$ denotes the log-monodromy operator. As an immediate application, exactness of the sequence \eqref{eq:Clemens--Schmid} now implies that all the 3-cocycles on $V_z$ which are invariant under monodromy must come from 3-cocycles in $\cV$. This result is also known as the local invariant cycle theorem. 

Most importantly for our purposes, the Clemens--Schmid exact sequence is in fact an exact sequence of mixed Hodge structures, and hence also provides a relation between the two kinds of weight filtrations and associated graded spaces we have encountered. To see this, note that we can first write \eqref{eq:Clemens--Schmid} as
\begin{equation}
\label{eq:Clemens--Schmid_short}
    0\to H^3(V_0)\stackrel{\iota^*}{\to}H^3(V_z)\cap\mathrm{ker}\,N\stackrel{N}{\to} 0\,,
\end{equation}
where we have used the fact that $H^3(\cV)\cong H^3(V_0)$, see also \cite{Morrison:1984}. Then the fact that the map $\iota^*$ is a strict morphism of Hodge structures of type (0,0) implies a similar sequence for the corresponding weight filtrations\footnote{The map being of type (0,0) simply means that it does not shift the indices of the weight filtrations.}
\begin{equation}
\label{eq:sas_W}
    0\to W_\ell \to W_\ell(N)\cap\mathrm{ker}\,N\to 0\,.
\end{equation}
Let us briefly clarify the notation to avoid confusion. In this relation, the filtration $W_\ell$ denotes the ``geometric'' weight filtration which we constructed in Appendix \ref{subsec:MHS_MV}, see in particular equation \eqref{eq:weight-filtration_Tyurin}. In contrast, the filtration $W_\ell(N)$ denotes the ``algebraic'' weight filtration that is constructed via the log-monodromy operator $N$, as explained in Section \ref{app:hodgestr}. Since the sequence \eqref{eq:sas_W} is very short, we simply find the isomorphism
\begin{equation}
    W_\ell(N)\cap\mathrm{ker}\,N\cong W_\ell\,.
\end{equation}
In other words, the monodromy invariant part of the monodromy weight filtration $W_\ell(N)$ is isomorphic to the geometric weight filtration $W_\ell$ that was computed above. In fact, since $N$ acts as $N W_\ell(N)\subset W_{\ell-2}(N)$, and $W_1(N)=0$, we see that $W_3(N)\subset\mathrm{ker}\,N$, so that
\begin{equation}
    W_{\ell}(N)\cong W_{\ell}\,,\qquad \text{for $\ell\leq 3$}\,.
\end{equation}
Of course, this implies the same relation for the graded spaces, i.e.
\begin{equation}\begin{aligned}
    \mathrm{Gr}_\ell H^3(V_z)&\cong \mathrm{Gr}_\ell H^3(V_0)\,,\qquad \text{for $\ell\leq 3$}\,,\\
    &\cong \begin{cases}
        H^3(X_1)\oplus H^3(X_2)\,, & \ell=3\,,\\
        H^2(Z)/(\mathrm{im}(i^*)+\mathrm{im}(j^*))\,, & \ell=2\,.
    \end{cases}
\end{aligned}\end{equation}
This gives us a full geometric understanding of the lower half of the corresponding Hodge--Deligne diamond depicted in Figure \ref{fig:II}, which is sufficient for our purposes.\footnote{The attentive reader may wonder whether there is also a direct geometric interpretation of the space $\mathrm{Gr}_4$. The answer is yes. In terms of the weight filtration, one can identify $W_4=H^3_{\mathrm{log}}(V_0)$, where the latter denotes the logarithmic de Rham cohomology group associated to $V_0$. The reason for including such logarithmic forms is that the fourfold $\mathcal{V}$ is non-compact. This is explained in detail in Section 5 of \cite{Griffiths:1973}. }

\bibliography{papers_Max}
\bibliographystyle{JHEP}

\end{document}